%% file: ms.tex
\documentclass[]{aa}
\usepackage{graphicx}
\usepackage{subcaption}
\usepackage{mwe}
\usepackage{txfonts}
\usepackage{hyperref}
\usepackage{url}
\usepackage{xcolor}
\hypersetup{
colorlinks,
linkcolor={red!80!black},
citecolor={blue!80!black},
urlcolor={blue!80!black}
}

\def\cm3{cm$^{-3}$}
\def\kms{km~s$^{-1}$}

\def\lsun{L$_{\odot}$}
\def\rsun{R$_{\odot}$}

\def\msun{M$_{\odot}$}

\def\one{\ts {\,\sc i}}
\def\two{\ts {\,\sc ii}}
\def\three{\ts {\,\sc iii}}
\def\four{\ts {\,\sc iv}}
\def\five{\ts {\sc v}}

\def\beq{\begin{equation}}
\def\eeq{\end{equation}}

\def\lesssim{\mathrel{\hbox{\rlap{\hbox{\lower4pt\hbox{$\sim$}}}\hbox{$<$}}}}
\def\gtrsim{\mathrel{\hbox{\rlap{\hbox{\lower4pt\hbox{$\sim$}}}\hbox{$>$}}}}

\def\one{{\,\sc i}}
\def\two{{\,\sc ii}}
\def\three{{\,\sc iii}}
\def\four{{\,\sc iv}}
\def\five{{\sc v}}

\newcommand{\code}[1]{\texttt{#1}}

\newcommand{\voned}{\code{V1D}}

\newcommand{\mesa}{\code{MESA}}
\newcommand{\cmfgen}{\code{CMFGEN}}

\def\ergs{erg\,s$^{-1}$}
\def\foe{10$^{51}$\,erg}

\newcommand{\iso}[2]{\ensuremath{^{#1}\rm{#2}}}

\def\aj{AJ}
\def\pasp{PASP}
\def\pasa{PASA}
\def\apj{ApJ}
\def\apjs{ApJS}
\def\apjl{ApJL}
\def\aap{A\&A}

\def\aaps{A\&AS}
\def\mnras{MNRAS}
\def\nat{Nature}

\def\apss{Astrophysics and Space Science}

\def\nifs{\iso{56}Ni}

\def\cofs{\iso{56}Co}

\def\caiidoub{[Ca\two]\,$\lambda\lambda$\,$7291,\,7323$}
\def\caiitrip{Ca\two\,$\lambda\lambda\,8498-8662$}

\def\oidoub{[O\one]\,$\lambda\lambda$\,$6300,\,6364$}

\begin{document}

   \title{A sequence of Type Ib, IIb, II-L, and II-P supernovae from binary-star progenitors of varying initial separation}
   \titlerunning{Type Ib to II-P from binary progenitors}

\author{
Luc Dessart\inst{\ref{inst1}}
\and
Claudia P. Guti\'errez\inst{\ref{inst2},\ref{inst3}}
\and
Andrea Ercolino\inst{\ref{inst4}}
\and
Harim Jin\inst{\ref{inst4}}
\and
Norbert Langer\inst{\ref{inst4},\ref{inst5}}
}

\institute{
Institut d'Astrophysique de Paris, CNRS-Sorbonne Universit\'e, 98 bis boulevard Arago, F-75014 Paris, France\label{inst1}
\and
Institut d'Estudis Espacials de Catalunya (IEEC), Gran Capit\`a, 2-4,
Edifici Nexus, Desp. 201, E-08034 Barcelona, Spain\label{inst2}
\and
Institute of Space Sciences (ICE, CSIC), Campus UAB, Carrer de Can
Magrans, s/n, E-08193 Barcelona, Spain\label{inst3}
\and
Argelander Institut f\"{u}r Astronomie, Auf dem H\"{u}gel 71, DE-53121 Bonn, Germany\label{inst4}
\and
Max-Planck-Institut f\"{u}r Radioastronomie, Auf dem Hf\"{u}rgel 69, DE-53121 Bonn, Germany\label{inst5}
 }

   \date{Received; accepted}

  \abstract{
Over the last decade, evidence has accumulated that massive stars do not typically evolve in isolation but instead follow a tumultuous journey with a companion star on their way to core collapse. While Roche-lobe overflow appears instrumental for the production of a large fraction of supernovae (SNe) of Type Ib and Ic, variations in the initial orbital period $P_{\rm init}$ of massive interacting binaries may also produce a wide diversity of case B, BC, or C systems, with preSN stars endowed from minute to massive H-rich envelopes. Focusing here on the explosion of the primary, donor star, originally of 12.6\,\msun, we use radiation-hydrodynamics and nonlocal thermodynamic equilibrium time-dependent radiative transfer to document the gas and radiation properties of such SNe, covering from Type Ib, IIb, II-L to II-P. Variations in $P_{\rm init}$ are the root cause behind the wide diversity of our SN light curves, with single-peak, double-peak, fast-declining or plateau-like morphologies in the $V$ band. The different ejecta structures, expansion rates, and relative abundances (e.g., H, He, \nifs) are conducive to much diversity in spectral line shapes (absorption vs emission strength, width) and evolution. We emphasize that H$\alpha$ is a key tracer of these modulations, and that He\one\,7065\,\AA\ is an enduring optical diagnostic for the presence of He. Our grid of simulations fare well against representative SNe Ib, IIb, and II-P SNe, but interaction with circumstellar material, which is ignored in this work, is likely at the origin of the tension between our Type II-L SN models and observations (e.g., of SN\,2006Y). Remaining discrepancies in our model rise time to bolometric maximum call for a proper account of both small-scale and large-scale structures in core-collapse SN ejecta.  Discrepant Type II-P SN models, with a large plateau brightness but small line widths, may be cured by adopting more compact red-supergiant star progenitors.
}

   \maketitle

\section{Introduction}

Over the last several decades, the community has assembled a large dataset of core-collapse supernova (SN) light curves whose characteristic morphology includes single peak, double peak, fast declining or plateau events \citep{anderson_2pl,stritzinger_sesn_18,modjaz_rev_19}.\footnote{These morphologies generally refer to $V$-band and not bolometric light curves, which differ when there is significant UV flux.} Such variations have been understood as arising from different evolutionary channels and in particular single vs  binary star evolution (e.g., \citealt{podsiadlowski_92}; \citealt{yoon_ibc_10}; \citealt{eldridge_sn2_18}; \citealt{schneider_ibc_21}; \citealt{laplace_evol_21}). Roche-lobe overflow (RLOF) allows for huge mass loss rates in stars of modest luminosity and weak winds, opening a channel for the production of Type IIb and Ib  SNe\footnote{Type Ic are unlikely to arise from lower-mass massive stars in binary systems because, once stripped of the H-rich shell through RLOF, the stripping of the progenitor He-rich shell must occur through wind mass loss, which is strong enough only in higher mass Wolf-Rayet stars (see, for example, \citealt{yoon_wr_17}).} from stars of moderate mass, which would have otherwise exploded as Type II-P SNe had they been single. The consideration of binary-star evolution has also become a necessity when investigating core-collapse SN progenitors since a high fraction of massive stars reside in binary systems with an initial orbital separation tight enough for interaction and RLOF at some stage in their life (e.g., see \citealt{sana_bin_12}).

The strict segregation of events in separate SN types hides the complexity inherent to SN light curves which tend to show a continuum in-between morphological types. For example, the exact dichotomy between slow (aka, SNe II-P) and fast decliners (aka, SNe II-L) is hard to define \citep{anderson_2pl} and the wealth of information encoded in spectra is typically reduced to the presence or absence of certain lines or ions. The diversity in spectral morphology is indeed considerable among Type II SNe \citep{gutierrez_pap1_17}, even when considering a single line like H$\alpha$ \citep{gutierrez_ha_14}. This profile morphology is generally considered a feature, but it can provide critical information on the processes controlling the SN radiation characteristics  and in particular in lifting degeneracies \citep{HD19}. Clear examples of this are the detection of short lived Type IIn signatures, which provide robust evidence for ejecta interaction with circumstellar material (CSM) in the early aftermath of shock breakout \citep{yaron_13fs_17}. Type II SNe in which H$\alpha$ develops a strong and broad emission line with weak or no absorption likely interact with CSM, covering from extreme cases like SN\,1998S \citep{leonard_98S_00,fassia_98S_00,fassia_98S_01} to weaker ones like SN\,2013ej \citep{bose_13ej_15,yuan_13ej_16,HD19}.

Similarly, the dichotomy between Type IIb and Type Ib SNe rests upon the residual H-rich envelope that survives at the progenitor surface at the time of core collapse. Depending on the mass transferred to a companion in a binary and the adopted wind mass loss rates, this extended and tenuous envelope may range from as little as 0.001\,\msun\  in tight binaries \citep{yoon_ibc_10} up to several solar masses in the widest binaries that effectively evolve as single stars \citep{yoon_ib_iib_17,gilkis_arcavi_22,ercolino_bin_23}. The exact threshold H mass separating Type IIb from Type Ib SNe is inferred from observations to be around a few 0.01\,\msun\ \citep{hachinger_13_he,gilkis_arcavi_22}, although theoretically, $M$(H) values as low as 0.001\,\msun\ are sufficient to produce a Type IIb SN \citep{dessart_11_wr}. Previous studies have documented the influence of mass loss on a massive star, in particular in terms of H-rich envelope mass at core collapse. \citet{snec} showed the impact of an enhanced red-supergiant (RSG) mass loss on the final preSN star and the corresponding Type II SN light curves, covering from Type II-P to Type II-L SNe. Interaction with CSM lies behind the early-time luminosity boost of Type II SNe \citep{gonzalez_gaitan_2p_15,moriya_13fs_17,morozova_2l_2p_17,yaron_13fs_17,forster_csm_18}. \citet{HD19} presented similar simulations describing both light curves and spectra. They argued that the excess luminosity of fast-declining Type II as well as  the spectral morphologies required interaction --- a low $M$(H) cannot alone explain the shape of fast-declining Type II SNe and their large luminosity. \citet{eldridge_sn2_18} discussed the broad diversity of Type II SN light curves that arise when including interacting binaries rather than single stars alone. Hence, the impact of mass loss on the preSN star, and in particular on the residual H-rich envelope, as well the potential presence of that stripped material in the direct vicinity of the exploding star, play a major role in determining the SN type and properties.

In this work, we present explosion and radiative-transfer calculations for massive stars in a binary system with an initial orbital period varying from 562 to 2818\,d, employing the binary-star evolution models of \citet{ercolino_bin_23}. Although similar work has been done by, for example, \citet{gilkis_arcavi_22}, we document here the gas, photometric, and spectroscopic properties of the resulting SNe, finding that this sample of models cover the whole range from Type Ib, IIb, II-L (or fast-declining Type II)\footnote{Our models with such fast declining light curves are qualitatively similar to Type II-L but typically underluminous at peak relative to events like SN\,1979C \citep{panagia_79c_80}. The terminology should thus be taken loosely rather than strictly.} to Type II-P SNe. In particular, through this sample of progenitors, we reveal how the progressive changes in progenitor H-rich envelope impact the photometric and spectral properties during the photospheric phase.

In the next section, we present the numerical setup for our calculations. We then present the results of our radiative transfer simulations, focusing first on photometric properties (Section~\ref{sect_phot}) and then on spectroscopic properties (Section~\ref{sect_spec}). In Section~\ref{sect_halpha}, we discuss in more detail the properties obtained for H$\alpha$, in particular the evolution of its strength and width in our set of models.  Section~\ref{sect_he} discusses the evolution of He\one\ lines and in particular the properties of He\one\,7065\,\AA, which is found to be present in all models at late times and may thus offer an alternative to test for the presence of He in Type Ic SNe. Section~\ref{sect_obs} then presents an extensive comparison of our model results with prototypical SNe of Type Ib (Section~\ref{sect_ib}), IIb (Section~\ref{sect_iib_11dh} and \ref{sect_iib_93J}), II-L (i.e., fast-declining fast-evolving SNe; Section~\ref{sect_iil}) and II-P (Section~\ref{sect_iip}).

 \section{Numerical approach}
\label{sect_setup}

  Our numerical exploration is a three-step process consisting of stellar evolution models carried out until core collapse with \mesa\ \citep{mesa1,mesa2,mesa3,mesa4,mesa5}, followed by radiation hydrodynamics of the explosion phase until homologous expansion with \voned\ \citep{livne_93,dlw10a,dlw10b}, and finally the remapping of the resulting SN ejecta into the radiative transfer code \cmfgen\ \citep{hm98, HD12} to compute the evolution of the properties of the gas and the radiation (bolometric and multi-band light curves as well as spectra covering from the far-UV to the far-IR).

\input{table_init.tex}

\begin{figure*}
   \centering
    \begin{subfigure}[b]{0.49\textwidth}
       \centering
       \includegraphics[width=\textwidth]{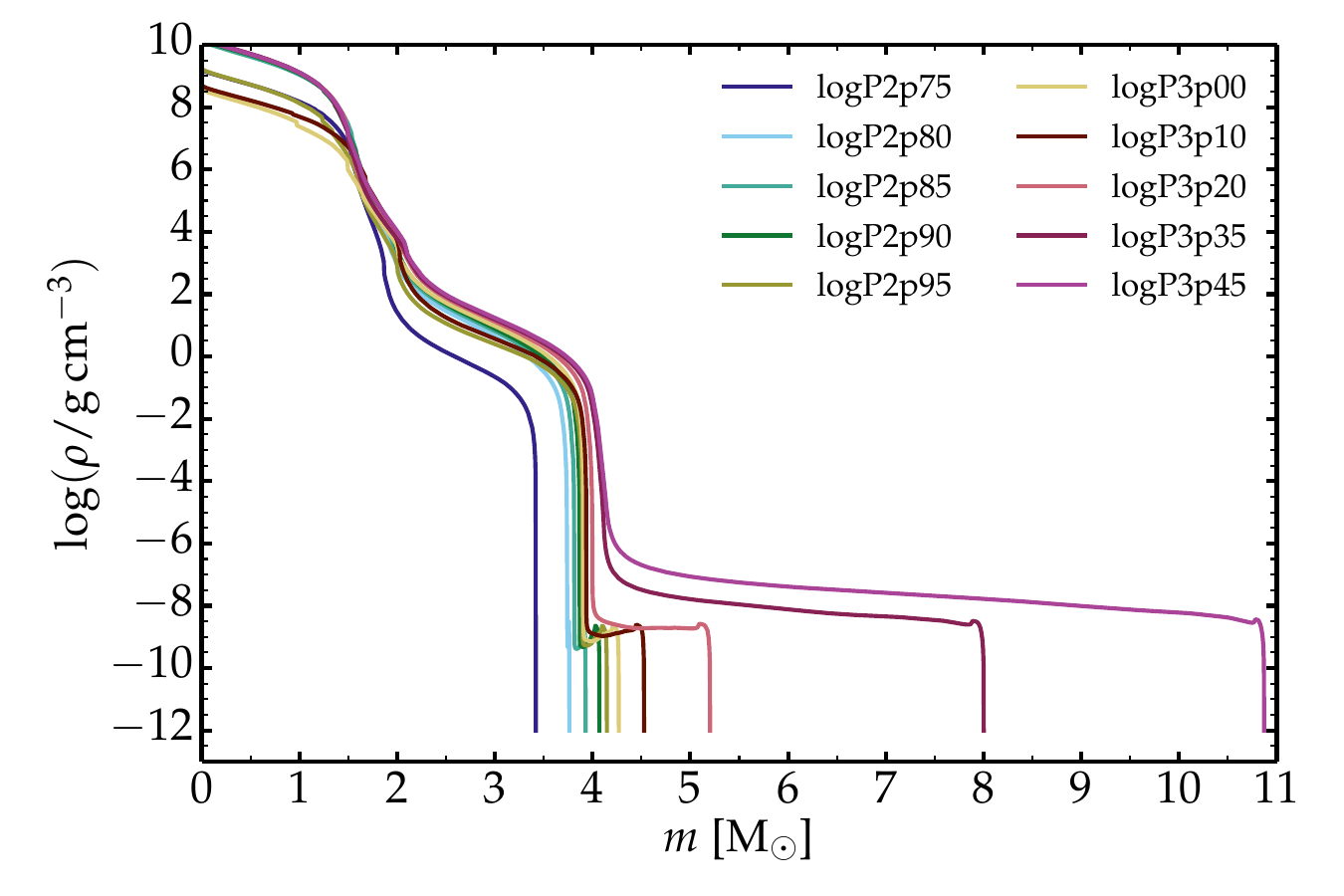}
    \end{subfigure}
    \hfill
    \centering
    \begin{subfigure}[b]{0.49\textwidth}
       \centering
       \includegraphics[width=\textwidth]{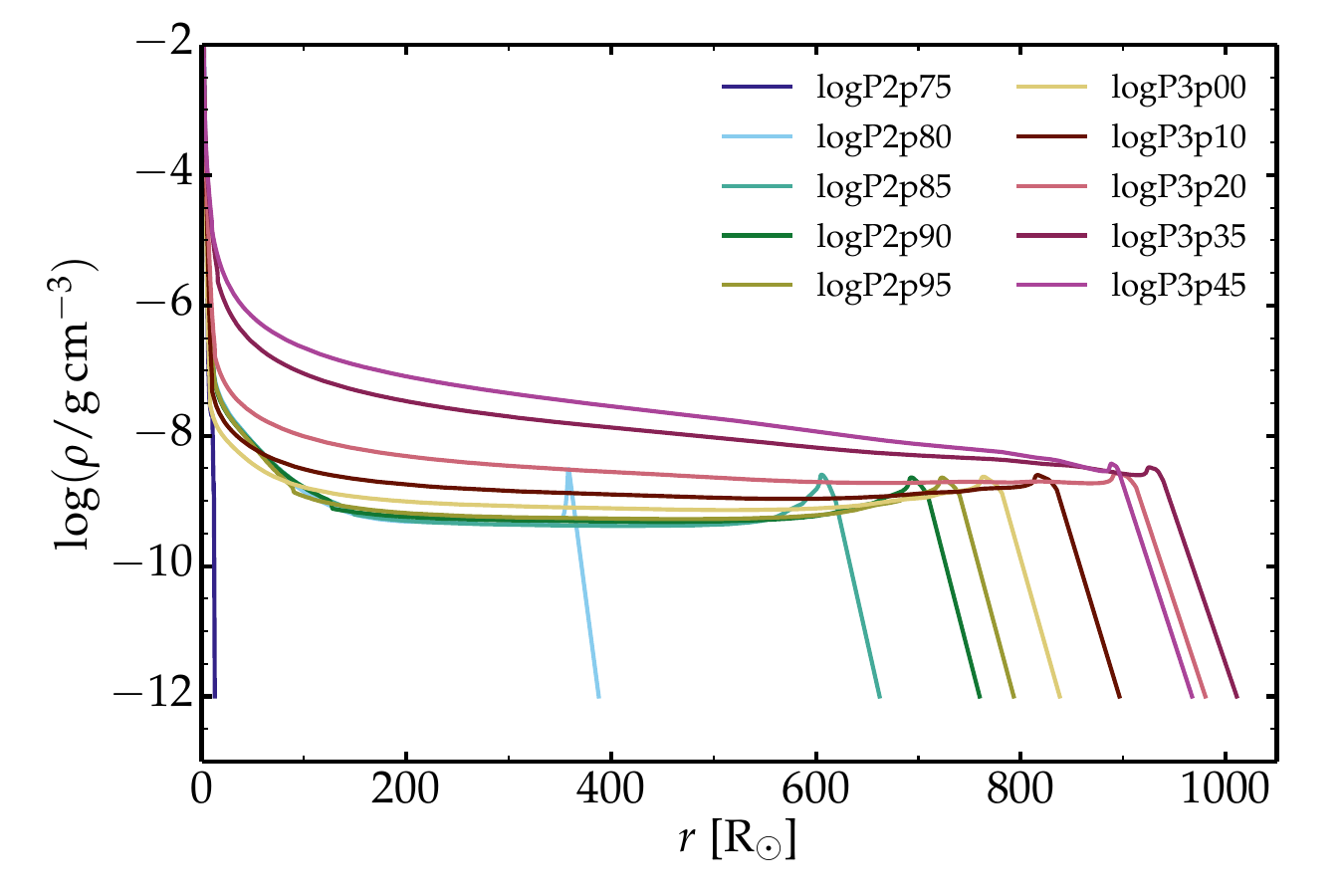}
    \end{subfigure}
\vskip\baselineskip
    \centering
    \begin{subfigure}[b]{0.49\textwidth}
       \centering
       \includegraphics[width=\textwidth]{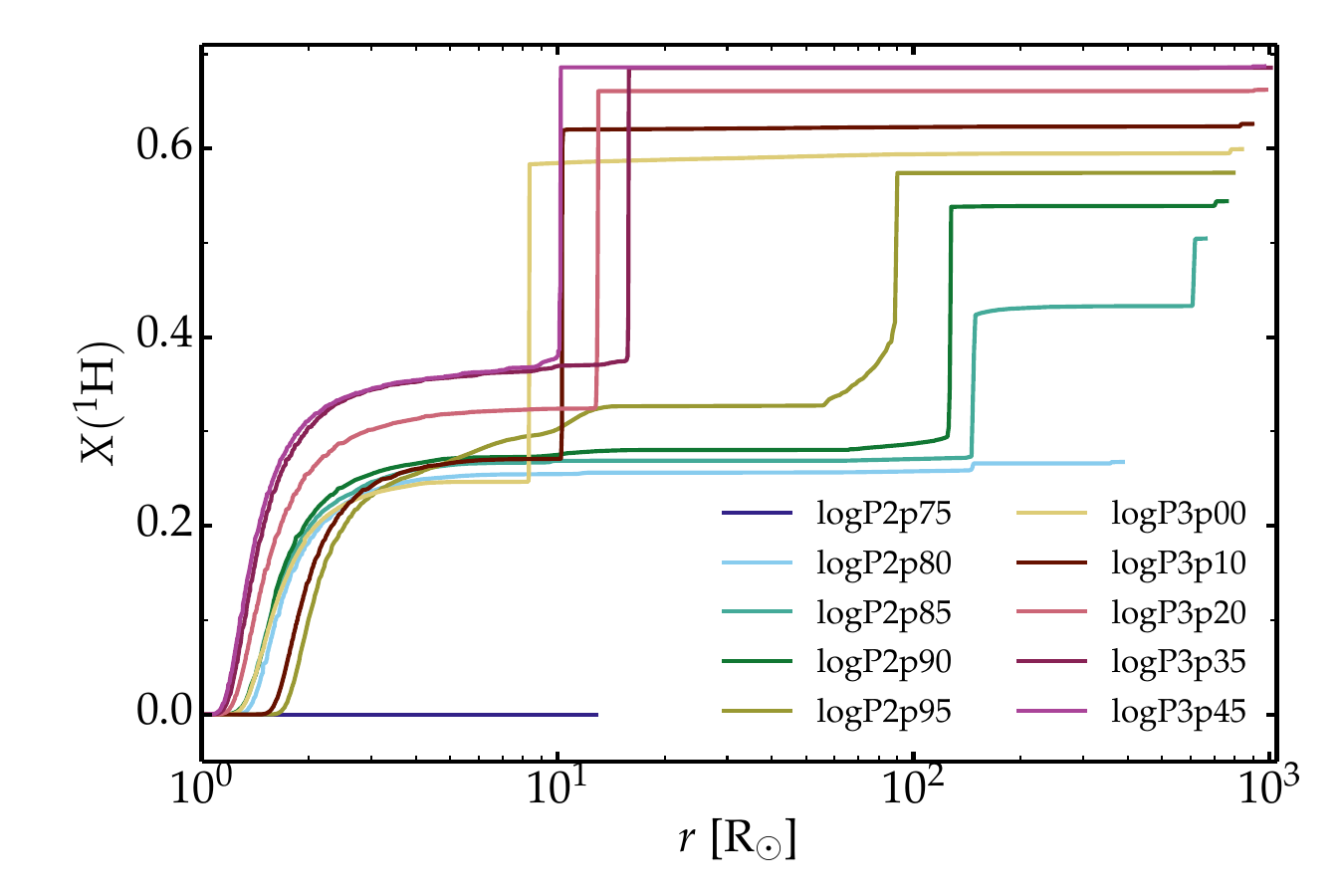}
    \end{subfigure}
    \hfill
    \centering
    \begin{subfigure}[b]{0.49\textwidth}
       \centering
       \includegraphics[width=\textwidth]{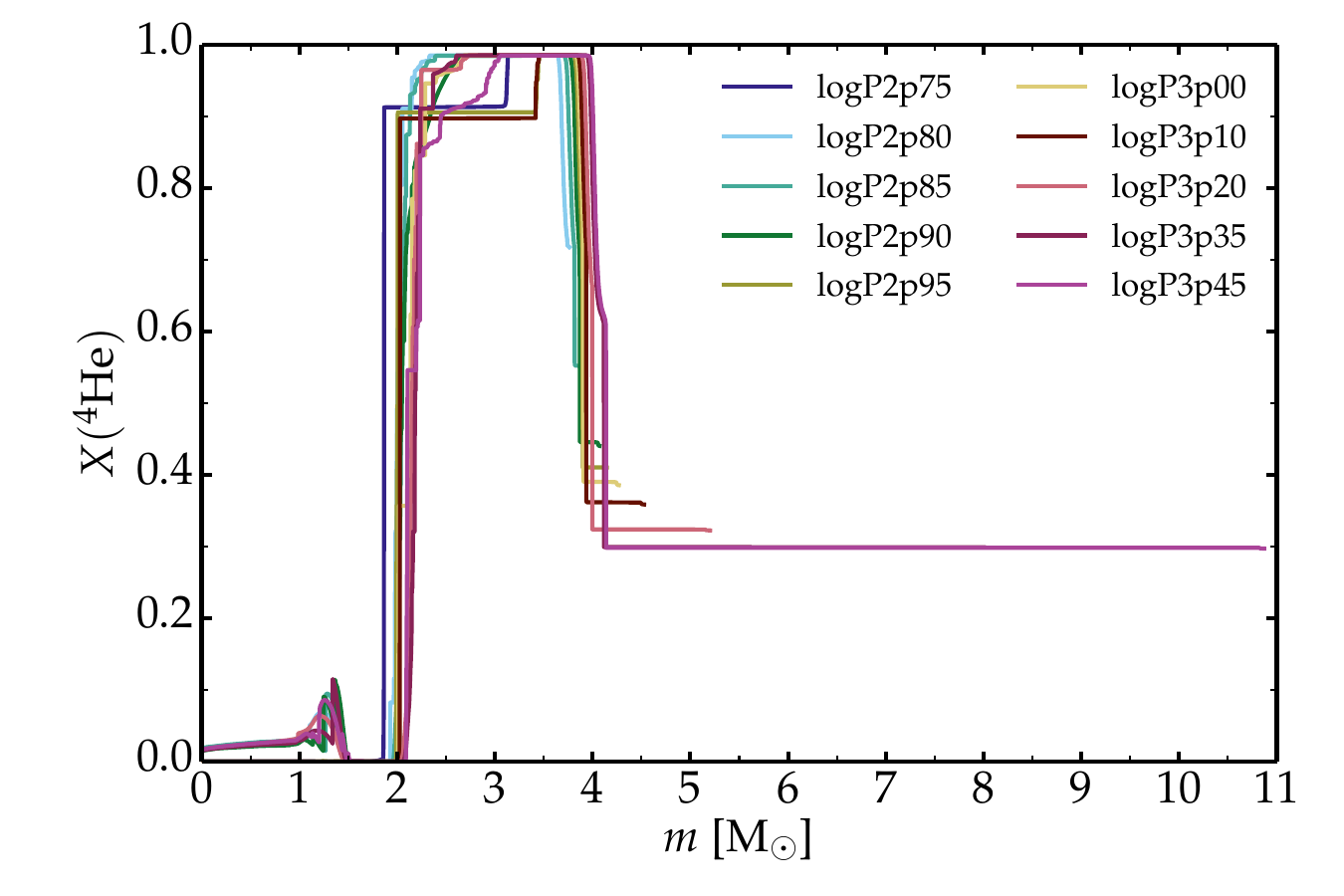}
    \end{subfigure}
\caption{PreSN structure for our model set. We show the density versus mass (top-left) and radius (top-right), the H mass fraction versus radius (bottom-left), and the He mass fraction versus mass. [See Section~\ref{sect_mesa} for discussion.]
\label{fig_mesa}
}
\end{figure*}

 \subsection{PreSN evolution}
 \label{sect_mesa}

  Our work starts from a subset of the binary-star evolution models presented in \citet{ercolino_bin_23}, namely the models for a binary system with a primary star mass of 12.6\,\msun, a system mass ratio of 0.95, and a range of initial binary period from 562\,d (logP2p75) up to 2818\,d (logP3p45) -- our nomenclature is to refer to the models by the log of their initial period in days since our models differ initially only in that parameter. A mixing-length parameter $\alpha_{\rm MLT}$ of 1.5 was chosen to keep in line with previous stellar-evolution simulations (see Section~\ref{sect_iip} for further discussion). Full details about the \mesa\ parameters used are given in \citet{ercolino_bin_23}.

  With the adopted range of initial orbital separations, mass transfer takes place at different stages in the evolution of the primary star. For the shortest initial periods (i.e., tightest initial orbits), the primary H-rich envelope is severed early in the life of the primary star and leads to a case B mass transfer. For the longest initial periods (i.e., widest initial orbits), the primary never fills its Roche lobe and the star effectively evolves as a single star. In between these two extremes, RLOF starts at different ages and lasts different durations, producing a range of case B, case BC, and case C systems. The ultimate result from this set of progenitors are preSN stars with essentially the same core (the He-core masses for our set covers from 3.42 to 3.98\,\msun, with most cluttering around 3.8\,\msun) but widely different H-rich envelope masses covering from 0 (model logP2p75) up to 6.86\,\msun\ (model logP3p45). As is well known (see, e.g., \citealt{woosley_94_93j}; \citealt{dlw10b}; \citealt{snec}; \citealt{HD19}) and already discussed in \citet{ercolino_bin_23}, the high-brightness phase of Type II SNe being primarily controlled by the H-rich envelope mass of the progenitor (and not of its metal-rich core; \citealt{d19_sn2p}), one expects a wide diversity of SN outcomes covering from Type Ib (model logP2p75) to Type IIb (cases with $M$(H-env) of order 0.1\,\msun) to fast-declining Type II (cases with $M$(H-env) of order 1\,\msun)  and more standard Type II-P (cases with $M$(H-env) of several solar masses).

  \citet{ercolino_bin_23} also emphasize the large wind mass loss rates and mass transfer rates for case BC and case C systems within O\,(10\,000\,yr) prior to core collapse. These conditions are prime for producing interacting SNe. However, in this study, we will ignore this circumstellar material (we thus assume that all these preSN models explode successfully and in a vacuum) in order to document the fundamental properties of the exploding stars alone. This is a useful first step to consider before incorporating the extra complexity associated with such CSM. We present various properties of our selected sample of binary star models in Table.~\ref{tab_mesa} and illustrate some of their important properties at core collapse in Fig.~\ref{fig_mesa}.

\begin{figure*}
   \centering
    \begin{subfigure}[b]{0.49\textwidth}
       \centering
       \includegraphics[width=\textwidth]{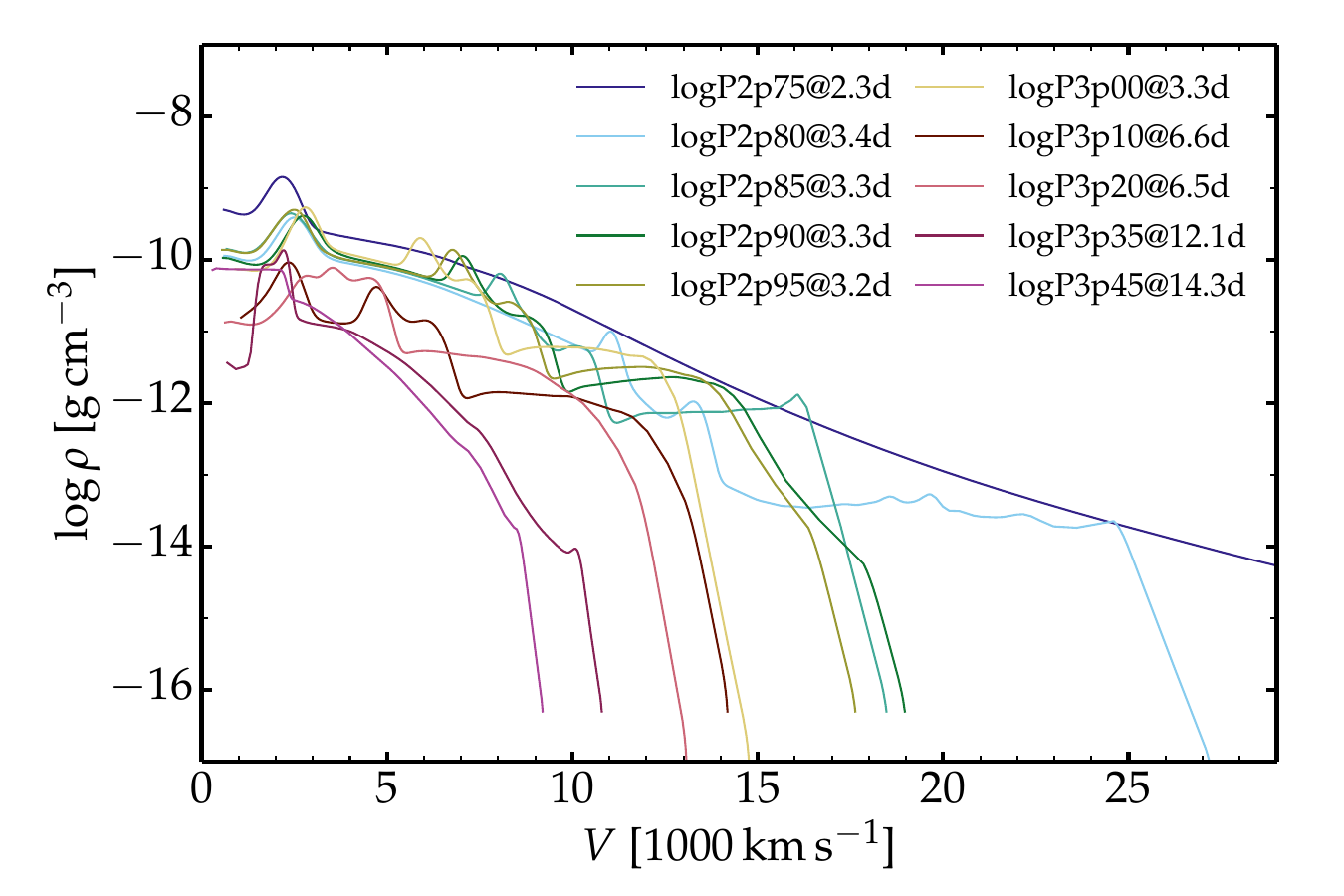}
    \end{subfigure}
    \hfill
    \centering
    \begin{subfigure}[b]{0.49\textwidth}
       \centering
       \includegraphics[width=\textwidth]{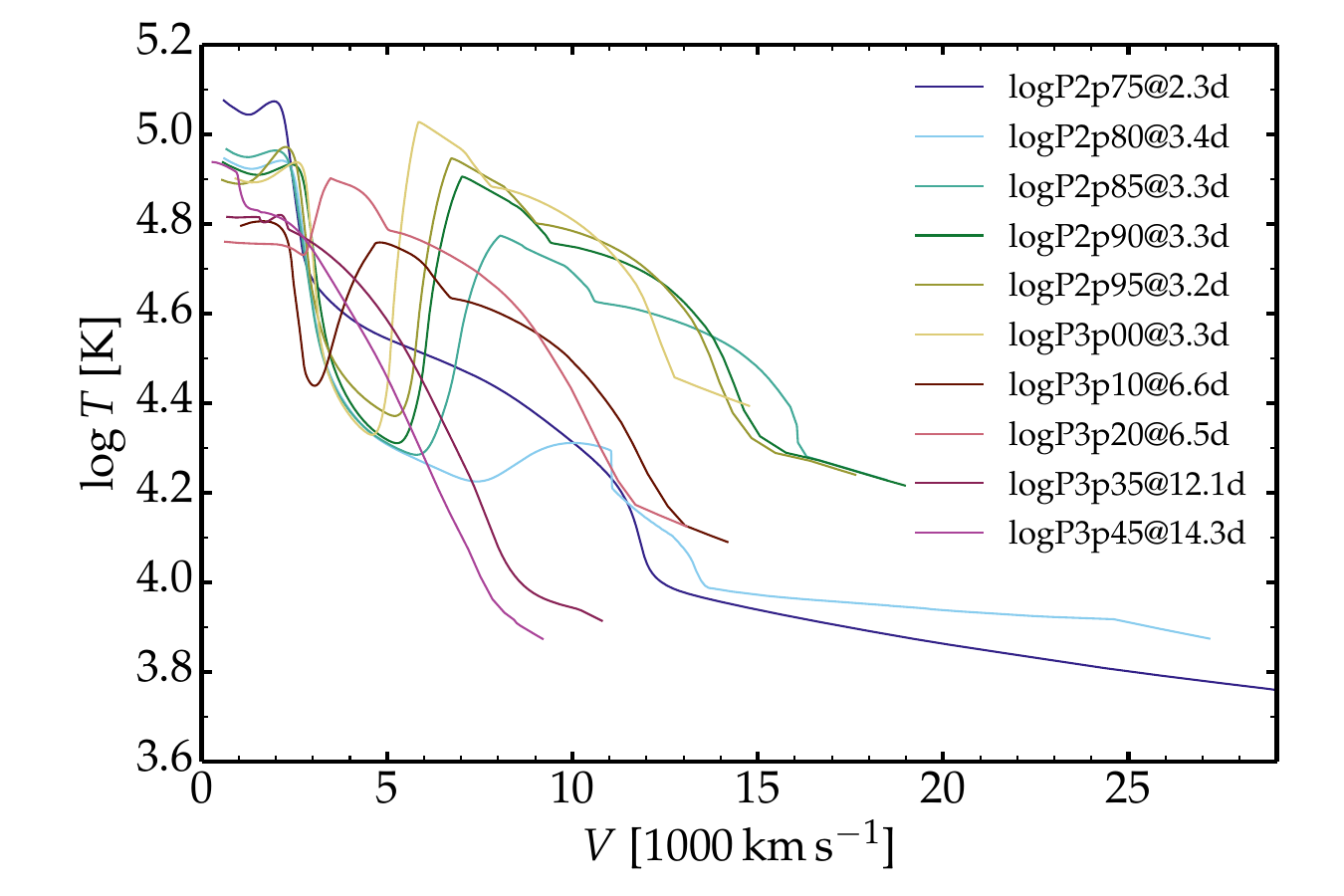}
    \end{subfigure}
\caption{
Initial ejecta structure for our \cmfgen\ simulations. We show the density (left) and the temperature (right) versus velocity for the ejecta that result from the explosion of the primary star in our set of binary progenitor models. These profiles are used as initial conditions for the radiative transfer calculations. The initial SN age, at which homologous expansion is essentially reached, occurs earlier for more compact progenitors and covers from 2.3 (model logP2p75) to 14.3\,d (model logP3p45). The ejecta extend to about 40\,000\,\kms\ for model logP2p75 (cut at 29\,000\,\kms\ here for better visibility).
\label{fig_voned}
}
\end{figure*}

\begin{figure}
\includegraphics[width=0.49\textwidth]{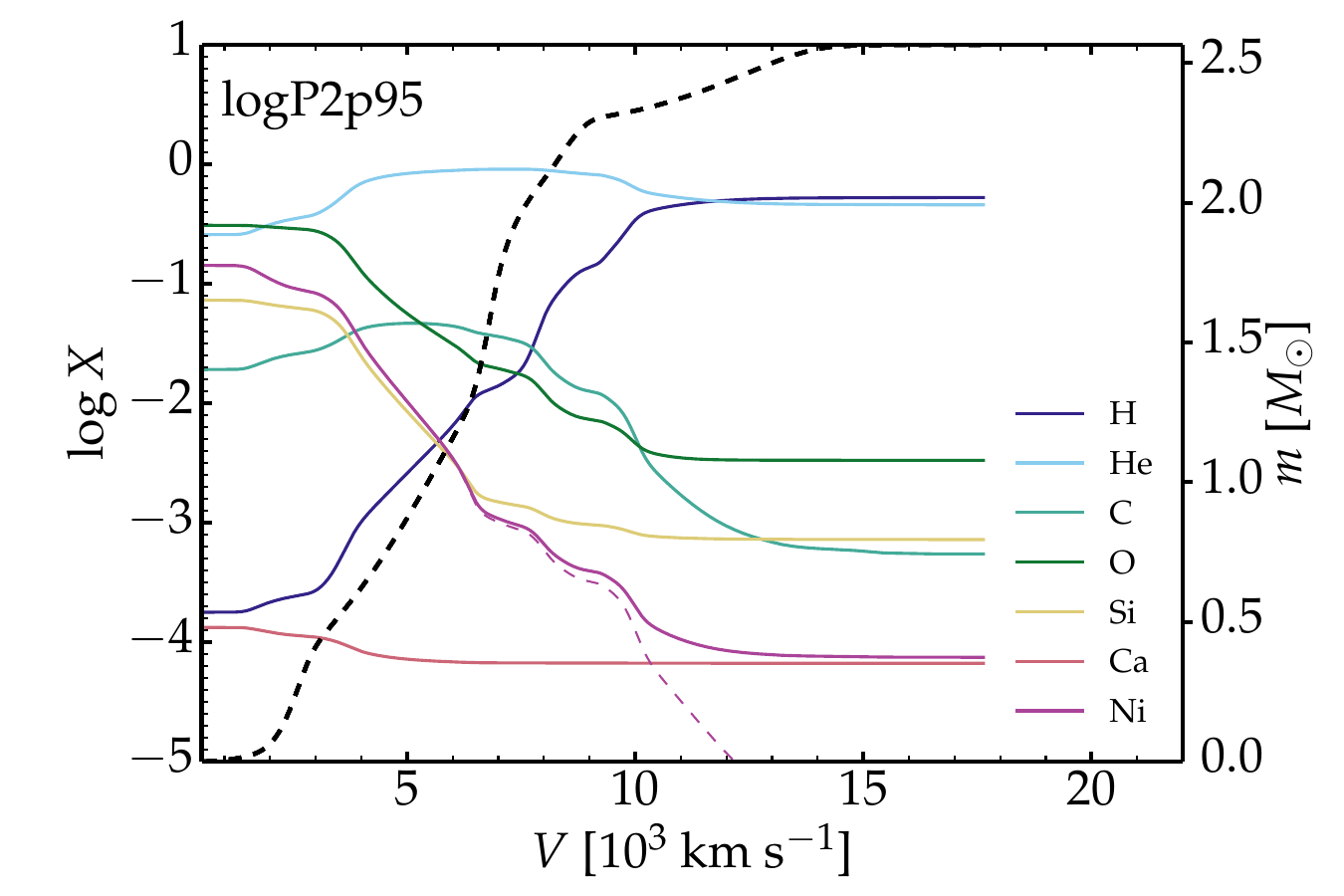}
\caption{
Composition profile of model logP2p95 at 3.5\,d used as initial conditions for the \cmfgen\ calculations. In the case of Ni, we show both the total mass fraction from all Ni isotopes (solid) and the mass fraction of \nifs\ alone (dashed). The right axis corresponds to the Lagrangian mass (black dashed line).
\label{fig_comp}
}
\end{figure}

\subsection{Explosion phase and evolution to homologous expansion}
\label{sect_voned}

The next stage was to model the explosion of the primary star once it had reached its evolutionary endpoint.  All models terminated after silicon core exhaustion but not all reached the nominal endpoint with a maximum core infall velocity of 1000\,\kms. This issue is not critical for our work since we artificially trigger the explosion and do not focus on the exact post-shock nucleosynthesis nor on the remnant mass. In practice,  at a mass cut of 1.53\,\msun\ (1.60\,\msun\ for models logP2p90 and logP2p95),\footnote{For consistency with previous similar simulations with \cmfgen, we choose our mass cut at the location where the entropy jumps above 4\,k$_{\rm B}$\,baryon$^{-1}$ -- other choices are possible, as long as they remain reasonable (i.e., shifts of order 0.1\,\msun). The exact value of the mass cut has little importance for the present study.} we deposit, within a narrow region of 0.05\,\msun, an energy corresponding to the binding energy of the overlying layers (about $-2 \times 10^{50}$\,erg) and the desired kinetic energy $E_{\rm kin}$ of the ejecta at infinity.  Although \voned\ solves for the explosive nucleosynthesis during the simulation (each progenitor yields a specific value but with little scatter given the very similar core structures for all models in our sample), we reset the \nifs\ mass to be the same in all models. Here, guided by the inferred values for the Type IIb SN\,1993J \citep{woosley_94_93j}, we adopt $M$(\nifs)$=$\,0.09\,\msun\ and $E_{\rm kin}=$\,10$^{51}$\,erg (the actual values in the final ejecta differ a little because of the additional energy for nuclear burning as well as the various adjustments we make to the ejecta structure -- see below). The exact value is unimportant as long as it is essentially the same one for all models in our sample -- this will facilitate the interpretation of photometric and spectroscopic properties.

In all cases, the preSN structure was adjusted to ensure that the density at the outermost grid point was low enough to allow for shock breakout under optically-thin conditions (or at least close to that). Hence, progenitors were extended with an atmosphere having a density scale height of 1\% of the stellar radius and going down to a minimum density of 10$^{-12}$\,g\,cm$^{-3}$ (see top-right panel in Fig.~\ref{fig_mesa}).\footnote{Other choices corresponding to a mass loaded atmosphere, a slowly accelerating wind, or even enhanced mass loss rates are largely irrelevant here since we focus on the long-term evolution of the SN ejecta and radiation rather than on the immediate phase following shock breakout.}  Although the \voned\ simulations were run until one year after explosion, we use the \voned\ results for each model at the onset of homologous expansion, or at about 2\,d after explosion, whatever came first. The initial density and temperature structures used for the \cmfgen\ calculations are shown in Fig.~\ref{fig_voned}. In contrast to the smooth density and temperature structure obtained for  the compact H-deficient logP2p75 model, most models in our sample exhibit strong density variations with depth, with multiple dense shells. To reduce the sharp density jumps obtained in a Lagrangian code like \voned, we apply a gaussian smoothing before remapping in \cmfgen\ (Fig.~\ref{fig_voned} shows the profiles after smoothing). These adjustments are mostly done for numerical convenience but also to get rid of artifacts arising from the imposed spherical symmetry. For a more extended discussion on shock propagation in progenitors with extended envelopes, we refer the reader, for example,  to \citet{woosley_94_93j} or \citet{d18_ext_ccsn} for the 1-D case and to \citet{gabler_3dsn_21} for the multi-dimensional case.

Finally, to avoid having all the \nifs\ present in the innermost ejecta layers, we apply a boxcar smoothing (width of 0.1\,\msun, run once through the model at 100\,s after the explosion trigger). This leads to chemical mixing between all shells and the presence of \nifs\ out to velocities of a few thousand \kms\ (see illustration of the composition for model logP2p95 in Fig.~\ref{fig_comp}). Table~\ref{tab_cmfgen_init} summarizes the ejecta properties of our model set used as initial conditions for the \cmfgen\ calculations.

\input{table_cmfgen_init.tex}

\subsection{Radiative transfer with \cmfgen}
\label{sect_cmfgen}

  Starting from the ejecta models described in the previous section (see also Table~\ref{tab_cmfgen_init} and Fig.~\ref{fig_voned}), we solve the nonlocal thermodynamic equilibrium (NLTE) time-dependent radiative transfer with  \cmfgen\ \citep{HD12} and compute the evolution of the gas and radiation until the nebular phase.  Species that have a relatively high abundance or have been understood to impact SN radiation are included in our model atom. We treat H, He, C, N, O, Ne, Na, Mg, Al, Si, S, Ar, Ca, Sc, Ti, Cr, Fe, Co, and Ni -- species with heavier nuclei than Ni are not included because they are not covered by our nuclear network (the corresponding atomic data is also scarce or inexistent). We use the updated atomic data described in \citet{blondin_21aefx_23} and include the following ions:  H\one\ (26,36), He\one\ (40,51), He\two\ (13,30), C\one\ (14,26), C\two\ (14,26), C\three\ (62,112), N\one\ (44,104), N\two\ (23,41), N\three\ (25,53), O\one\ (21,51), O\two\ (54,123), O\three\ (44,86), Ne\one\ (78,155), Ne\two\ (22,91), Ne\three\ (32,80), Na\one\ (22,71),  Mg\one\ (39,122), Mg\two\ (31,80), Mg\three\ (31,99), Al\two\ (26,44), Al\three\ (17,45), Sc\one\ (26,72), Sc\two\ (38,85), Sc\three\ (25,45), Si\one\ (100,187), Si\two\ (31,59), Si\three\ (33,61), S\one\ (106,322), S\two\ (56,324), S\three\ (48,98), Ar\one\ (56,110), Ar\two\ (134,415), Ar\two\ (32,346), K\one\ (25,44),  Ca\one\ (76,98), Ca\two\ (21,77), Ca\three\ (16,40), Ti\two\ (37,152), Ti\three\ (33,206), Cr\two\ (28,196), Cr\three\ (30,145), Cr\four\ (29,234),  Fe\one\ (44,136), Fe\two\ (228, 2696), Fe\three\ (96,1001), Fe\four\ (100,1000), Fe\,\five\ (47,191), Co\two\ (44,162), Co\three\ (33,220), Co\four\ (37,314), Co\,\five\ (32,387),  Ni\one\ (56,301), Ni\two\ (27,177), Ni\three\ (20,107), Ni\four\ (36,200), and Ni\,\five\ (46,183). The numbers in parenthesis correspond to the number of super levels and full levels employed (for details on the treatment of super levels, see \citealt{hm98}). With our model atom, we treat 11\,939 levels and a total of 830\,000 bound-bound transitions. These are mostly coming from iron-group elements, in particular Fe, Co, and Ni (and of these three elements, Fe is the most important followed by Ni).

Besides the initially stored radiative energy, we account at all times for the radioactive decay power from the two-step \nifs\ decay chain -- all other isotopes are stable (the only exceptions are \iso{44}Ti and \iso{57}Ni but because of their low abundance or long lifetime, they are irrelevant at the times considered here; see, for example, \citealt{d23_interaction}). We treat non-thermal processes for high-energy electrons as described in \citet{li_etal_12_nonte}. Nonlocal $\gamma$-ray energy deposition is computed by solving the radiative-transfer equation using a grey, absorption-only opacity equal to 0.06\,$Y_{\rm e}$\,cm$^{2}$\,g$^{-1}$, where $Y_{\rm e}$ is the electron fraction.

The grid employs 80 to 100 grid points uniformly spaced on a logarithmic optical-depth scale (the exact number depends on model and SN age). Remapping the grid is necessary at the start of every timestep as well as during the calculation in order to track recombination fronts, if present. To cover the full evolution to the nebular phase, a total of 30 to 40 timesteps were computed (differences arise because the start time is not the same for all sequences and also because some ejecta take much longer to turn nebular). The main \cmfgen\ output of interest for this study is the emergent flux, which is calculated at each time step in the observer's frame from the far-UV to the far-IR. This spectrum can be shown directly or used to compute the bolometric luminosity $L_{\rm bol}$ or the absolute magnitude in various filters.

In this work, to refer to the models in our set, we will repeatedly call them as members of a sequence ``from'' logP2p75 ``up'' to logP3p45 because  this forms a sequence of increasing initial period for the binary system from which we extracted the exploding primary star. This sequence is a natural one since it is also one of increasing ejecta mass or H mass (see Table~\ref{tab_mesa}). This is a convenient shorthand.

\begin{figure*}
   \centering
    \begin{subfigure}[b]{0.49\textwidth}
       \centering
       \includegraphics[width=\textwidth]{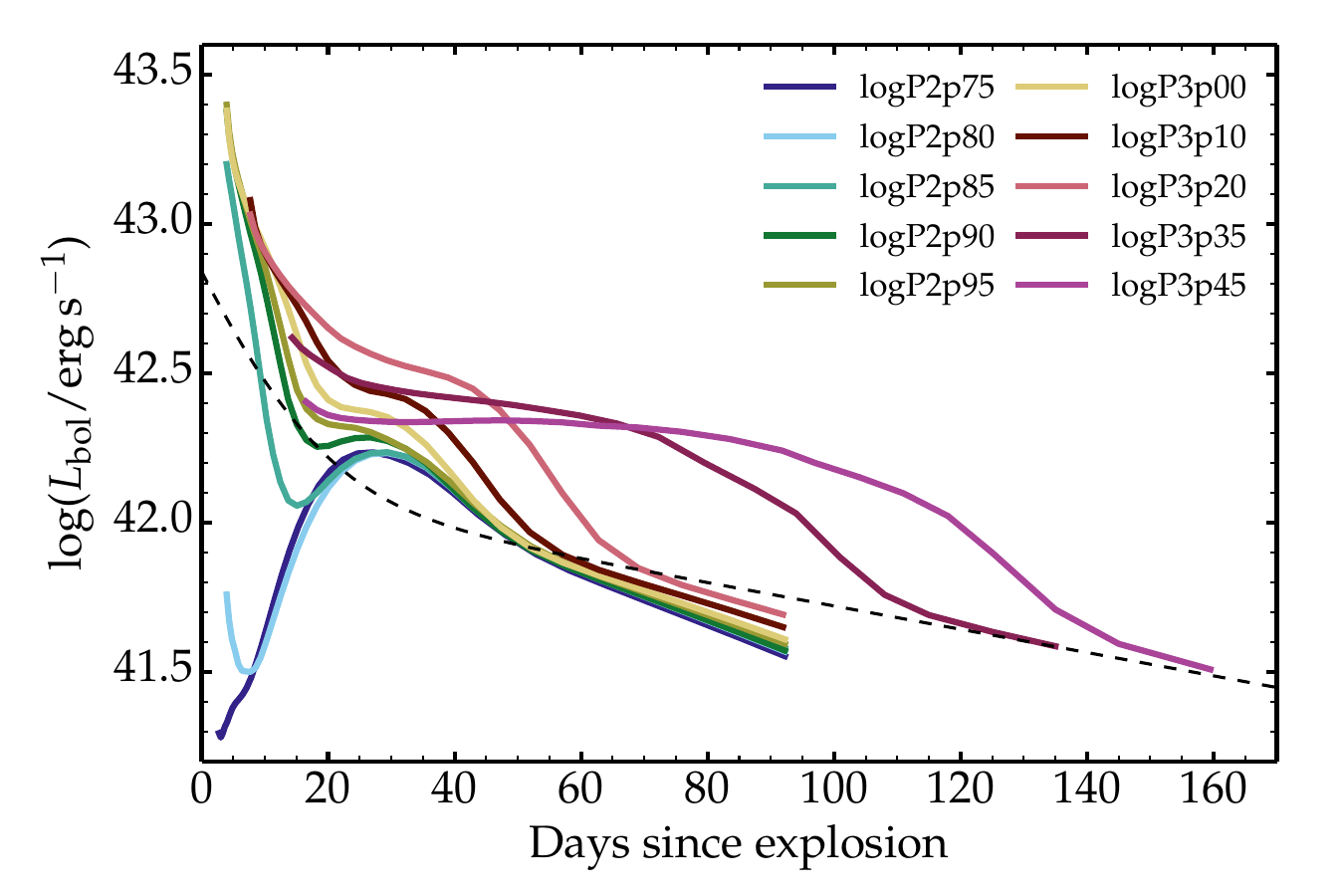}
    \end{subfigure}
    \hfill
    \centering
    \begin{subfigure}[b]{0.49\textwidth}
       \centering
       \includegraphics[width=\textwidth]{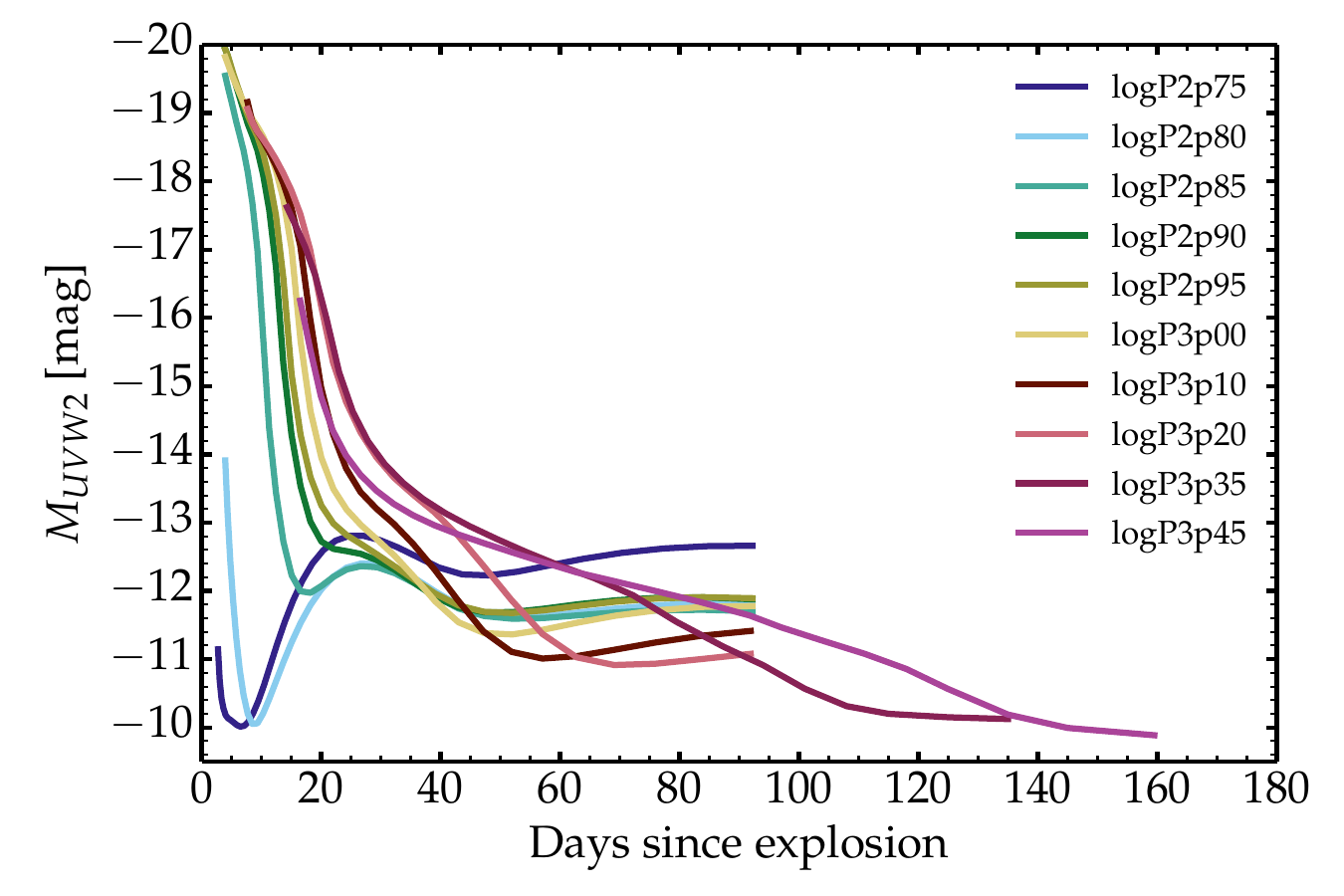}
    \end{subfigure}
    \vskip\baselineskip
   \centering
    \begin{subfigure}[b]{0.49\textwidth}
       \centering
       \includegraphics[width=\textwidth]{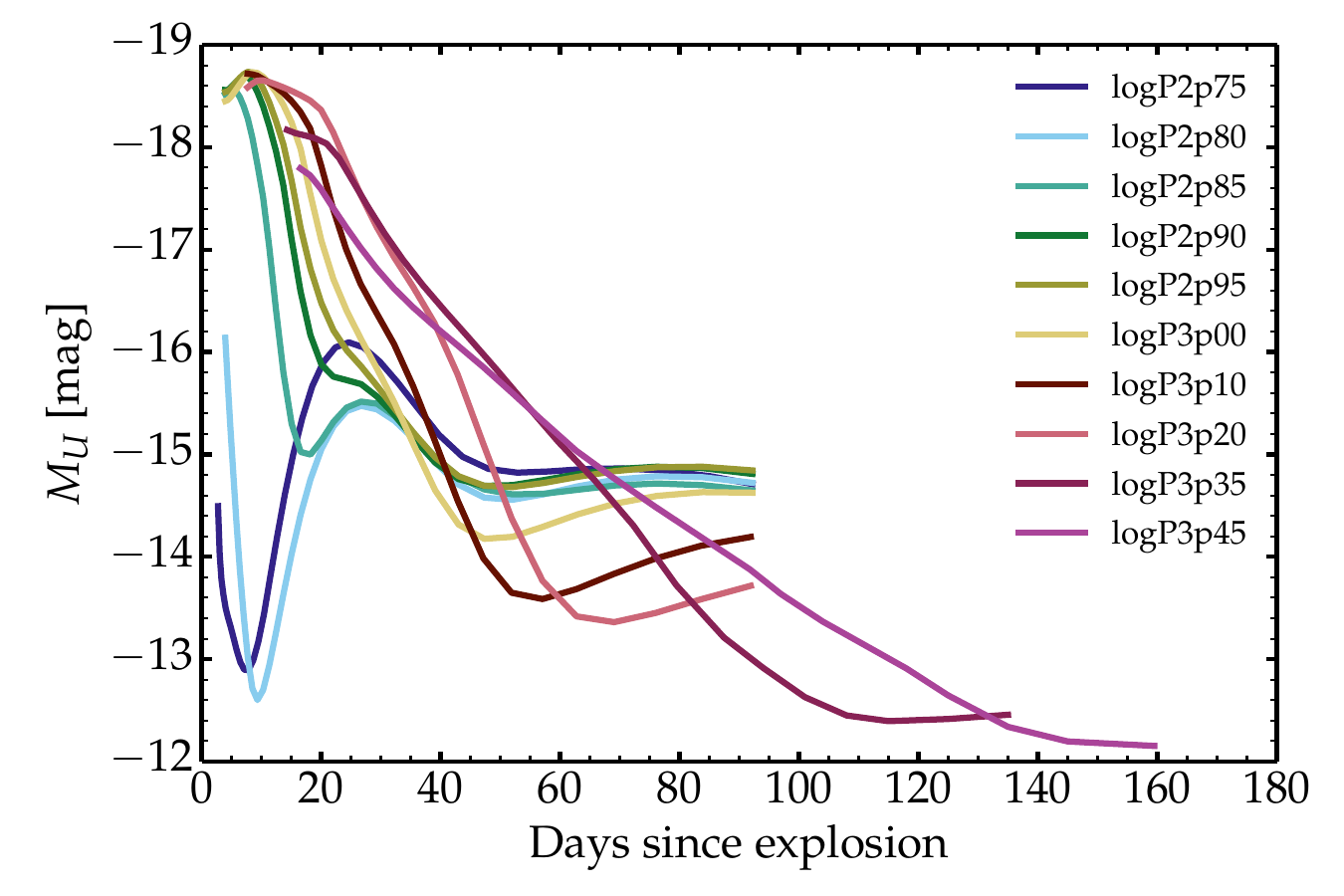}
    \end{subfigure}
    \hfill
    \centering
    \begin{subfigure}[b]{0.49\textwidth}
       \centering
       \includegraphics[width=\textwidth]{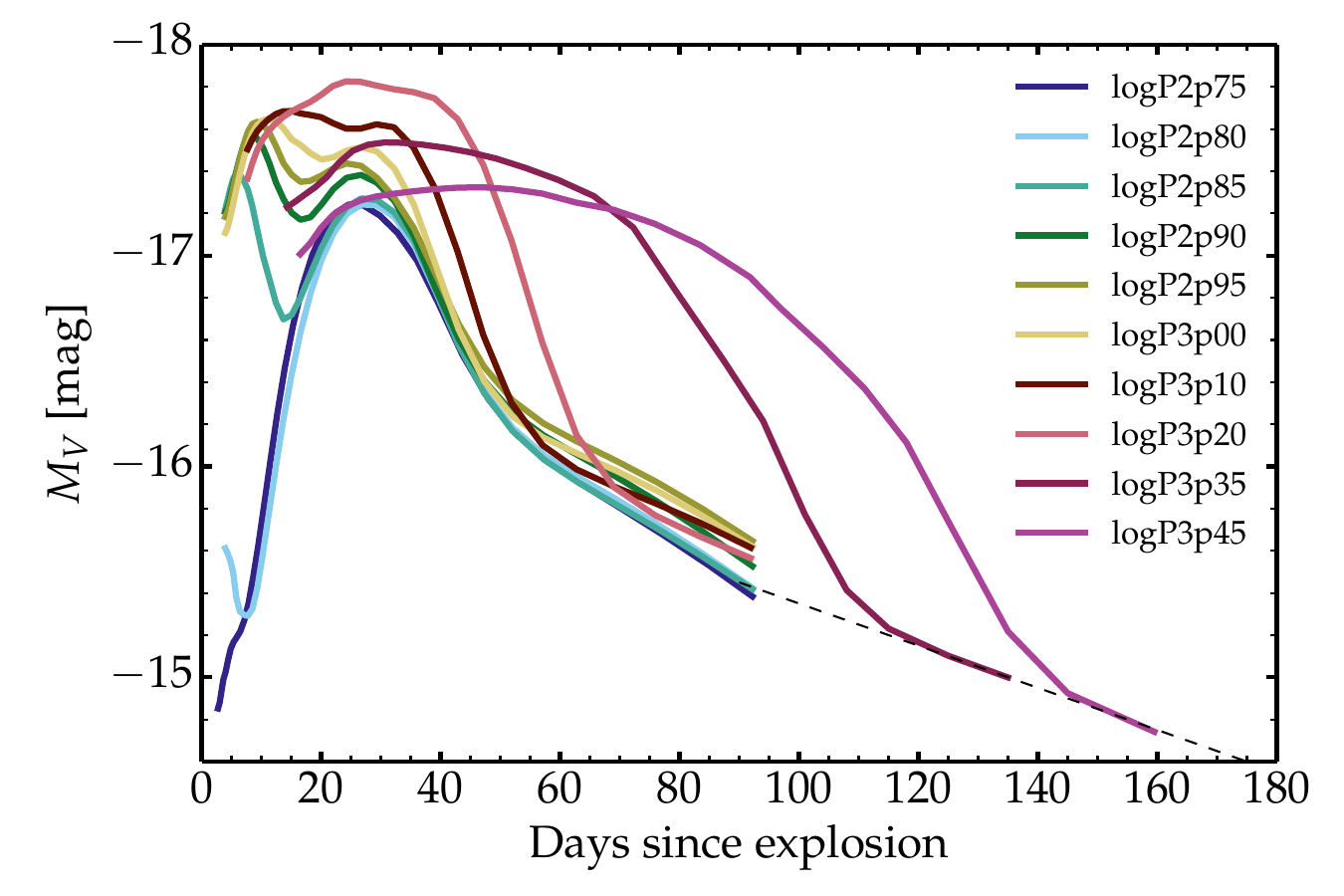}
    \end{subfigure}
\caption{Photometric properties for our model set. We show the bolometric light curves (top left) followed by the $UVW2$ (top right), $U$ (bottom left), and $V$-band (bottom right) light curves. In the top-left panel, the dashed line corresponds to the emitted power from 0.09\,\msun\ of \nifs. In the bottom-right panel, the dashed curve gives the predicted decline rate for full $\gamma$-ray trapping and constant color, which essentially holds in the more massive ejecta models logP3p35 and logP3p45.
\label{fig_phot}
}
\end{figure*}

\section{Results: Diversity of model light curves}
\label{sect_phot}

Figure~\ref{fig_phot} presents the photometric evolution for our model set from 2.3$-$14.3\,d until 92.2$-$159.5\,d after explosion, essentially stopped when the ejecta optical depth drops to near unity. The bolometric light curves (top-left panel) exhibit different morphologies. Model logP2p75, whose progenitor has no residual H-rich envelope, shows a single, \nifs-powered  luminosity peak (we exclude here the shock breakout burst, modeled with \voned\ but not with \cmfgen) followed by a steep decline from maximum, and eventually a quasilinear decline at nebular times with a slope that is steeper than predicted for full $\gamma$-ray trapping. Models logP2p80, logP2p85, logP2p90 show two separate phases in their light curves, with a first high and slowly declining luminosity down to a minimum at 10$^{41.5}-10^{42.3}$\,\ergs, and a subsequent evolution that lines up with that obtained for model logP2p75.\footnote{SNe from He-giant star explosions also yield a double-peak light curve in the $V$ band and this arises from the presence of an extended He-rich H-deficient envelope \citep{d18_ext_ccsn}.} Models logP2p95, logP3p00, logP3p10, and logP3p20 show a prolonged decline of the initial high, post-breakout luminosity, with no clear separated peak, and eventually merge with the light curve obtained for model logP2p75 at 50$-$70\,d after explosion. The late-time decline in those models is flatter for greater ejecta masses. Models logP3p35 and logP3p45 exhibit a clear plateau in their light curve, which lasts until the end of the optically-thick phase at 100$-$120\,d after explosion. At a qualitative level, this set of light curves thus covers that obtained for Type Ib or Ic SNe (e.g., logP2p75) from stripped compact progenitors, for Type IIb SNe from progenitors with an extended but low-mass H-rich envelope (e.g., models  logP2p80, logP2p85, logP2p90), for fast-declining Type II-L SNe  (e.g., models  logP2p95, logP3p00, logP3p10, and logP3p20), and for more standard Type II-P SNe (e.g., models logP3p35 and logP3p45).

Such a diversity of light curves is not a new prediction of our work. Single-peaked light curves are typical of stripped-envelope SNe and are well understood in the context of moderate mass ejecta (say 3$-$5\,\msun) powered by the radioactive decay of O\,(0.1\,\msun) of \nifs\ (see., e.g., \citealt{ensman_woosley_88}, \citealt{dessart_11_wr}, \citealt{bersten_08D_13}). The extended high luminosity phase prior to that main peak arises from progenitors with extended O\,(0.1\,\msun)  H-rich envelopes (see, .e.g, \citealt{woosley_94_93j}, \citealt{blinnikov_94_93j}, \citealt{bersten_etal_12_11dh}, \citealt{d18_ext_ccsn}). Our results for higher but still moderate, O\,(1\,\msun) H-rich envelopes have been obtained before in the context of massive stars partially stripped of their envelope because of wind mass loss (see, e.g., \citealt{blinnikov_bartunov_2l_93}; \citealt{snec}, \citealt{HD19}). Finally, standard RSG progenitors are notorious for producing a plateau in their light curves, a feature known since the 70s (see, e.g., \citealt{grassberg_71}). What is new though is that our model set based on the same primary star but different initial system orbital separation can reproduce the full sequence of light curves from Type Ib (or Ic), to Type IIb, Type II-L, and Type II-P SNe. This property arises in part from the fact that the residual H-rich envelope in our progenitors varies from zero to $\sim$\,7\,\msun\ while the metal rich core remains essentially the same for the whole set, leading to an explosion with the same ejecta kinetic energy and \nifs\ mass (the adopted values for $E_{\rm kin}$ and $M_{\rm ej}$ could have been different, but these values should be the same for the whole set). For a given explosion energy, the potential presence of a loosely bound H-rich envelope (whose binding energy is about $-10^{47}$\,erg) does not affect the ejecta kinetic energy at infinity, although it strongly impacts the ejecta structure, the SN light curves and the spectra.

Models that stand between logP2p75 (progenitor with no residual H-rich envelope) and logP3p35/logP3p45 (standard RSG star progenitors with a massive H-rich envelope) are hybrid cases in which the photospheric phase is powered by the release of shock-deposited energy (whose original budget is essentially set at the time of shock breakout) and by continuous radioactive decay. In model logP2p85, the light curve is composed of two, essentially separate components. The first one arises from the power released from the shocked H-rich envelope, which lasts about 10\,d, and the second one from the \nifs\ heating of the underlying metal-rich ejecta (more diversity could have been produced by varying the explosion energy, the \nifs\ mass, the chemical mixing etc). Because essentially the same \nifs\ mass is used (and expected for such progenitors) in our set, the model sequence from logP2p75 to logP3p20 yields progressively more luminous light curves since an increasing amount of shock-deposited power within the increasingly massive progenitor envelope is added to the contribution from radioactive decay power (which is comparable in all models; see Table~\ref{tab_cmfgen_init}). More massive ejecta eventually break this luminosity sequence because this increase in ejecta mass reduces the ejecta expansion rate ($E_{\rm kin}/M_{\rm ej}$ is smaller) and increases the ejecta optical depth (which inhibits radiative diffusion and the recession of the photosphere). In models logP3p35 and logP3p45, the \nifs\ is more deeply embedded and contributes to the light curve at later times.  The release of shock-deposited energy over the first 30$-$50\,d exceeds the decay power in all relevant models (i.e., those whose preSN star has a massive H-rich envelope).  Finally, the diversity in late-time decline rate reflects the greater efficiency of higher mass ejecta at trapping $\gamma$-rays.  For example, at 50\,d after explosion, model logP2p75 absorbs only 88\% of the emitted decay power whereas model logP3p45 absorbs 100\% of that emitted power. In the same order, this fraction is 59\% and 99\% at 92\,d after explosion.

Figure~\ref{fig_phot} also shows the model light curves in the $UVW2$, $U$, and $V$ bands. Each of these light curves differs significantly from the bolometric light curves because of the evolution of the color, with the noticeable exception of model logP2p75. Because of its small progenitor radius ($R_\star=$\,12.1\,\rsun), the ejecta model logP2p75 undergoes significant cooling so that the model essentially evolves at constant photospheric, or equivalently color temperature, at all times covered here (i.e., 2\,d and beyond after explosion). Model logP2p75 radiates predominantly at optical wavelengths at all times and consequently its $V$-band light curve follows closely the bolometric light curve. In all other models, the presence of a residual H-rich envelope in the preSN star comes with a large progenitor radius of about 400 to 900\,\rsun. All these models first exhibit colors more typical of RSG star explosions (the flux peaks in the UV; top-right panel of Fig.~\ref{fig_phot}) before shifting to the optical band as the outer H-rich ejecta layers become optically thin (and the \nifs-powered ejecta dominates the radiative display) or until the H-rich ejecta layers recombine. This transition occurs over a few days in models logP2p80, over 10$-$20\,d in models logP2p85$-$logP3p00 and even longer in the rest of the model set. This hybrid powering source between early and late times is what causes the double-peak $V$-band light curves in models logP2p80 to logP2p90 (see also discussion about this process in \citealt{woosley_94_93j}; \citealt{bersten_etal_12_11dh}; \citealt{nakar_piro_14}; \citealt{d18_ext_ccsn}; \citealt{park_iib_23}).

Another feature of interest in the models is the $V$-band rise time, which is of order 20$-$50\,d in models logP3p35 and logP3p45. This is a typical result for explosions of RSG stars with a very large radius (see, e.g., \citealt{KW09}; \citealt{DH11_2p}; \citealt{snec}). The peak in the $V$-band light curve occurs in those models when the spectral energy distribution essentially peaks in the optical, hence at the onset of the recombination phase. Prior to that, the bulk of the flux falls in the UV and the optical brightness is small. We will return to this property in Section~\ref{sect_iip}.

\begin{figure}
\includegraphics[width=0.49\textwidth]{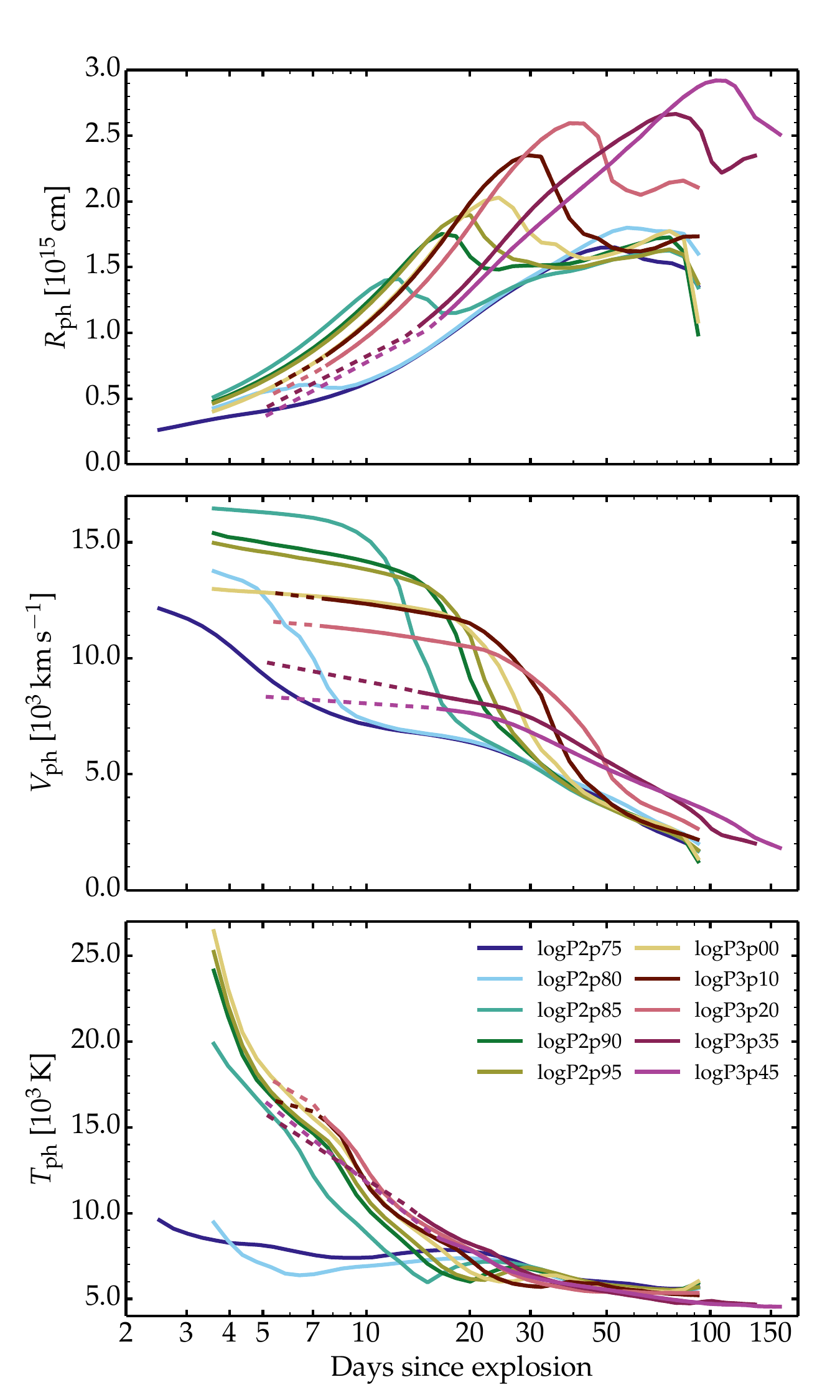}
\caption{Evolution of photospheric properties for our model set. From top to bottom, we show the evolution (the $x$-axis is shown in logarithmic scale to better reveal the rapid changes at early times) of the radius (top), velocity (middle), and gas temperature (bottom) at the location where the inward-integrated electron scattering optical depth is equal to 2/3. The dashed lines correspond to early epochs for models logP3p10 to logP3p45 in which we computed single snapshots based on \voned\ inputs.
\label{fig_at_phot}
}
\end{figure}

We can also connect these photometric properties with the evolution of properties at the photosphere, which we show for all models in Fig.~\ref{fig_at_phot}. At early times, there is a marked dichotomy in photospheric temperature $T_{\rm ph}$ with high values up to 20$-$25\,kK in models with large radii and a sizable H-rich envelope and much cooler temperatures of $\lesssim$\,10\,kK in those from more compact progenitors (models logP2p75 and logP2p80) -- this offset is the cause lying behind the color offset between models (i.e., UV bright vs optically bright at early times). This offset disappears after about 20\,d when the photosphere is at about 7\,kK -- in models logP3p45 and analogs, this corresponds to the recombination phase (plateau-like in the $V$ band) of the SN. In all models, the photospheric velocity $V_{\rm ph}$ decreases steadily at all times. In the lighter ejecta models, the recession is fast so that despite the relatively larger outer ejecta velocity (see Fig.~\ref{fig_voned}), they have relative low $V_{\rm ph}$ values at 5\,d after explosion, comparable to values obtained for more massive, slower moving ejecta -- the offset in $E_{\rm kin}/M_{\rm ej}$ is compensated by a greater ionization and optical depth. Models with the highest photospheric velocities at early times (up to 16\,000\,\kms) are those with intermediate-mass H-rich envelopes (i.e., logP2p85) since they have a combination of a relatively high $E_{\rm kin}/M_{\rm ej}$ and a relatively high optical-depth from the outer H-rich ejecta. Models that exhibit a steep density drop-off in their outer ejecta (e.g., logP3p45; see Fig.~\ref{fig_voned}) have a very slow evolving $V_{\rm ph}$ as long as the photosphere remains in those external layers. This arises because that density cliff represents a large jump in optical depth at a fixed velocity. This velocity also gives the maximum velocity one may record from any spectral line in these models (e.g., about 9\,000\,\kms\ in model logP3p45, which is much smaller than typically observed in SNe II-P -- see Section~\ref{sect_iip}). Because of our assumption of homologous expansion (which is warranted), the photospheric radii yield an information equivalent to that of the velocity since $R_{\rm ph}$ is just $V_{\rm ph}$ multiplied by the SN age. Not surprisingly, differentiating between the models of our set (whose progenitors differ strongly in H-rich envelope mass) is best done at early times, when the photosphere probes the outer ejecta layers.

\begin{figure*}
    \centering
    \begin{subfigure}[b]{0.48\textwidth}
       \centering
       \includegraphics[width=\textwidth]{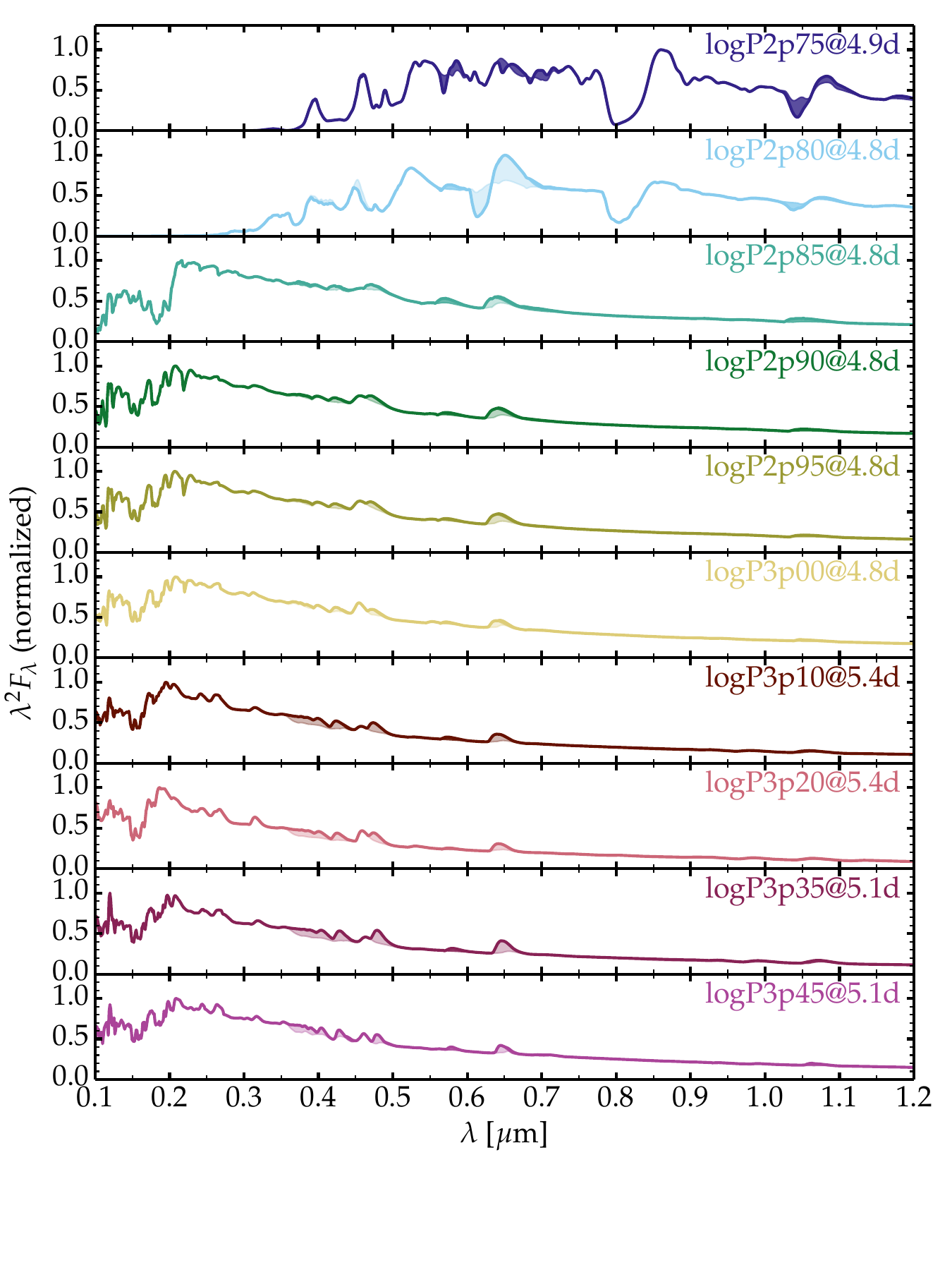}
    \end{subfigure}
    \hfill
    \centering
    \begin{subfigure}[b]{0.48\textwidth}
       \centering
       \includegraphics[width=\textwidth]{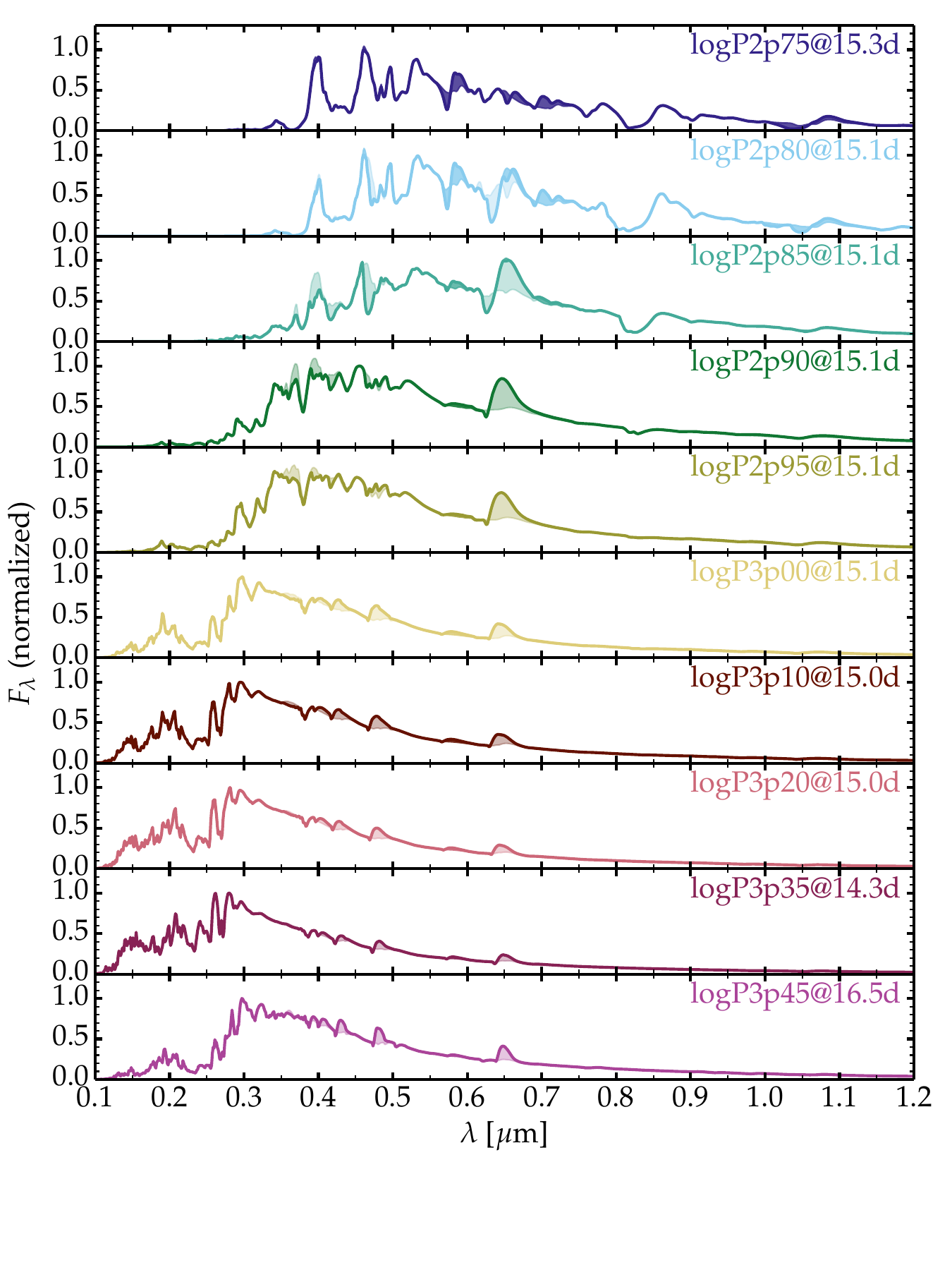}
    \end{subfigure}
    \vskip\baselineskip
    \centering
    \begin{subfigure}[b]{0.48\textwidth}
       \centering
       \includegraphics[width=\textwidth]{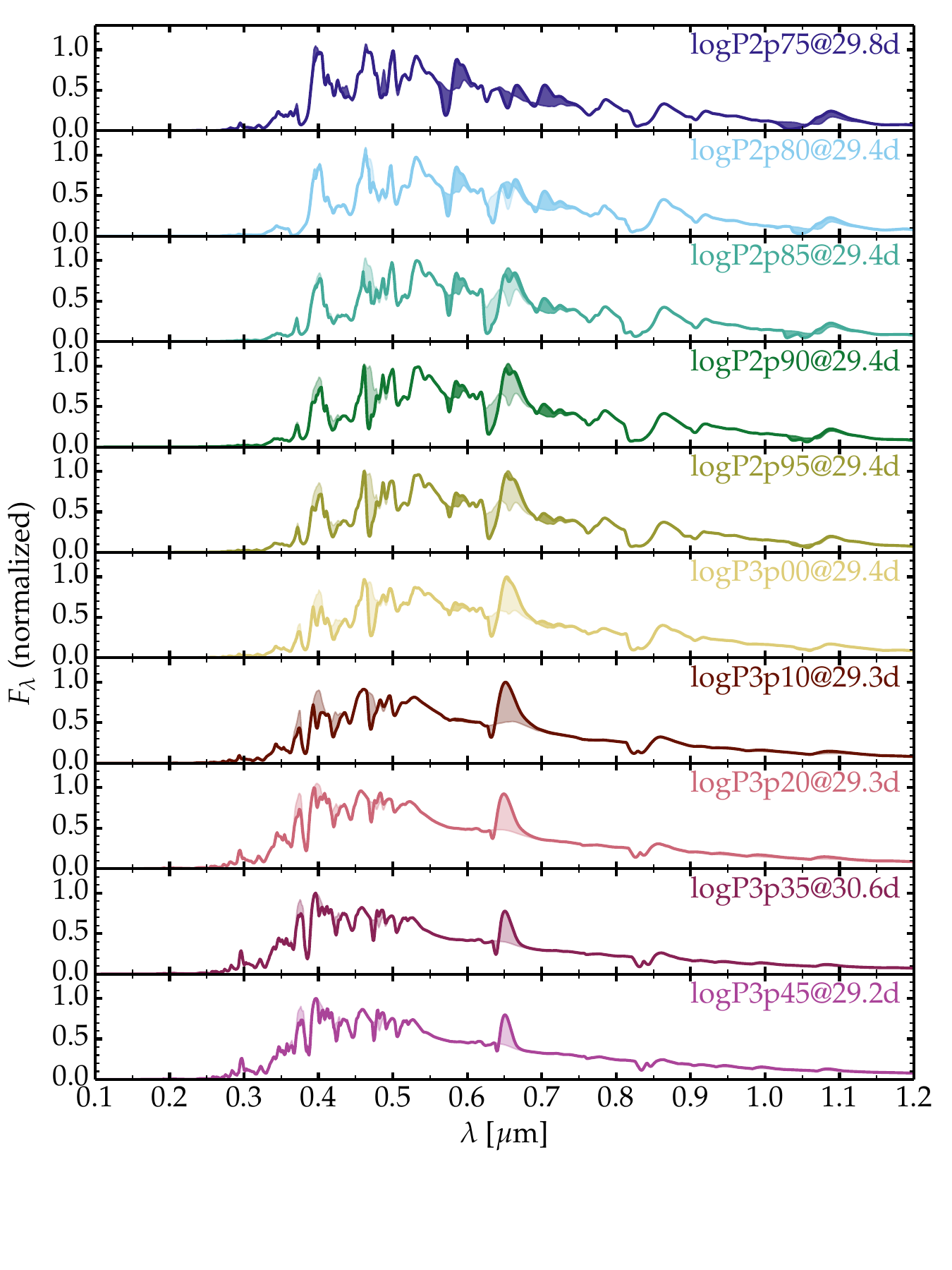}
    \end{subfigure}
    \hfill
    \centering
    \begin{subfigure}[b]{0.48\textwidth}
       \centering
       \includegraphics[width=\textwidth]{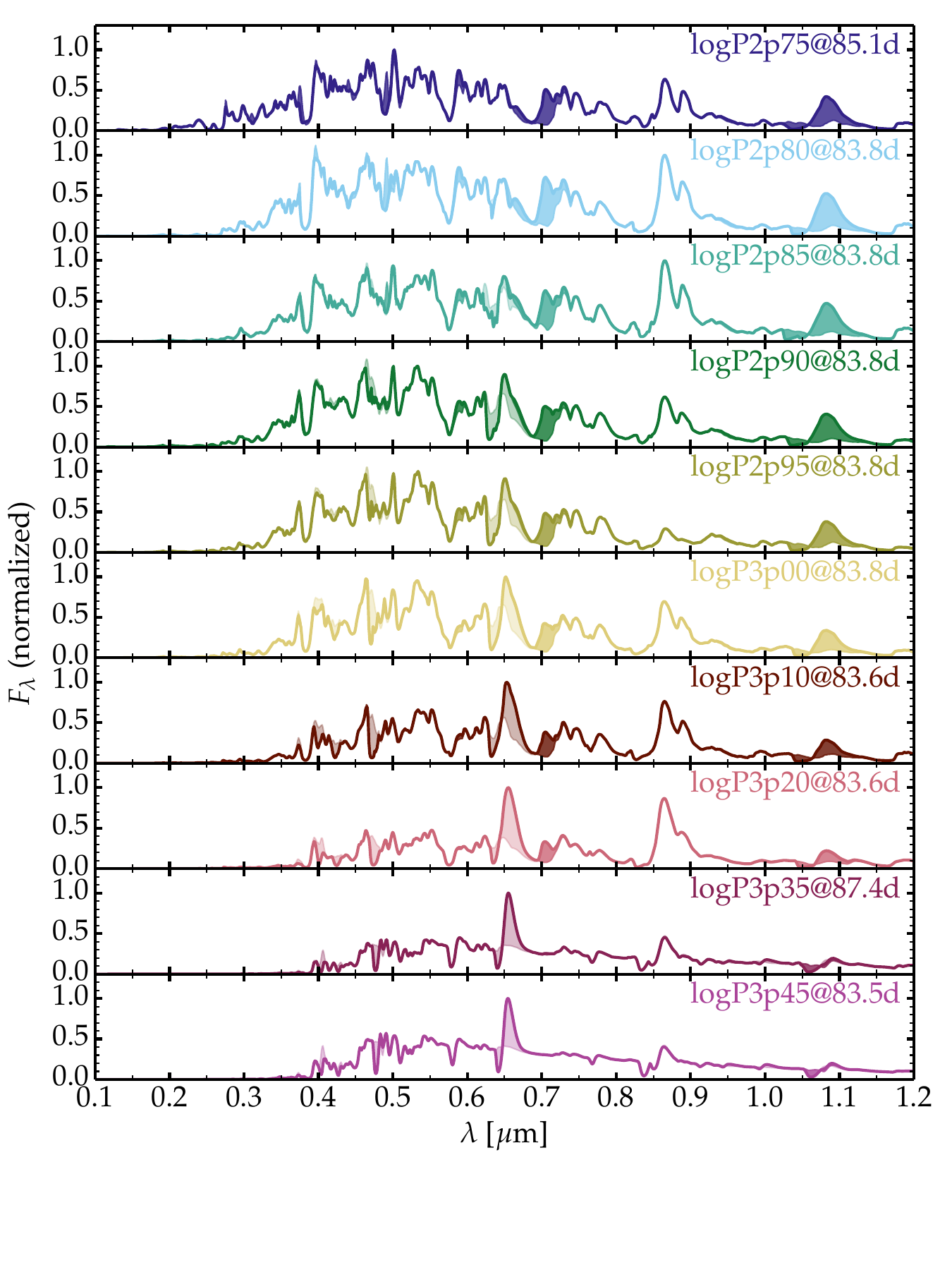}
    \end{subfigure}
    \vspace{-1cm}
    \caption{Spectral properties of our model set at about 5 (top-left), 15 (top-right), 30 (bottom-left), and 85\,d (bottom-right) after explosion. We stack our ten models from top to bottom in order of increasing H content, using the same color coding as before. For the montage at 5\,d, we show the quantity $\lambda^2 F_\lambda$ in order to reduce the contrast between UV and optical ranges. We also shade the flux associated with bound-bound transitions of H\one\ (light shade) and He\one\ (dark shade).}
\label{fig_spec_4epochs}
\end{figure*}

\section{Results: Diversity of evolution of model spectra}
\label{sect_spec}

A richer information is conveyed by the spectral evolution since it contains information on photometry (and in particular color) as well as on the temperature, ionization, composition, and kinematics of the ejecta. In Fig.~\ref{fig_spec_4epochs}, we show the spectra for our model set at about 5, 15, 30, and 85\,d after explosion.\footnote{For the models logP3p10, logP3p20, logP3p35, and logP3p45 at 5\,d (top-left panel of Fig.~\ref{fig_spec_4epochs}), we computed the spectra assuming steady state and fixing the temperature to the value obtained in the \voned\ calculation. This is not ideal but is used nonetheless to illustrate the spectral properties for these models prior to the phase of homologous expansion -- the outer ejecta layers are in fact close to homologous expansion at that time.} There are numerous differences between models and between epochs which reflect the different ejecta structure and composition.

At 5\,d after explosion (top-left panel of Fig.~\ref{fig_spec_4epochs}), the first striking spectral difference is that models logP2p75 and logP2p80 are optically bright and UV faint whereas all other models have the bulk of their flux emerging in the UV, and the more so the bigger the progenitor radius (which covers from about 600 to 900\,\rsun). The luminosity in those bluer models is much greater (see top-left panel of Fig.~\ref{fig_phot}). The other striking difference is the change in H\one\ (e.g., H$\alpha$) or He\one\ (e.g., He\one\,10830\,\AA) line widths and strengths, which are large in models logP2p75 (H-free model) and logP2p80, and weak in models with a greater H content. In the absence of a sizable H-rich envelope, the outer ejecta layers expand much faster and become much more transparent and cold, causing this red color, as well as the broader and stronger lines. In models from progenitors with a sizable H-rich envelope, a greater shock-deposited energy is radiated away and less turned into kinetic energy so the lines are systematically narrower. The weakness of the lines arises from the steep density profile in the outer ejecta (where the spectrum forms at such times; see Fig.~\ref{fig_voned}) as well as the relatively high ionization. Most optical lines are from the H\one\ Balmer series. In addition, SN ejecta from more compact RSG progenitors show spectra with stronger He\one\ lines whereas those from more extended RSG stars show stronger He\two\ lines (e.g., He\two\,4686\,\AA\ in the optical). This reflects a shift to higher temperature and ionization at the photosphere, rooted in the difference in progenitor radius. In the optical, these spectra exhibit weak lines but are not featureless (they might appear so in observations with low quality signal). Most strong lines reside in the UV. This includes resonant transitions from CNO elements, Al, as well as Fe (in particular the forest of lines from Fe\three; see for example the discussion of models tailored to the observations of the Type II-P SN\,2022acko at $\sim$\,5\,d after explosion; \citealt{bostroem_22acko_23}).

At 15\,d after explosion (top-right panel of Fig.~\ref{fig_spec_4epochs}), our model set shows a clear trend towards a bluer optical color as we progress to ejecta from progenitors with more massive and more extented H-rich envelopes. In model logP2p75, the spectrum has hardly changed (it already had a red color at 5\,d; see also top-left panel of Fig.~\ref{fig_spec_seq_ap1}) except for slightly stronger He\one\ lines and narrower line profiles (the photosphere has receded from $\sim$\,9000\,\kms\ to $\sim$\,7000\,\kms. In the rest of the model set, these He\one\ lines are progressively weaker. In lighter ejecta, He\one\ lines are influenced by non-thermal excitation by high-energy, Compton-scattered electrons \citep{lucy_91,swartz_ib_91}, while in heavier ejecta, the higher gas temperature and stronger photoionization are the main processes controlling their strength (the weakness of the lines in massive ejecta also results from the steeper density profile; Fig.~\ref{fig_voned}). Cooler ejecta in which the spectral energy distribution peaks around 4000\,\AA\ show clear signs of metal line blanketing in the optical (primarily associated with Ti\two\ and Fe\two), while the recombination phase has not yet started in bluer models ($T_{\rm ph}$ is about 10\,kK in model logP3p45 at that time). The most striking feature in our model set at this epoch is the strong decrease in H$\alpha$ absorption and emission strength, as well as in line width, as $M$(H) increases from models logP2p80/logP2p85 to logP3p45 -- this reflects the decrease in the volume over which the spectrum forms and little the actual H/He abundance ratio (the models with the stronger H$\alpha$ line here are those that are the more He enriched -- see Fig.~\ref{fig_mesa}). The more massive ejecta (with more H), have a lower $E_{\rm kin}/M_{\rm ej}$ and therefore yield narrower lines (best seen in H\one\ lines and in particular H$\alpha$).

At 30\,d after explosion (bottom-left panel of Fig.~\ref{fig_spec_4epochs}), all models exhibit a similar spectral morphology, with a spectral energy distribution peaking in the optical, which reflects their similar photospheric temperature of about 7000\,K. All models show signs of metal-line blanketing in the optical. He\one\ lines are strongest in model logP2p75 and decrease in strength for models with higher ejecta mass -- He\one\ lines are absent beyond model logP3p00 because the influence of \nifs\ is too weak and the He mass fraction smaller (even though the total He mass is greater in more massive ejecta). H\one\ lines (of which H$\alpha$ is the most visible Balmer transition) are present in models logP2p80 to lopP3p45 but in lighter models the lower H content yields a weaker feature. H$\alpha$ is strongest at this phase in intermediate models like logP2p90 and logP2p95. We also see a trend of decreasing strength for O\one\,7774\,\AA\ (likely a composition effect since the O-rich layers are revealed at this time in lighter ejecta) or the Ca\two\ NIR triplet (reflecting the lower ionization  and more extended ejecta in lighter models).

At 85\,d after explosion (bottom-right panel of Fig.~\ref{fig_spec_4epochs}), the physical conditions prevailing for the different models differ fundamentally since lighter ones are optically thin (e.g., logP2p75) while others are optically thick (e.g., logP3p45). With their greater He and \nifs\ content (i.e., relative to $M_{\rm ej}$), lighter models are slightly bluer in the optical although strong signs of metal-line blanketing are present in all models. The He\one\,10830\,\AA\ shows a clear trend of decreasing strength from logP2p75 to logP3p45. In the optical, the strongest He\one\ line is due to the transition at 7065\,\AA\ (see further discussion in Section~\ref{sect_he}). H$\alpha$ is strongly affected by overlapping lines (e.g., due to Fe\two) in lighter models. In heavier models, H$\alpha$ is weaker and narrower, primarily because of the steeper density gradient and slower velocity at the photosphere.

To summarize, the evolution of each model and the differences between models reflect the differences in $M_{\rm ej}$ (all models in our set have essentially the same $E_{\rm kin}$ so the ejecta density is typically greater and the expansion rate smaller in higher mass ejecta), in the relative H abundance (i.e., $M$(H)/$M_{\rm ej}$), which also determines how external in the ejecta H is present, in the relative He abundance (i.e., $M$(He)/ $M_{\rm ej}$), and in the relative \nifs\ abundance (i.e., $M$(\nifs)/$M_{\rm ej}$), which determines how strong non-thermal processes can impact elements like He. We show the more complete spectral evolution of all ten models in Figs.~\ref{fig_spec_seq_ap1}$-$\ref{fig_spec_seq_ap3}.

\begin{figure}
\includegraphics[width=0.49\textwidth]{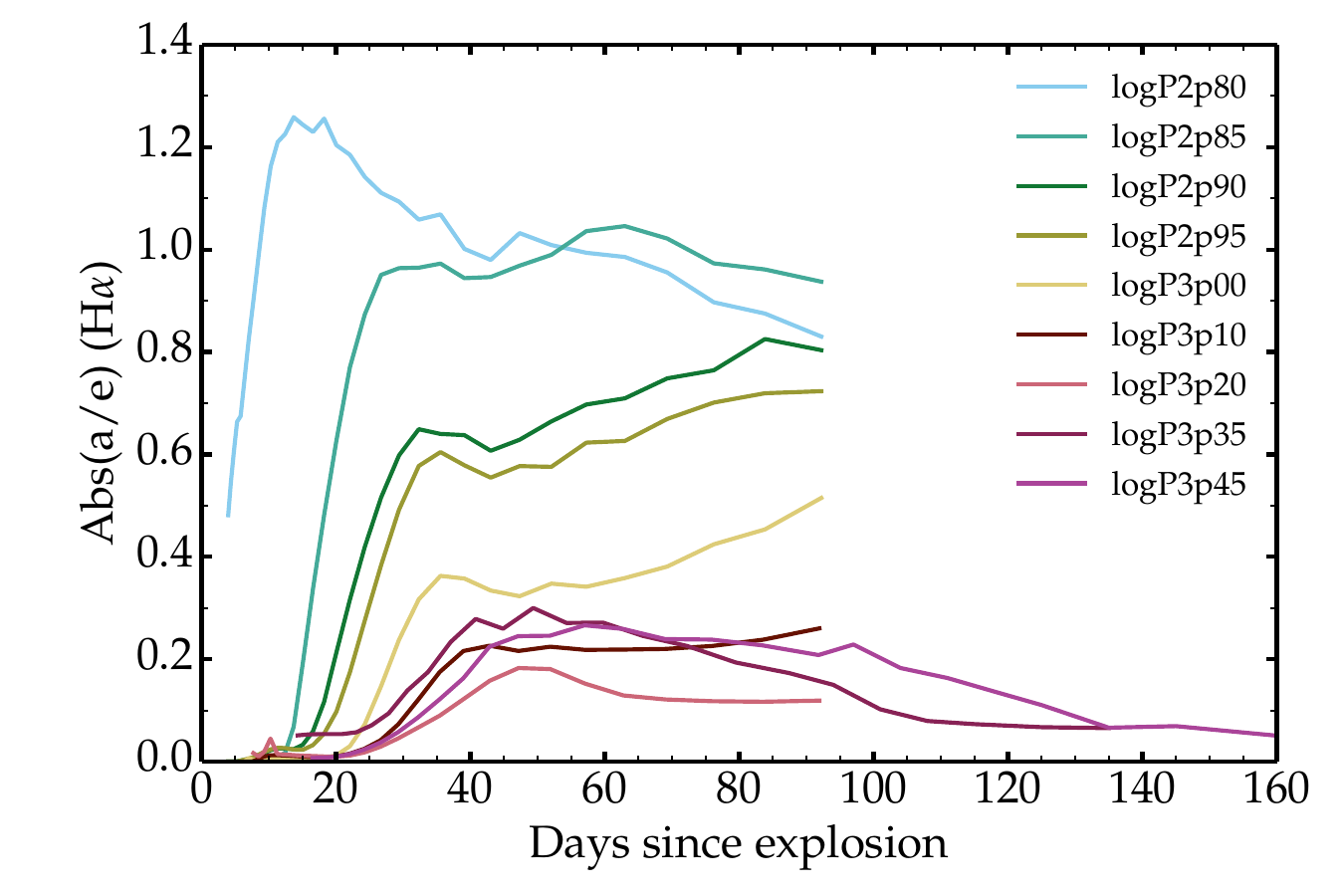}
\caption{Evolution of the flux ratio associated with the absorption and emission parts of H$\alpha$. This ratio is computed by integrating the flux over the absorption and emission parts of the H$\alpha$ profiles (see Fig.~\ref{fig_ae_halpha_indiv} for time sequences of these profiles).
\label{fig_ae}
}
\end{figure}

\section{Results: H$\alpha$ properties}
\label{sect_halpha}

Much discussion surrounds the H$\alpha$ transition in core-collapse SNe. It is the strongest line in optical spectra of SNe II at most epochs and is also the diagnostic line used to differentiate Type IIb from Type Ib SNe. There is a consensus that not much residual H is needed to produce a clear H$\alpha$ line, with values ranging from as low as 0.001\,\msun\ \citep{dessart_11_wr}, around 0.01\,\msun\ \citep{hachinger_13_he}, all the way to a few 0.1\,\msun\ \citep{benami_iib_15,d18_ext_ccsn}. Objects with double-peak $V$-band light curves or with a large bolometric luminosity for days after shock breakout (e.g., SN\,1993J; \citealt{richmond_93j_94}) are understood to arise from progenitors with an extended, H-rich envelope with a mass of order 0.1\,\msun\ \citep{blinnikov_94_93j,nomoto_93j_93,podsiadlowski_93j_93,woosley_94_93j,nakar_piro_14,d18_ext_ccsn}. The photometric signature is much more elusive, or even absent if the H-rich envelope mass is lower (see, e.g., \citealt{dessart_11_wr,bersten_etal_12_11dh}).

In binary progenitors, the amount of residual hydrogen of the primary star at core collapse is a function of the post-RLOF stellar wind mass loss. It also strongly depends on the initial orbital separation. For example, very low values of $M$(H) were obtained for the short-period, tight binary systems studied by \citet[that work also employed weak mass loss rates that facilitated the survival of H at the progenitor surface until core collapse]{yoon_ibc_10}. SN radiative-transfer  simulations based on such progenitors indicated that H$\alpha$ would be present, although only for a few days after shock breakout \citep{dessart_11_wr}. Other studies that included wider orbits yielded a broader range of residual H (see, for example, \citealt{yoon_ib_iib_17},  \citealt{gilkis_arcavi_22}, \citealt{ercolino_bin_23}).

The radiative-transfer simulations in this work are based on a subset of ten binary progenitor models from \citet{ercolino_bin_23}. Although limited to a primary star of 12.6\,\msun\ with an initial binary mass ratio of 0.95, they cover initial periods ranging from 562 up to 2818\,d and final $M$(H) of 0, 0.02, 0.06, 0.12, 0.15, 0.22, 0.37, 0.80, 2.68, and 4.64 (these values correspond to about 40 to 70\,\% of the progenitor H-rich envelope mass since the H/He mass-fraction ratio covers 0.69/0.30 in model logP3p45 down to 0.27/0.72 in model logP2p80). With our fixed explosion energy of \foe\ and moderate, ad-hoc chemical mixing, this preSN chemical stratification in mass leads to a stratification in velocity space. As indicated in Table~\ref{tab_cmfgen_init}, the minimum velocity bounding 99\,\% of the H decreases from 7120 (model logP2p80) to 781\,\kms\ (model logP3p45). This implies that no absorption or emission can occur in H$\alpha$ below that threshold (in the emergent H$\alpha$ profile, absorption and emission can be seen below this velocity because of the change in projected velocity for non-zero impact parameters and this may be aggravated by departures from spherical symmetry; see Section~5 of \citealt{D05_epm}). In those models, the depth at which H is present depends also on the adopted mixing whose exact magnitude is not robustly known and probably varies a great deal in nature.

Figure~\ref{fig_spec_4epochs} and Figs.~\ref{fig_spec_seq_ap1}$-$\ref{fig_spec_seq_ap3} show that H$\alpha$ is visible until about 50\,d in model logP2p80, until about 70\,d in model logP2p85, and at least until 90\,d in all other models with a higher $M$(H). In models logP3p35 and logP3p45, H$\alpha$ is visible throughout the nebular phase, as in typical RSG star explosions since in those events, H is mixed inwards while \nifs\ is mixed outwards, so that H and \nifs\ overlap in velocity space (see Table~\ref{tab_cmfgen_init} and specifically the values for $V_{99,{\rm H}}$ and $V_{99,^{56}{\rm Ni}}$). This means that in all models (except H-deficient model logP2p75), H$\alpha$ is visible until the nebular phase.

Figure~\ref{fig_ae} shows in more detail the evolution of the H$\alpha$ profile characteristics by comparing the flux ratio of the absorption and emission parts. To make these measurements on the H$\alpha$ profile, we computed the spectrum accounting for all bound-bound transitions and subtracting from it the spectrum obtained when all but H\one\ transitions are included. Over the H$\alpha$ profile region (shown for all H-rich models in Fig.~\ref{fig_ae_halpha_indiv}), the flux is zero far away from the line (at Doppler velocities exceeding the maximum ejecta velocity), negative in the blue-shifted part (corresponding to the P-Cygni trough), and positive as we move closer to line center and into the so-called ``red-wing". Besides the notorious blue-shift of peak emission \citep{DH05a,anderson_blueshift_14}, we see that both emission and absorption persist at all epochs in all cases. The absorption component is typically weaker than the emission part. This is related to complicated radiative transfer effects (i.e., how the line and continuum source functions compare, how the ionization varies with depth above the photosphere), but it is also the result of a simple geometric effect since the emission arises potentially from a large volume (this volume is greater for flatter density profiles) while the absorption is fundamentally limited to the column of material falling onto the radiating disk (roughly limited by an impact parameter equal to the photospheric radius). This geometric effect dominates at late times in models like logP3p45. In model logP2p85 or logP2p90, the H$\alpha$ absorption extends from about 10\,000\kms\ out to 25\,000\,\kms, hence essentially from the photosphere out to the outermost ejecta layers at large velocity. The maximum range of velocities is much more limited in more massive H-rich ejecta (see Fig.~\ref{fig_voned}) so there is little space between the photosphere and the outermost ejecta layers to accommodate a broad absorption.

Figure~\ref{fig_ae_halpha_indiv} also reveals the intrinsic morphology and characteristics of the H$\alpha$ profile. For example, in model logP2p80, the line covers from $-$\,25\,000 to $+$\,20\,000\,\kms\ at early times and eventually narrows down to exhibit at late times a narrow absorption at about $-$\,11\,000\,\kms\ together with a weak flat-topped emission bounded by a similar velocity. This velocity of 11\,000\,\kms\ is where the H mass fraction drops to about 0.1 as we progress inwards from the outermost ejecta layers (the absorption exhibits two minima at 11\,000 and 13\,000\,\kms, which correspond to the location of two maxima in the density profile; see Fig.~\ref{fig_voned}). In model logP2p85, the morphology is analogous but the velocities are smaller, the peak blueshift even more marked at early times, the flat-topped profile at late times is narrower, and the P-Cygni absorption exhibits multiple minima reflecting the presence of density extrema at about 8000, 10\,000, and 16\,000\,\kms. Such extended troughs are present at late times in models with $M$(H) between 0.06 and 0.37\,\msun\ (progenitors with a H-rich envelope mass between 0.14 and 0.59\,\msun). These troughs are narrow and weak at late times in models logP3p20, logP3p35, and logP3p45 because of the limited range in ejecta velocity and the large emitting volume of the H-rich ejecta (this emission fills in the absorption part).

\section{Results: Evolution of Helium lines}

\label{sect_he}

Our set of simulations also exhibit interesting features and paradoxes in relation to He lines. As shown earlier in Fig.~\ref{fig_spec_4epochs} and in Figs.~\ref{fig_spec_seq_ap1}$-$\ref{fig_spec_seq_ap3}, He lines are present in all models at early times. Models logP2p75 and logP2p80 have cool and partially ionized photospheres at essentially all epochs past 2\,d so that the excitation of He\one\ lines is caused by collisions with high-energy electrons themselves produced through scattering with $\gamma$-rays emitted by decaying \nifs\ and \cofs\ isotopes. In these models, He\one\ lines (e.g., He\one\,5875, 6678, 7075, and 10830\,\AA) persist through this process throughout the evolution (i.e., thus out to at least 90\,d). In models whose progenitors have a residual H-rich envelope greater than 0.1\,\msun, the presence of He lines at early times is caused by the large photospheric temperature and photoionizing flux. That temperature is large enough to produce He\two\ lines (in particular He\two\,4686\,\AA; this line is obvious in all models beyond logP2p90)  or only He\one\ lines (e.g., at 5875\,\AA; model logP2p85 at $\sim$\,5\,d). In models with intermediate $M$(H) values, the transition from photoionization to non-thermal excitation as the process behind the production of He\one\ lines occurs after about 20\,d. In more massive ejecta, the He\one\ lines are essentially absent in the optical after the onset of the recombination phase, apart from He\one\,10830\,\AA\ in the NIR, which persists until late times. It is thus paradoxical, though not surprising, that the ejecta with the largest amount of He are those that show weaker and only short-lived He lines.

Surprisingly, the He\one\,7065\,\AA\ line is predicted in all models once they turn nebular -- it strengthens as the continuum optical depth drops. This line is present earlier on (i.e., during the photospheric phase) as we progress from model logP3p45 to logP2p75 (i.e., in order of increasing $M$(He)/$M_{\rm ej}$).  It is much more obvious than the weak He\one\,6678\,\AA\ but also He\one\,5875\,\AA, which overlaps with Na\,\one\,D. Although not the main focus of this work, this He\one\,7065\,\AA\ line offers an interesting diagnostic to test for the presence of He in the ejecta. The detection of He\one\ lines  is the sole criterion for differentiating Type Ib from Type Ic SN and this task is trivial when such He\one\ lines are strong. However, if they are weak, one is often limited to searching absorptions related to those He\one\ transitions (see, e.g., \citealt{liu_snibc_15} or \citealt{williamson_ibc_19}), and such absorptions only probe for the presence of He within the column of material falling onto the photodisk.\footnote{The alternative is to use the NIR and in particular search for the unblended He\,\one\,20581\,\AA\ line; see, for example, \citet{shahbandeh_nir_sesn_22}.} In an aspherical explosion where most of the He may reside along other lines of sight, this absorption may be weak or even absent, complicating the identification of He\one\ lines. In contrast, when searching for signs of He at later times, after bolometric maximum and as the ejecta become optically thin, a line like He\one\,7065\,\AA\ would be ideal since it does not suffer from overlap with Na\,\one\,D or Fe\two\ lines, and it is also predicted as a strong emission line.

\input{table_obs.tex}

\section{Comparison to observations}
\label{sect_obs}

In this section, we compare a few models (using both light curves and multiepoch spectra) from our set to well observed, prototypical  core-collapse SNe of various kinds, that is of Type Ib (i.e., iPTF13bvn; Section~\ref{sect_ib}), Type IIb (i.e., SNe\,2011dh and 1993J; Section~\ref{sect_iib_11dh} and \ref{sect_iib_93J}), Type II-L (aka fast decliners; i.e., SN\,2006Y; Section~\ref{sect_iil}), and Type II-P (i.e., SN\,2017eaw; Section~\ref{sect_iip}). The origin of the photometric and spectroscopic data that we use together with some of their inferred characteristics (i.e., redshift, distance modulus, extinction, and explosion date) are presented in Table~\ref{table_obs}.  This comparison is mostly qualitative and focused on assessing whether the models are fundamentally compatible with observations (e.g., if a slight shift in kinetic energy could resolve a discrepancy) or whether there is something fundamentally inadequate about the model (e.g., there is a missing physics, process, or ``ingredient'' in the model). The comparison is also limited to photospheric epochs (or at most the onset of the nebular phase). At later times, a different approach is used in \cmfgen\ simulations in order to treat chemical mixing in the ejecta \citep{DH20_shuffle}, and this approach has been applied to Type II SNe from 9 to 29\,\msun\ single-star progenitors \citep[specifically, 9.0, 9.5, 10.0, 10.5, 11.0, 12., 12.5, 13.5, 14.5, 15.2, 15.7, 16.5, 17.5, 18.5, 20.1, 21.5, 25.2, and 26.5\,\msun]{D21_sn2p_neb} as well as  from 2.6 to 12\,\msun\ He-star progenitors \citep[corresponding to a zero-age main sequence mass between 13.85 up to 35.74\,\msun; the specific He-star models have masses of 2.6, 2.9, 3.3, 3.5, 4.0, 4.5, 5.0, 6.0, 7.0, 8.0, and 12.0\,\msun, including progenitor models evolved with different mass loss prescriptions]{dessart_snibc_21}.

To address some of the mismatches and to give a sense of how ejecta properties impact the observables, we also include models computed in previous work using a similar numerical approach with \mesa, \voned, and \cmfgen. Namely, we additionally include :
\begin{enumerate}
\item The Type Ib model he4 from \citet{dessart_snibc_20}, with $M_{\rm ej}=$\,1.49\,\msun, $E_{\rm kin}=$\,0.75\,$\times$\foe, $M$(\nifs)$=$\,0.08\,\msun, and $M$(He)$=$\,1.13\,\msun. Strong chemical mixing was enforced in this model, as well as all others in that work. This model is used in Section~\ref{sect_ib}.
\item The Type IIb model 3p65Ax1 from \citet{D15_SNIbc_I}, characterized by moderate chemical mixing, and with $M_{\rm ej}=$\,2.22\,\msun, $E_{\rm kin}=$\,1.24\,$\times$\foe, $M$(H)$=$\,0.005\,\msun, $M$(He)$=$\,1.49\,\msun, and $M$(\nifs)$=$\,0.074\,\msun. This model is used in Section~\ref{sect_iib_11dh}.
\item  The Type II-L/II-P model x2p0 from \citep{HD19}. This model corresponds to a solar-metallicity single-star model for a 15\,\msun\ progenitor on the zero-age main sequence but evolved with a twice greater RSG mass loss rate relative to the nominal value in \mesa\ (as adopted in the corresponding version -- see \citealt{HD19} for details). The preSN progenitor has an H-rich envelope mass of 9.24\,\msun. The corresponding ejecta have $E_{\rm kin}=$\,1.2$\times$\,\foe,\ and $M$(\nifs)$=$\,0.036\,\msun. This model is used in Section~\ref{sect_iip}.
\end{enumerate}

\begin{figure}
    \centering
    \begin{subfigure}[b]{0.485\textwidth}
       \centering
       \includegraphics[width=\textwidth]{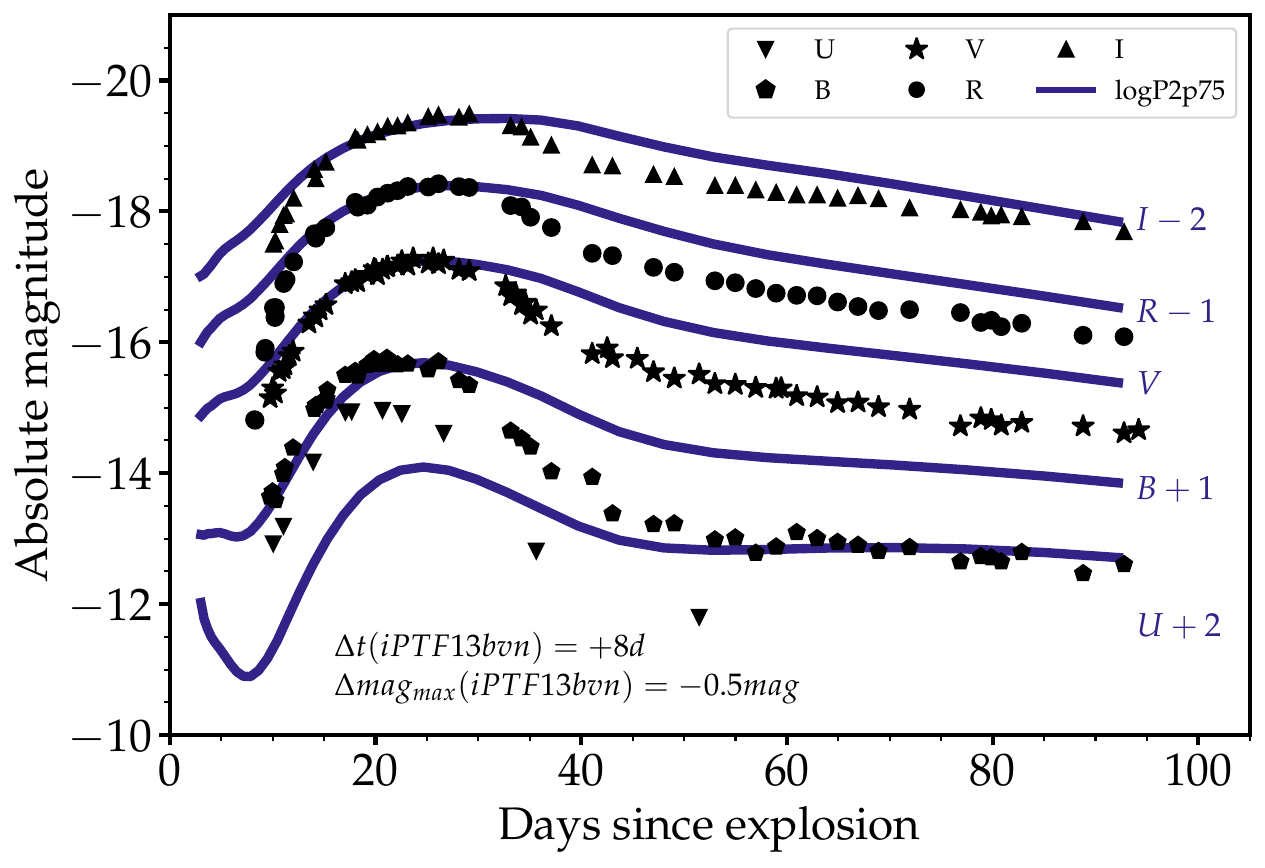}
    \end{subfigure}
    \centering
    \begin{subfigure}[b]{0.485\textwidth}
       \centering
       \includegraphics[width=\textwidth]{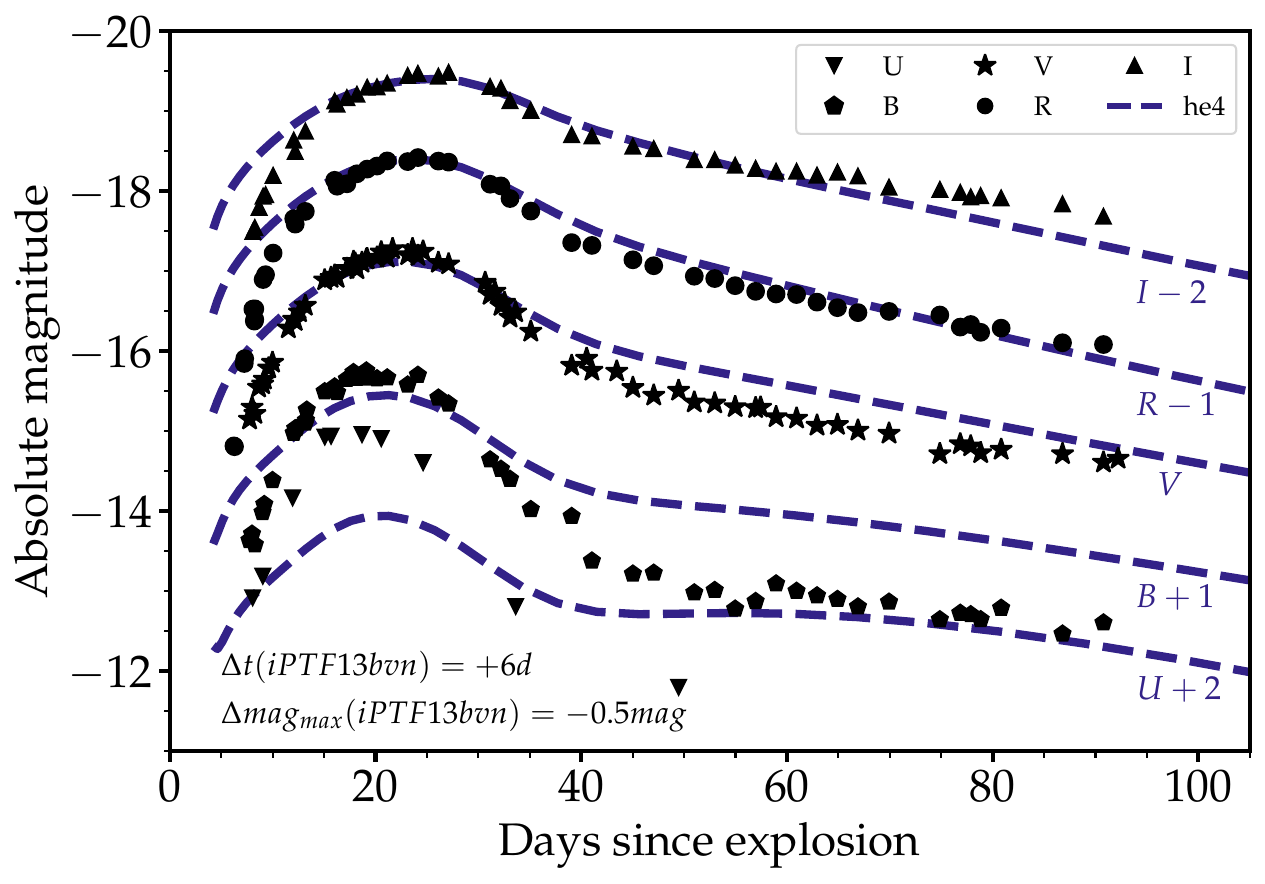}
    \end{subfigure}
    \centering
    \begin{subfigure}[b]{0.485\textwidth}
       \centering
       \includegraphics[width=\textwidth]{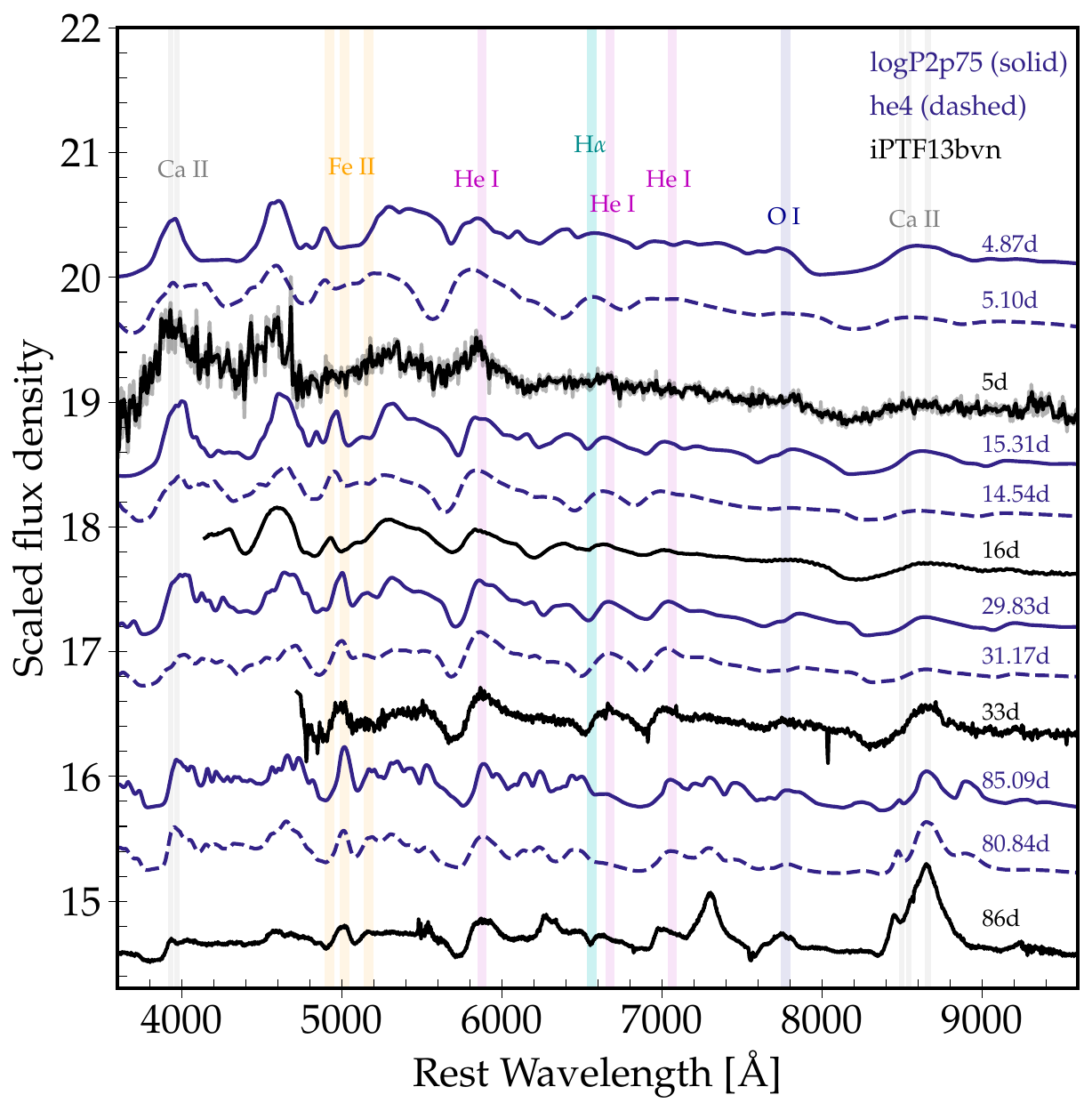}
    \end{subfigure}
    \caption{Comparison of model light curves and spectra with the observations of iPTF13bvn. Top: we show the photometric data (black symbols) shifted in time by $+8$\,d and in magnitude by $-0.5$\,mag as well as the results for model logP2p75 (solid line). With such shifts, data and model are in close agreement at $V$-band maximum.  Middle: Same as top but for the model he4 of \citet{dessart_snibc_20}, now with a shift in time of $+6$\,d. Bottom: Comparison of multiepoch spectra at about 5, 15, 30, and 85\,d after explosion between the observations of SN iPTF13bvn and models logP2p75 and he4 (no time shift has been applied to the observed spectra).}
\label{fig_13bvn}
\end{figure}

\subsection{Type Ib SN iPTF13bvn and model logP2p75}
\label{sect_ib}

The top panel of Fig.~\ref{fig_13bvn} compares the multiband optical light curves of type Ib SN iPTF13bvn with the results obtained for model logP2p75 whose H-deficient progenitor had a surface radius of 12\,\rsun. The model rises one week too slowly and is 0.5\,mag brighter than iPTF13bvn (both shifts were applied to the data in  Fig.~\ref{fig_13bvn}). The model stays too bright after maximum (i.e., it declines too slowly) and at late times in all bands except in the $U$ band. However, the overall shape of the observed light curves is reproduced, including the shorter rise time to maximum in bluer bands. The color at most epochs is also well matched (with the global magnitude shift of $-0.5$\,mag) although there is a marked offset in the $U$ band by about 1\,mag at all times (model is too faint early on and too bright later on).

The bottom panel of Fig.~\ref{fig_13bvn} compares the multiepoch spectra of type Ib SN iPTF13bvn with the results obtained for model logP2p75, specifically at about 5, 15, 30, and 85\,d. The model predicts essentially all observed lines. Overall, the model line widths are too small, with the exception of \caiitrip. The ejecta density of the model is also too high at late times because the model exhibits essentially no \caiidoub, which only appears when the material density is low enough for forbidden-line formation. In the optical spectrum at 85\,d, all model lines appear much narrower than observed.

Some of these discrepancies are reduced by using model he4 ($E_{\rm kin}=$\,0.75\,$\times$\,10$^{51}$\,erg, $M_{\rm ej}=$\,1.49\,\msun, $M$(\nifs)$=$\,0.08\,\msun) from \citet{dessart_snibc_20} -- see middle and bottom panels of Fig.~\ref{fig_13bvn}. The rise time is shorter by 2\,d, and this slight gain arises because of the lower ejecta mass and the 5\% higher $E_{\rm kin}/M_{\rm ej}$. The stronger chemical mixing causes a faster brightening (more \nifs\ at large velocities) and the faster decline (more $\gamma$-ray escape at late times), but the light curve is now too broad. The He\one\ lines are too broad at early times (which would argue against the adopted, strong mixing of \nifs), but well matched after 30\,d. Another way to reduce the discrepancy is to increase the explosion energy. This would shorten the rise time, brighten the peak, and steepen the fading at late times due to the enhanced $\gamma$-ray escape. The \nifs\ mass would need to be reduced since the peak would be too bright, but this might bring the late-time brightness too low. We leave this to future work.

Both models have roughly the same brightness as observed at the earliest times (i.e., without applying a time shift of the data by six or eight days) so the luminosity of the outermost ejecta layers is compatible with observations -- it is the rate at which radiation leaks from within the ejecta that is too small. In other words, the brightening rate is too slow. The simulations of stripped-envelope SNe with \cmfgen\ always yield rise times in excess of 20\,d unless the explosion energy is enhanced considerably, bringing in discrepancies with line width or late-time decline rates \citep{D15_SNIbc_I,D16_SNIbc_II} -- reducing the ejecta mass shortens the rise time but low-mass He-rich ejecta no longer match the nebular-phase spectra of typical SNe Ib, which systematically show strong \oidoub\ (see how weak \oidoub\ is in model he3p3 of \citealt{dessart_snibc_21}). Results from 1D grey, flux-limited diffusion, lagrangian, radiation hydrodynamics simulations yield a good match to the rise time and peak luminosity of iPTF13bvn with ejecta parameters similar to those employed here \citep{bersten_iPTF13bvn_14}. This suggests that different methods yield different diffusion models for the ejecta. There may be concerns that the opacities, and therefore the diffusion time, are overestimated in \cmfgen\ (e.g., because of the adoption of a turbulent velocity of 50\,\kms), but a similar concern holds for radiation hydrodynamics calculations since they use approximate opacity means that typically hold in the dense interior of stars (i.e., these codes might underestimate the opacities and the diffusion time).

A source of physical error in our \cmfgen\ simulations comes from the assumption of a smooth and spherically symmetric ejecta. In reality, the \nifs\ bubble effect \citep{woosley_87A_late_88} should lead to a strong expansion and rarefaction of the \nifs-rich regions and a compression of the surrounding gas, causing clumping on small scales and large voids, as predicted by long-term simulations of the neutrino-driven explosions \citep{gabler_3dsn_21} and also directly inferred from observations of SN remnants like Cas A (see., e.g., \citealt{milisavljevic_fesen_casA_15}). Small scale clumping can hasten the recombination of the ejecta material and speed up the recession of the photospheric layers, thereby releasing the energy stored from optically-thick layers \citep{d18_fcl}. In addition, larger scale inhomogeneities, arising in part from the \nifs-bubble effect,  can also reduce the effective optical depth of the ejecta and contribute to further enhance the release of stored energy from the ejecta \citep{dessart_audit_rhd_3d_19}. Both effects have been confirmed to reduce the rise time and enhance the peak brightness of stripped-envelope SNe  \citep{ergon_11dh_22,ergon_20acat_23}.

\begin{figure}
    \centering
    \begin{subfigure}[b]{0.485\textwidth}
       \centering
       \includegraphics[width=\textwidth]{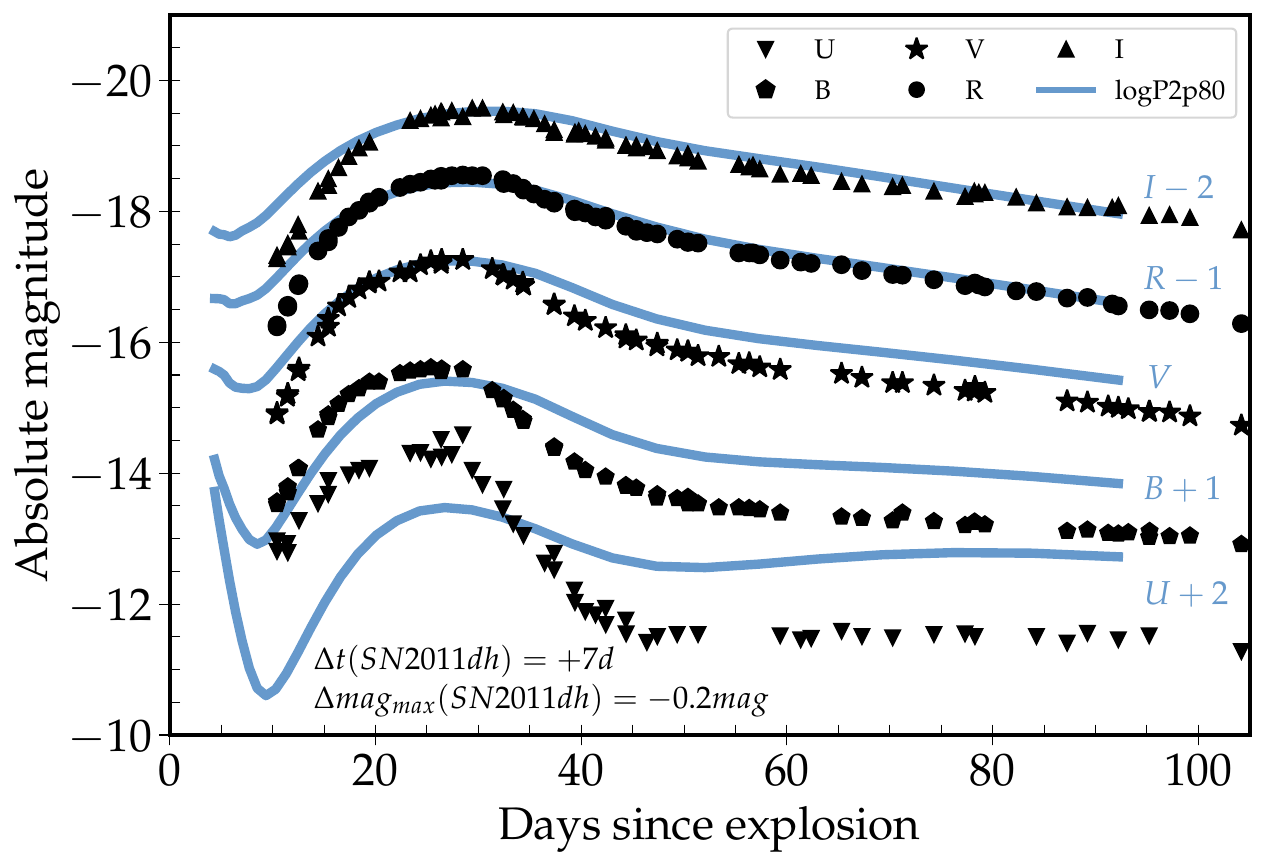}
    \end{subfigure}
   \centering
    \begin{subfigure}[b]{0.485\textwidth}
       \centering
       \includegraphics[width=\textwidth]{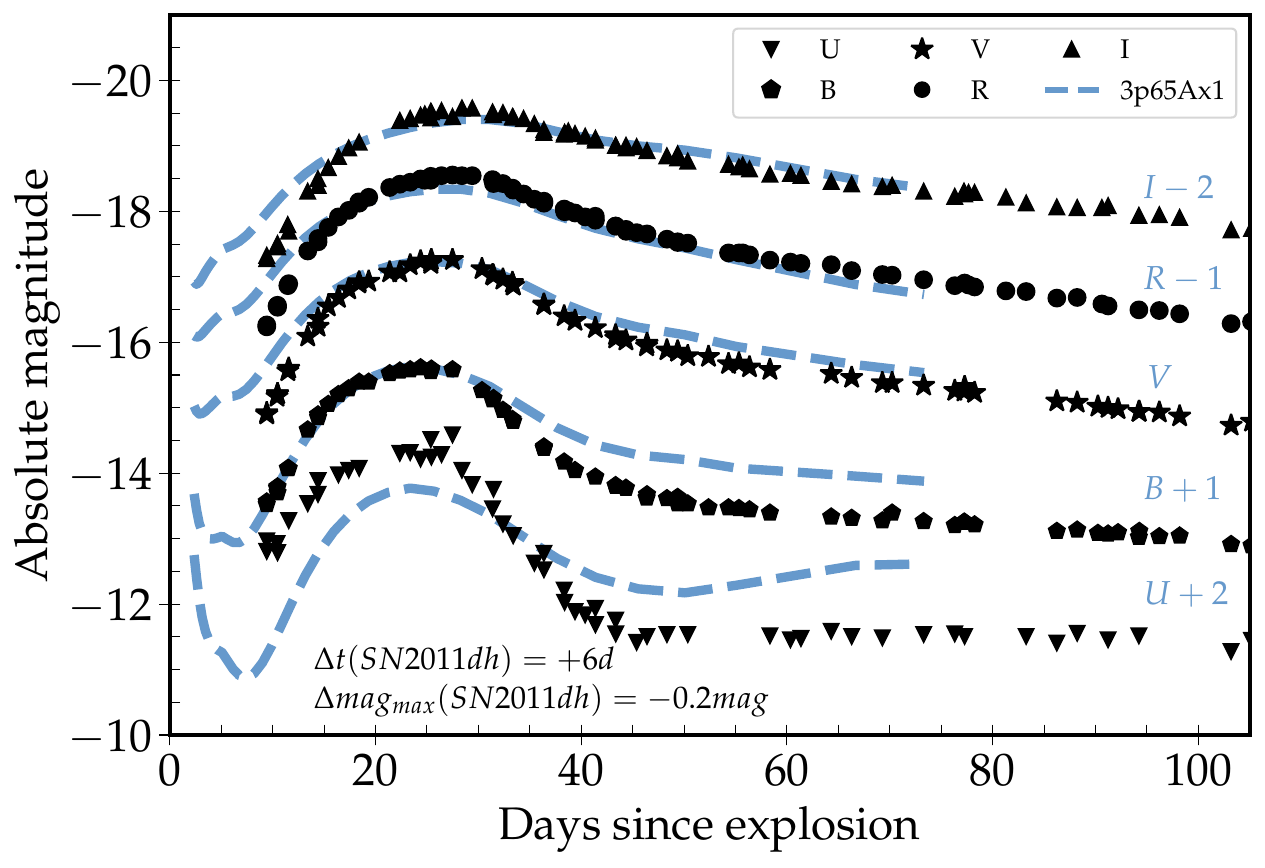}
    \end{subfigure}
    \centering
    \begin{subfigure}[b]{0.485\textwidth}
       \centering
       \includegraphics[width=\textwidth]{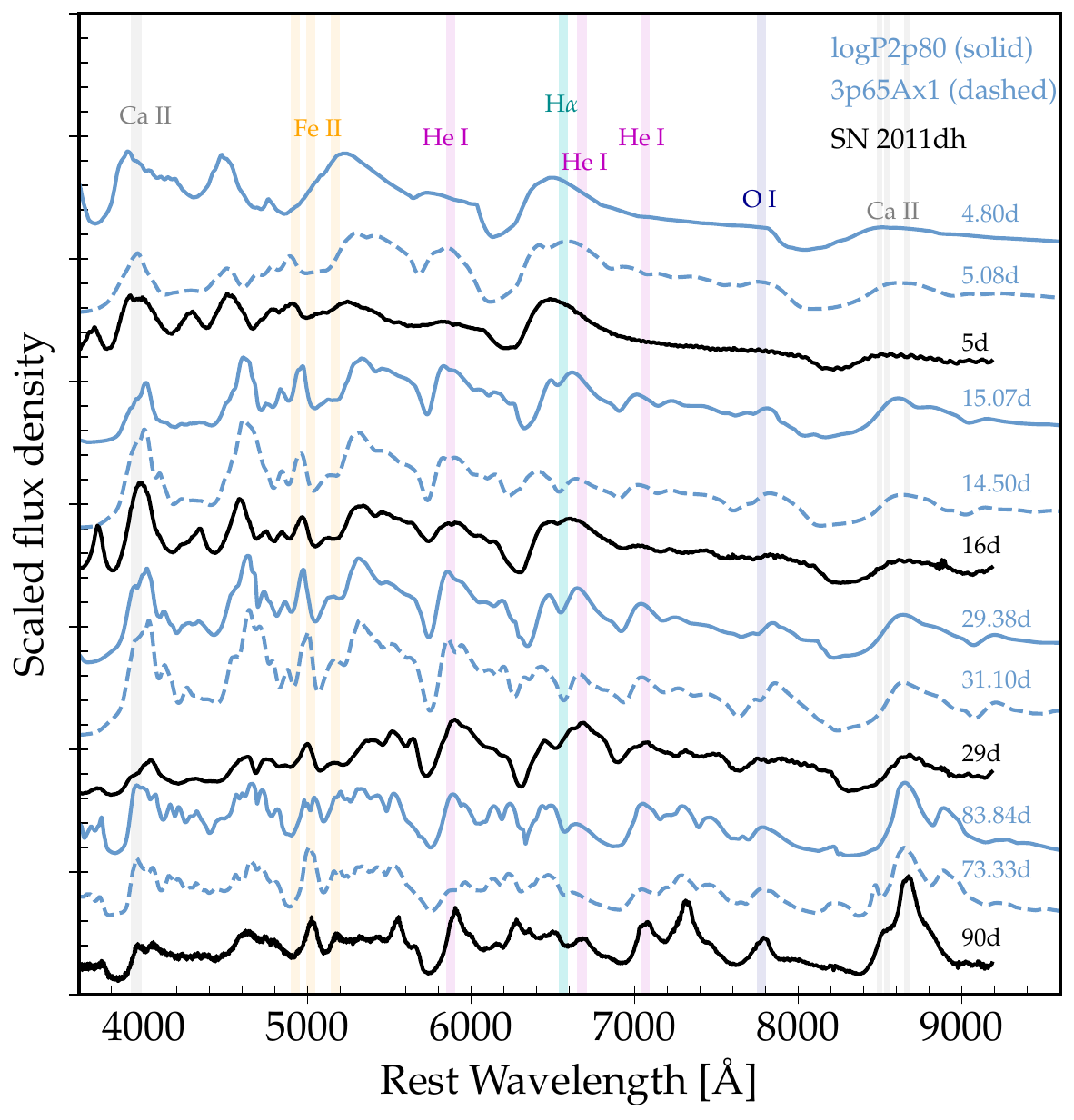}
    \end{subfigure}
    \caption{Comparison of model light curves and spectra with the observations of SN\,2011dh. Top: we show the photometric data (black symbols) shifted in time by $+7$\,d and in magnitude by $-0.2$\,mag as well as the results for model logP2p80 (solid line).  Middle: Same as top but for the model 3p65Ax1 of \citet{D15_SNIbc_I}, now with a shift in time of $+$6\,d. Bottom: Comparison of multiepoch spectra at about 5, 15, 30, and 85\,d after explosion for SN\,2011dh and models logP2p80 and 3p65Ax1.}
\label{fig_11dh}
\end{figure}

\subsection{Type IIb SN\,2011dh and model logP2p80}
\label{sect_iib_11dh}

The top panel of Fig.~\ref{fig_11dh} compares the multiband optical light curves of type IIb SN\,2011dh with the results obtained for model logP2p80, whose progenitor had a surface radius of 361\,\rsun\ and a H-rich envelope mass of 0.05\,\msun. The disagreements between model and observations are similar to those discussed in the previous section, namely that the model rises too slowly by about 7\,d and the model is too blue at late times. Here too the model and the observations have initially the same brightness so it is the brightening rate of the model on the rise to photometric maximum that is too slow. The model predicts a clear inflection in all bands at 5$-$10\,d, which is not obviously observed. Hence, the progenitor H-rich envelope is a little too massive and extended, yielding a luminosity that is too large during the first five to ten days after explosion. Omitting this brightness offset at early times and the rise time, the $R$-band light curve is very well reproduced by the model -- this good match would be degraded in the absence of the time shift applied to the data (i.e., it would require adjusting the ejecta kinetic energy and \nifs\ mass).

The bottom panel of Fig.~\ref{fig_11dh} compares the multiepoch spectra of type IIb SN\,2011dh with the results obtained for model logP2p80 (thick line). Overall, the spectral evolution is well matched by the model but there are offsets. The model is bluer at 5\,d and after maximum, but not blue enough around peak (though not by much). The H$\alpha$ line is too broad early on but too narrow at 15\,d, likely because we overestimated the amount of H mixed inwards to lower velocity (likely in addition to overestimating $M$(H-env)). The He\one\ lines are well matched in width and strength at all times, which suggests that the model has an adequate stratification of helium (and of \nifs, which is the source of excitation of He\one\ lines). \caiitrip\ is too broad at all times. Either the outer ejecta stretch out too much to large velocity or something quenches the ionization in those outer layers, or both.

In the middle and bottom panels of Fig.~\ref{fig_11dh}, we also show the photometric and spectroscopic results for model 3p65Ax1 of \citet{D15_SNIbc_I}. The smaller and lighter progenitor H-rich envelope lead to an ejecta with a faster drop in post-breakout luminosity. The weaker chemical mixing leads to a short post-breakout plateau before a brightening that is also too slow. However, this model reproduces better the spectral evolution of SN\,2011dh than model logP2p80, in particular concerning the width, strength, and survival time of H$\alpha$.

Most deficiencies of both models may be related to the ejecta clumping and porosity related to the \nifs\ bubble effect, which were both taken into consideration, and found to be important, in the study of \citet{ergon_11dh_22} -- see also \citet{jerkstrand_15_iib} for models supporting similar structure from nebular phase spectra. An interesting mismatch of our spectral models relative to SN\,2011dh at late times is that both \oidoub\ and \caiidoub\ appear too weak. The \oidoub\ would strengthen by lowering the ionization, as could occur if the O-rich material was clumped by the surrounding expanding \nifs-rich material, while that lower-density \nifs-rich material may facilitate the early emergence of the forbidden \caiidoub. This needs further study.

\begin{figure}
    \centering
    \begin{subfigure}[b]{0.485\textwidth}
       \centering
       \includegraphics[width=\textwidth]{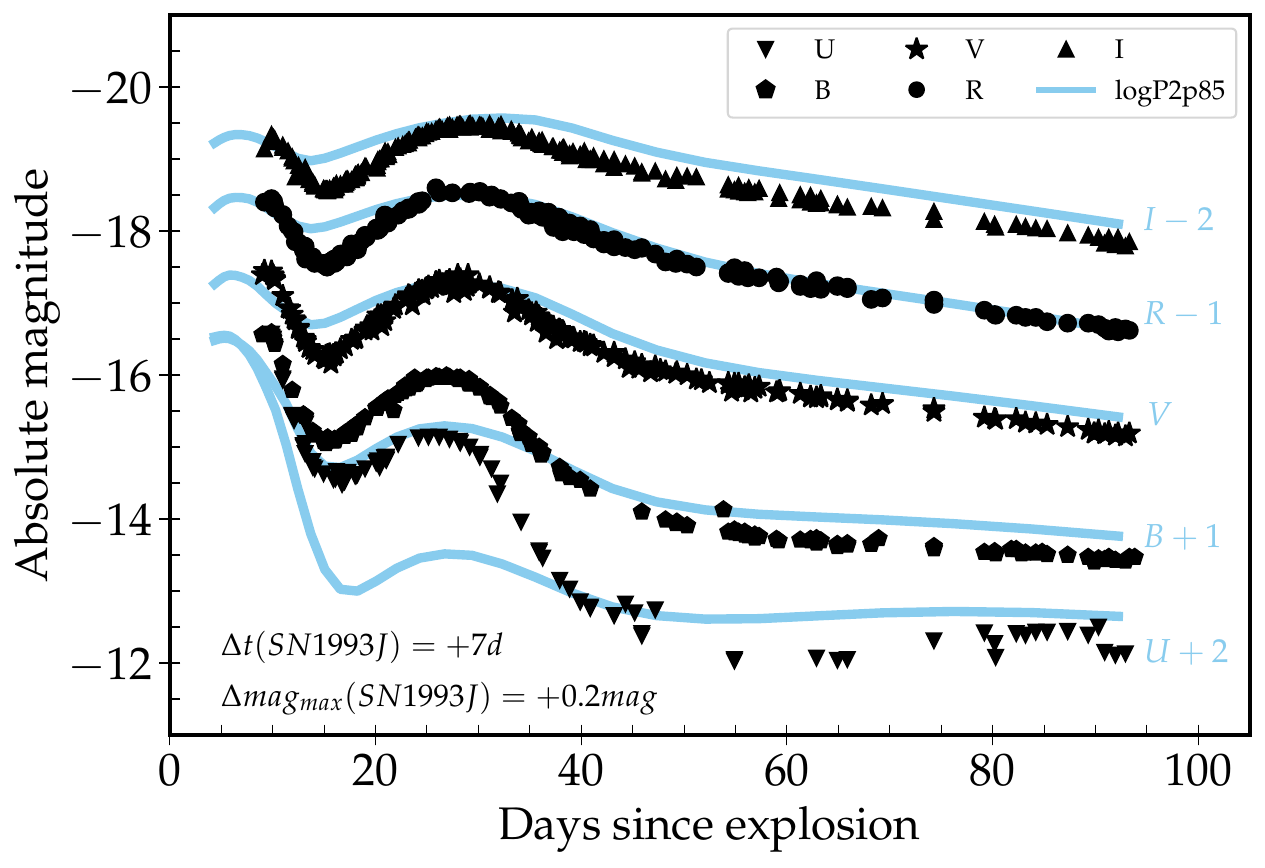}
    \end{subfigure}
    \centering
    \begin{subfigure}[b]{0.485\textwidth}
       \centering
       \includegraphics[width=\textwidth]{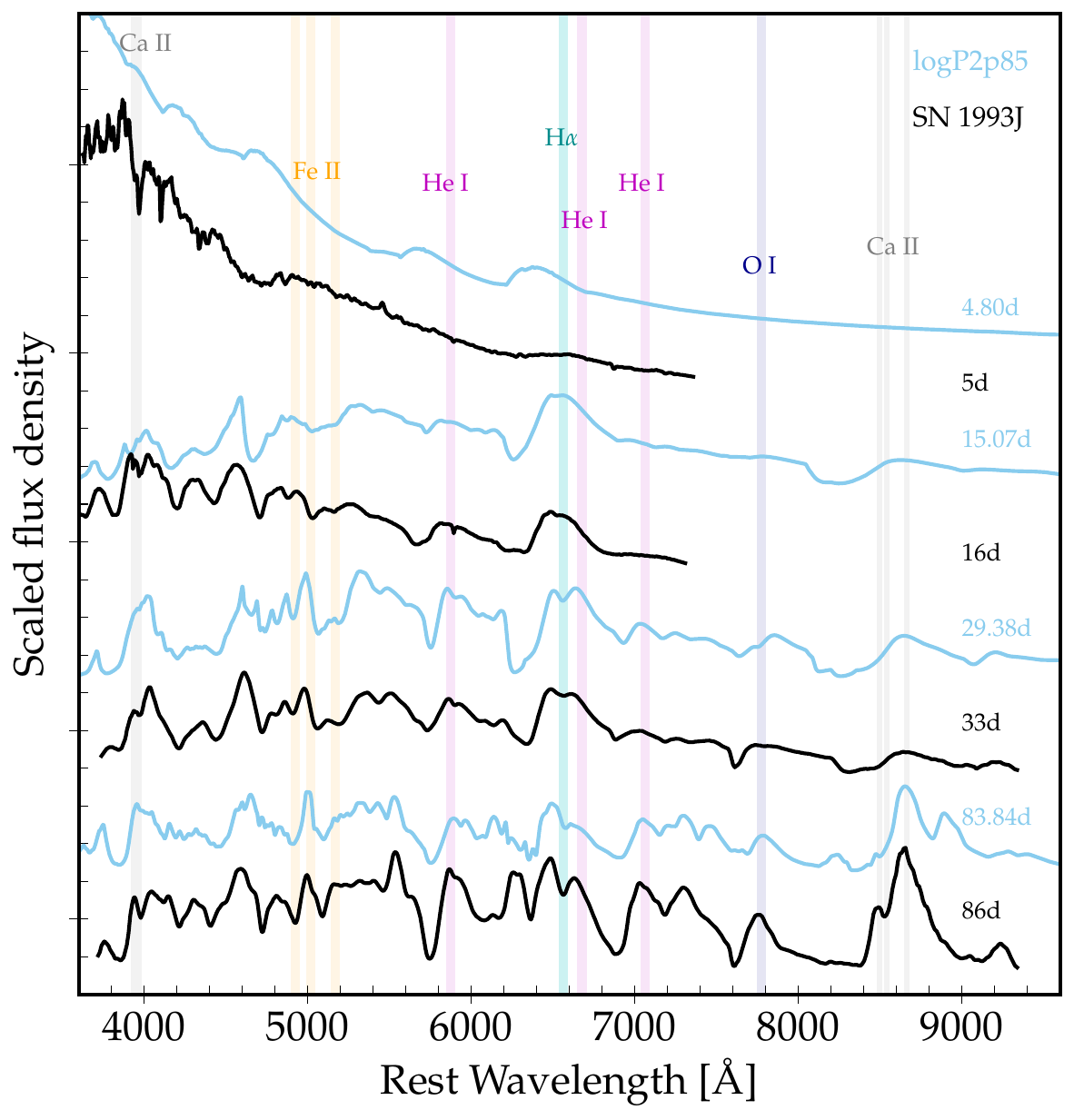}
    \end{subfigure}
    \caption{Same as Fig.~\ref{fig_11dh} but now for model logP2p85 and the observations of SN\,1993J.}
\label{fig_93J}
\end{figure}

\subsection{Type IIb SN\,1993J and model logP2p85}
\label{sect_iib_93J}

For the Type IIb SN\,1993J, we use for comparison the model logP2p85 whose progenitor had an H-rich envelope of 0.14\,\msun\ and a surface radius of 620\,\rsun. The model matches the two-peak light curve in all bands but only the timing of the first peak is well matched while the second peak occurs one week too late (the opposite impression is given in the top panel of Fig.~\ref{fig_93J} because the data was shifted by $+7$\,d to match the main \nifs-powered peak). The drop in brightness between the two peaks is too small, which suggests that the power released from the shocked envelope lasts for too long. The envelope may be too massive, too extended, perhaps both. As for the comparison to SN\,2011dh, the model matches well the (shifted) $R$-band light curve of SN\,1993J.

The spectral comparison shown in the bottom panel of Fig.~\ref{fig_93J} shows a good overall agreement for most lines of H\one, He\one, Ca\two\ or Fe\two. There are noticeable offsets at early times when the model does not match the observed quasi featureless optical spectrum. The model overpredicts the absorption strength of most lines (particularly for H$\alpha$). This may be caused by asphericity (see, for example, evidence of this from light-echo spectra of Cas A; \citealt{rest_casA_echo_11}) or from interaction with a wind, which would occasion  flux dilution from excess continuum radiation in the optical (see, for example, \citealt{d23_interaction}).

The overall agreement of model logP2p85 with SN\,1993J supports, but is also confirmed by, former modeling work of  SN\,1993J for which similar parameters were inferred \citep{blinnikov_94_93j,nomoto_93j_93,podsiadlowski_93j_93,woosley_94_93j}. However, when looking into the details of our mismatches, we argue that there is evidence here too for the effect of clumping and porosity, which were not taken into account in those older works nor in the present one.

\subsection{Type II-L SNe and models with intermediate $M$(H) values}
\label{sect_iil}

There does not appear to be any clear observational counterpart to our models with $M$(H) between 0.15 and 0.80\,\msun\ (models logP2p95 to logP3p20). They exhibit a fast-declining bolometric light curve, in principle qualitatively analogous to those of Type II-L SNe, but their $V$-band light curve exhibits two bright peaks that may even merge into a short-lived plateau of $\sim$\,50\,d before turning nebular. The first $V$-band peak is brighter than the main \nifs-powered peak, although the relative brightness of each peak could be modulated by varying the progenitor radius and the \nifs\ mass.

\begin{figure}
   \centering
    \begin{subfigure}[b]{0.485\textwidth}
       \centering
      \includegraphics[width=\textwidth]{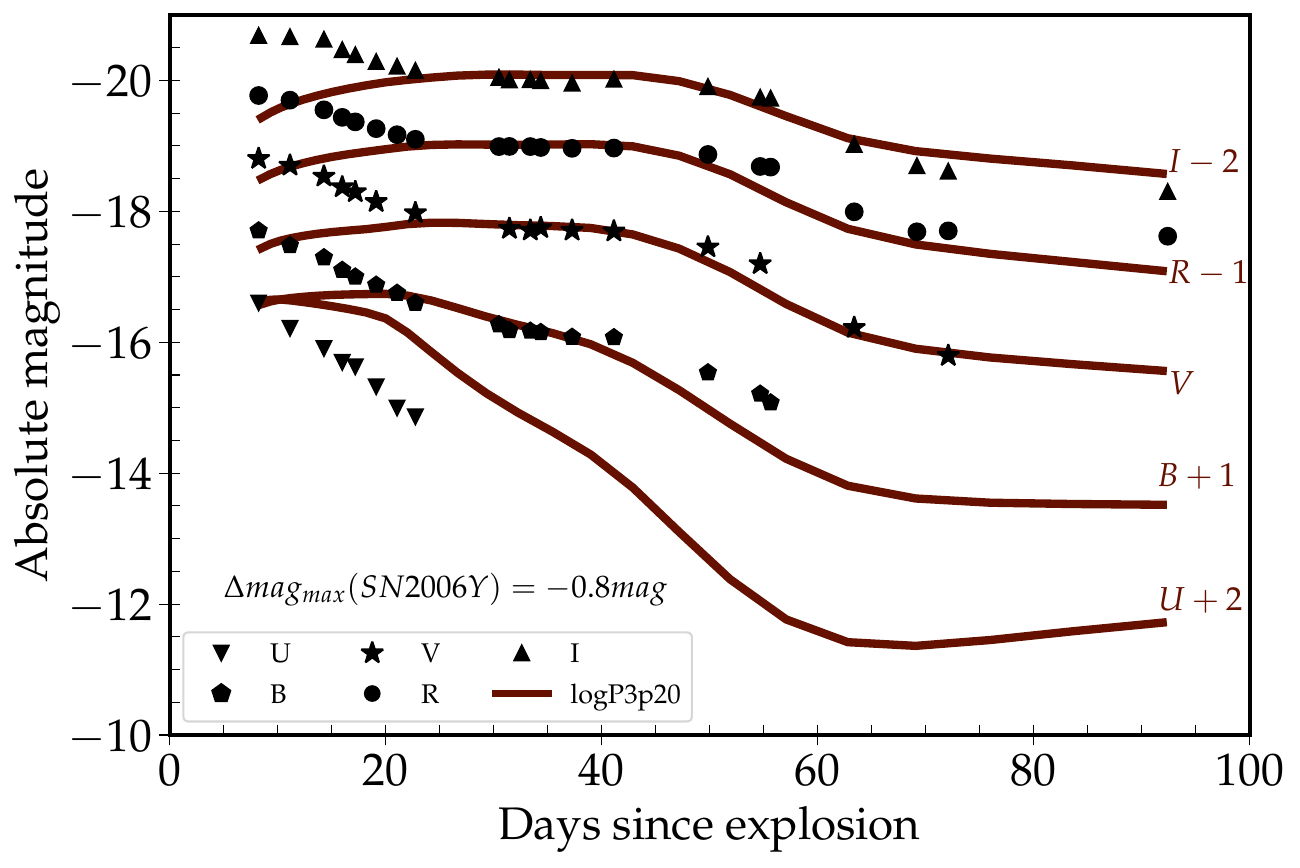}
    \end{subfigure}
   \centering
    \begin{subfigure}[b]{0.485\textwidth}
       \centering
       \includegraphics[width=\textwidth]{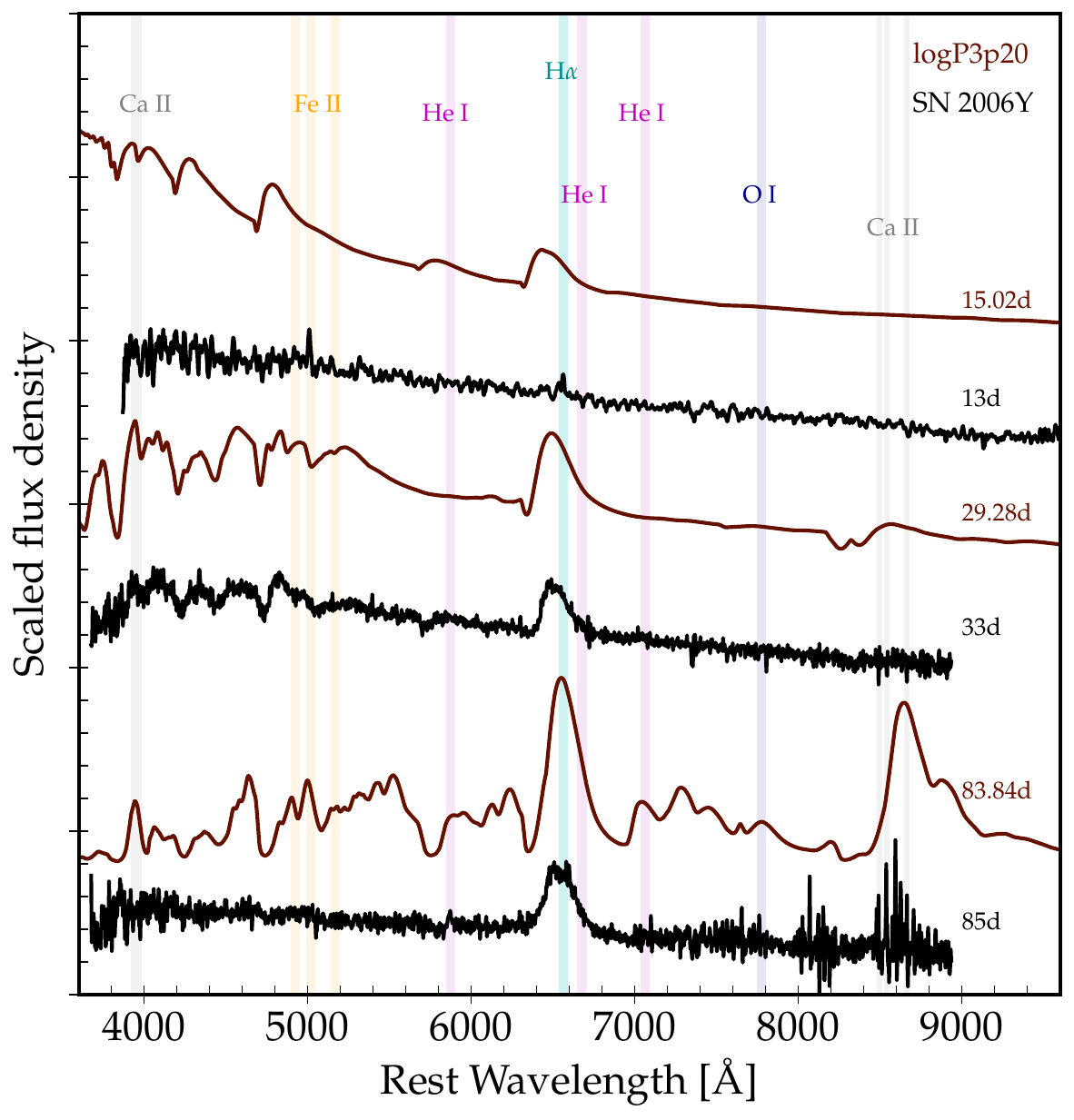}
    \end{subfigure}
    \caption{Same as Fig.~\ref{fig_93J} but now for model logP3p20 and the observations of SN\,2006Y.}
\label{fig_06Y}
\end{figure}

In the sample of \citet{anderson_2pl}, some Type II SNe exhibit short-lived plateau of about 50\,d, such as SN\,2006Y but this SN exhibits a very different spectral evolution from that of our models (see the comparison with model logP3p20 in the bottom panel of Fig.~\ref{fig_06Y}). Indeed, SN\,2006Y shows nearly featureless spectra early on, turning into a nearly-unchanging spectrum out to 100\,d, with weak signs of metal-line blanketing and a strong and broad H$\alpha$ line with little or no absorption -- the H$\alpha$ line in SN\,2006Y even broadens in time. Such spectral features in SN\,2006Y are indicative of ejecta interaction with CSM \citep{D16_2n,dessart_csm_22}, as was observed even more conspicuously in events like SN\,1998S \citep{leonard_98S_00,fassia_98S_00,fassia_98S_01}. The Type IIb SNe 2011fu and 2013df also exhibited a double-peak $V$-band light curve, with the first peak brighter than the second, \nifs-powered peak and a spectral evolution clearly suggestive of interaction, which also persisted until late times as is evidenced by the broad, and eventually  boxy H$\alpha$ profile \citep{morales_13df_14,morales_11fu_15}. Other evidence that interaction may take place at various levels in some Type IIb SNe comes from excess emission in the blue part of the optical or in the UV \citep{benami_iib_15}. This interaction may not swamp the radiation of the underlying SN, but it can considerably alter the ejecta structure and the SN spectral properties and evolution, and this is what complicates the comparison to the present set of models in which interaction was ignored.

It is interesting that the observational sample contains no obvious cases of Type II SNe with a short photospheric phase  and no sign of interaction, like in our model logP3p20. This suggests that the main sequence and RSG mass loss is strong enough to strip entirely the progenitor H-rich envelope and lead to a compact, H-deficient star at core collapse (then producing a stripped-envelope SN or a failed SN with no luminous radiative display and black-hole formation). Or such partially stripped stars (i.e., $M$(H) of order 1\,\msun) are systematically connected to Case C systems that transfer mass to the companion at or soon before core collapse \citep{ercolino_bin_23} and produce peculiar Type II SNe more closely connected to interacting SNe.

\subsection{Type II-P SN \,2017eaw and model logP3p45}
\label{sect_iip}

\begin{figure}
    \begin{subfigure}[b]{0.485\textwidth}
       \centering
       \includegraphics[width=\textwidth]{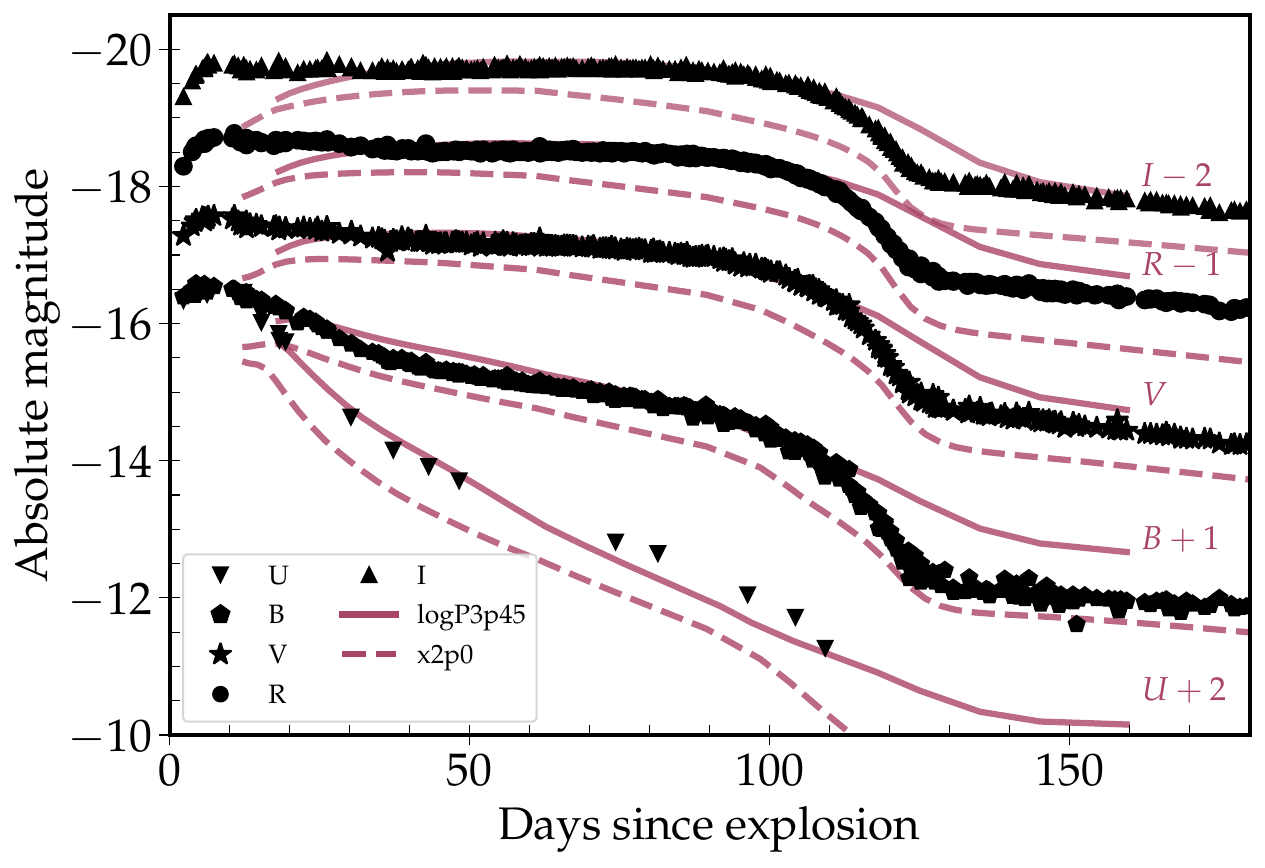}
    \end{subfigure}
    \centering
    \begin{subfigure}[b]{0.485\textwidth}
       \centering
       \includegraphics[width=\textwidth]{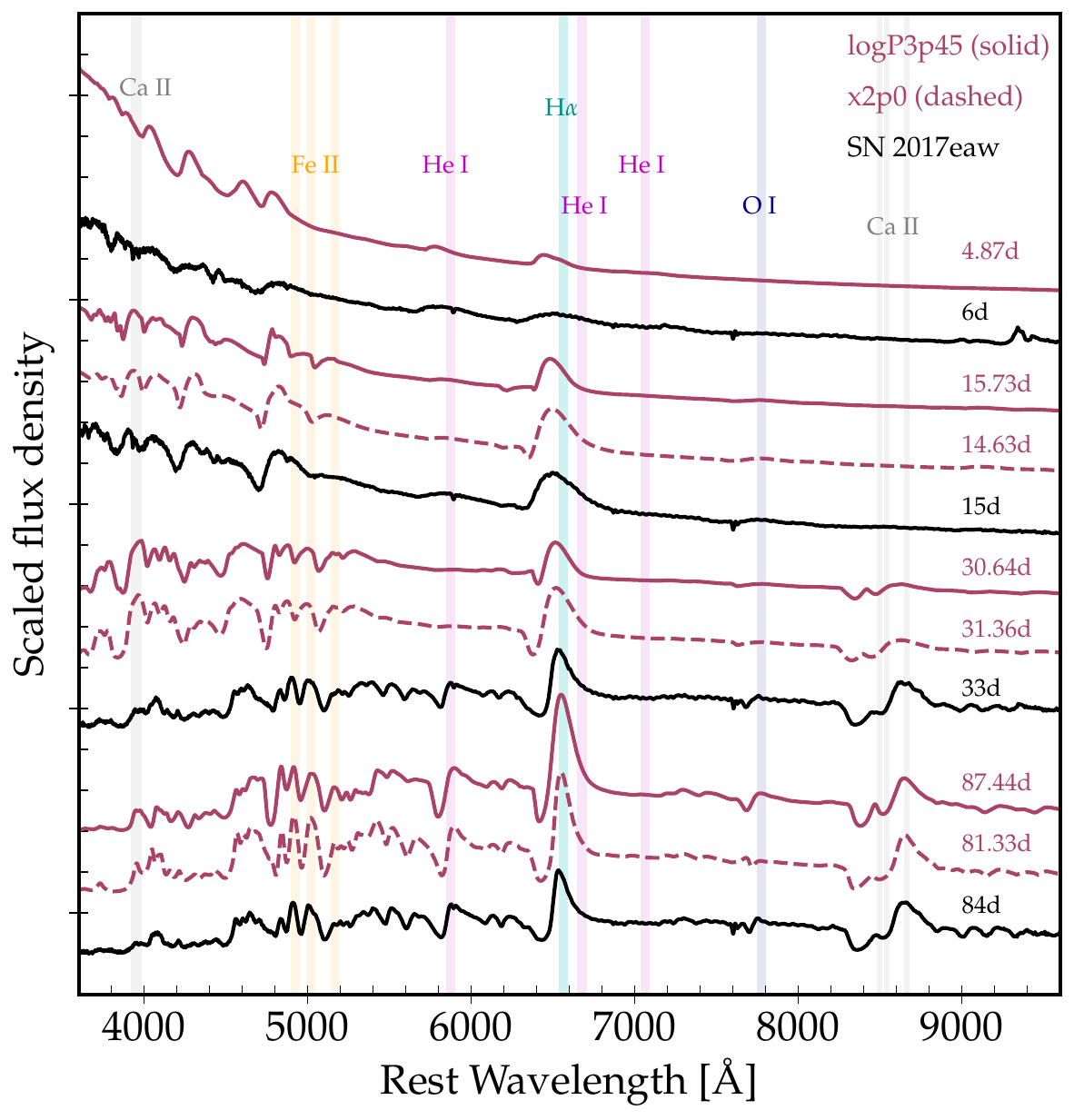}
    \end{subfigure}
    \caption{Same as Fig.~\ref{fig_06Y} but now for models logP3p45 and x2p0 and the observations of SN\,2017eaw.}
\label{fig_iip_17eaw}
\end{figure}

In this final comparison to observations, we consider model logP3p45 whose progenitor evolved as a single star, never experiencing RLOF. The progenitor lost mass nearly exclusively during the RSG phase through wind mass loss, dying as a RSG star with a massive H-rich envelope mass of 6.81\,\msun\ and external radius of 898\,\rsun. In Fig.~\ref{fig_iip_17eaw}, we compare the multi-band light curves and spectra of model logP3p45 with the data from SN\,2017eaw -- no shift has been applied in this figure, neither in time nor in magnitude.

The agreement between model logP3p45 and SN\,2017eaw is relatively good but there are several discrepancies in the light curves and spectra. The plateau brightness matches but the plateau duration is too long and the nebular luminosity is too large. Both could be resolved by having employed a lower \nifs\ mass (i.e., instead of the adopted value of 0.09\,\msun, a lower value of  0.045\,\msun\ should be used; \citealt{szalai_17eaw_19}). The early-time brightness is too faint in all bands and could be remedied by adding in some CSM in the direct environment of the SN \citep{morozova_2l_2p_17,moriya_13fs_17,d17_13fs}. Since no Type IIn signatures were observed at early times \citep{szalai_17eaw_19}, it suggests dense and confined CSM directly at the progenitor surface, but it is unclear whether this material implies a boost to the progenitor mass loss immediately before core collapse (see, e.g., \citealt{morozova_2l_2p_17} or \citealt{moriya_13fs_17}) or whether this is indicative of material ``stagnating" in the atmosphere of the RSG, in the transition zone between the hydrostatic surface and the accelerating regions at the base of the RSG wind \citep{d17_13fs, soker_csm_21, ercolino_bin_23}. An extra ingredient, not present in off-the-shelf stellar evolution models of Type II-P SN progenitors, is clearly needed to bring the $V$-band rise time of ``bare'' RSG star progenitors from  20$-$30\,d (as obtained here for model logP3p45 or long ago in \citealt{DH11_2p}) to less than a week (see also the sample of $V$-band rise times of \citealt{gonzalez_gaitan_2p_15} or the results from the HITS survey of \citealt{forster_csm_18}).

Another discrepancy of model logP3p45 is the underestimate of line widths relative to SN\,2017eaw. The model has narrower lines than observed, in particular at early times when the spectrum forms in the fastest moving layers of the ejecta. However, because the plateau brightness of the model is already well matched, one cannot increase $E_{\rm kin}$ since that would not just increase the line widths but also the plateau brightness. This is a recurrent problem of employing RSG star progenitors with a very large radius \citep{DH11_2p} and a solution to this problem was found by invoking a larger mixing length parameter \citep{d13_sn2p,mesa4,goldberg_sn2p_19}. There is independent evidence that RSG star radii may be smaller than typically assumed \citep{davies_rsg_13} and models of convective RSG star envelopes also support a greater mixing length parameter (see, e.g., \citealt{goldberg_3d_rsg_22}, and more recently \citealt{chun_rsg_18}). In Fig.~\ref{fig_iip_17eaw}, we show the results for the model x2p0 of  \citet{HD19} whose RSG progenitor radius was 582\,\rsun\ (rather than 898\,\rsun\ for model logP3p45). Model x2p0 would match better the plateau brightness with a slightly higher $E_{\rm kin}/M_{\rm ej}$ and $M$(\nifs) but this model is close to SN\,2017eaw  in terms of brightness, line widths, and spectral evolution, while model logP3p45 is too bright for its expansion rate.

\section{Conclusion}
\label{sect_conc}

We have investigated the final fate as a SN for the primary stars, initially of 12.6\,\msun,  in the large grid of binary-star evolution models computed by \citet{ercolino_bin_23}. Focusing on systems with an initial mass ratio of 0.95 and initial orbital periods from 562 to 2818\,d, we simulated the explosion with radiation hydrodynamics and solved the NLTE time-dependent radiative transfer for the corresponding SN ejecta from a few days after explosion to the nebular phase. Since the He-core properties of all models were essentially the same, a unique explosion energy of \foe\ and \nifs\ mass (i.e., $\sim$\,0.09\,\msun) was adopted for the whole set to be on par with the inferred properties of the Type IIb SN\,1993J. However, because of the wide range in residual H-rich envelope mass at the time of explosion, the SN models yielded widely different SN light curves and multiepoch spectra.

With our grid of models varying initially only in orbital separation, we recover a continuum of $V$-band light-curve morphologies from single- or double-peak to plateau-like, reflecting the observed types associated with stripped-envelope SNe  (case B mass transfer systems), the fast-declining Type II-L SNe (case BC and case C mass transfer systems) and  Type II-P SNe (primary stars without RLOF). This diversity reflects the range in preSN H mass (or H-rich envelope mass), from zero to several solar masses. Namely, the shortest period system produces a Type Ib SN with a single-peak \nifs-powered $V$-band light curve and spectra exhibiting unambiguous He\one\ lines (as observed for example in SN iPTF13bvn). Progenitors with intermediate $M$(H) values produce ejecta with double-peak light curves or light curves in which these two peaks merge into a short high-brightness plateau, with spectra exhibiting long-lived H\one\ and He\one\ lines (at the low $M$(H) end, these explosions capture the essence of events like Type IIb SNe 2011dh or 1993J).  And finally, progenitors with a massive H-rich envelope that evolve essentially in isolation and produce a plateau light curve typical of SNe II-P, with spectra exhibiting H\one\ lines at all epochs (as observed for Type II-P SNe like 2017eaw).

Although all case BC and case C systems in the study of \citet{ercolino_bin_23} underwent a major mass-transfer event shortly before, or were transferring mass at core collapse, we ignored ejecta interaction with CSM but find that it may be the dominant process affecting SNe from progenitors with intermediate H-rich envelope masses. Indeed, we found essentially no observed counterpart having the properties of models logP2p95 to logP3p20 (i.e., with $M$(H) between 0.15 to 0.8\,\msun), characterized by luminous double-peaked light curves or short plateaus and broad-lined spectra (relative to SNe II-P). Events like SN\,2006Y shows similar light curves but its spectral evolution suggest interaction dominates the whole evolution and such ejecta/CSM interaction is ignored in our work. Lacking observational counterparts to our intermediate $M$(H) models, we surmise that a large fraction of these case-BC and case-C systems are at the origin of interacting SNe like SN\,1979C or 1998S.

\begin{figure}
       \centering
       \includegraphics[width=0.49\textwidth]{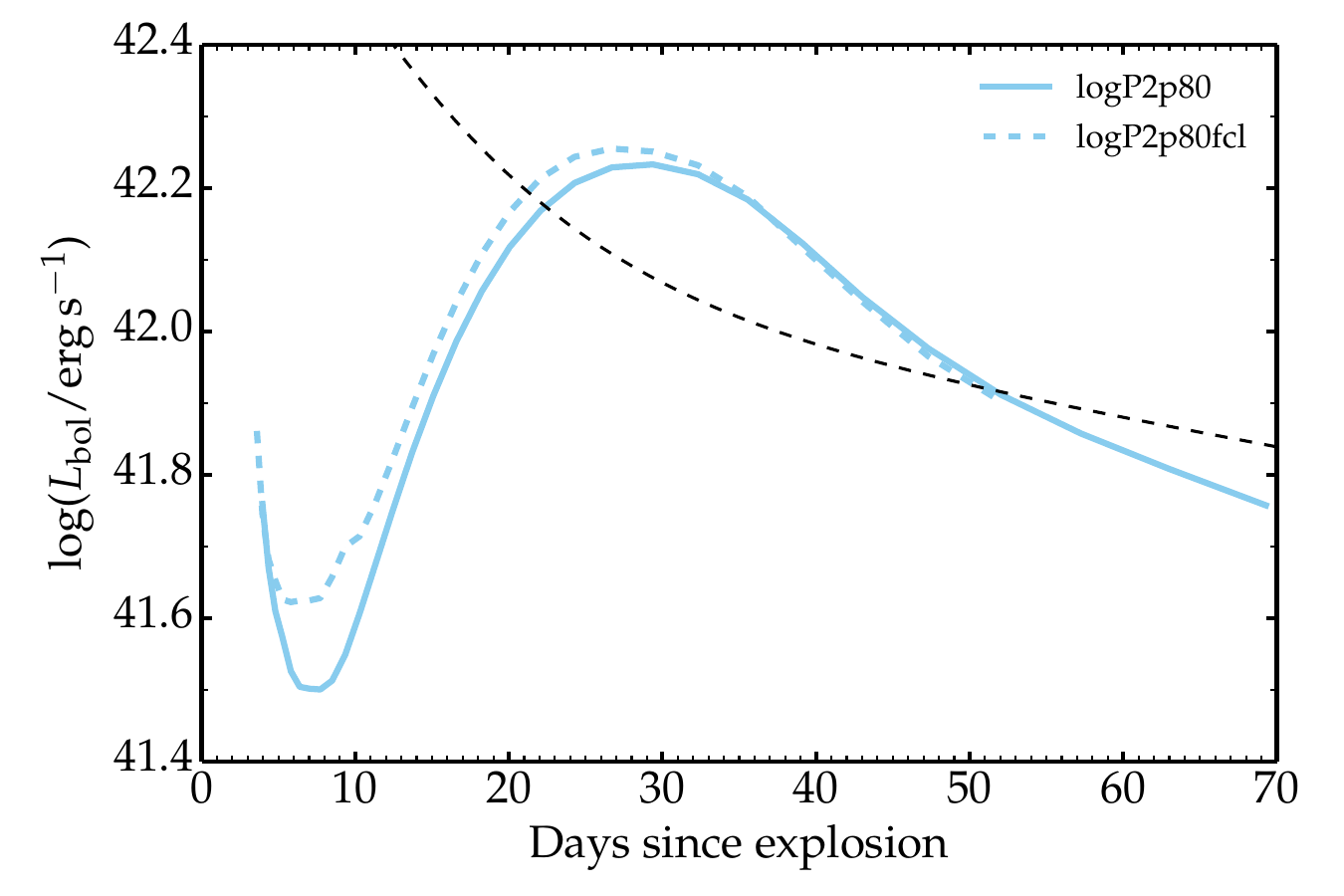}
    \caption{Comparison of the bolometric light curves for model logP2p80 (solid; smooth ejecta) and model logP2p80fcl (dashed; clumped ejecta). For the clumped model, we adopt the same ejecta as for model logP2p80 but introduce a volume filling factor for the material, with a value of 10\% up to about 12\,000\,\kms\ and rising steeply to asymptote at 100\% beyond.
\label{fig_fcl}
}
\end{figure}

Our grid of models also reveal a number of intrinsic properties of core-collapse SN progenitors and ejecta. As before in \citet{d13_sn2p}, we find that a RSG progenitor with a moderate radius of order 500\,\rsun\ yields a better match to the plateau brightness, the timing for the onset of recombination, and the line widths of Type II-P SNe. A larger radius as employed in logP3p45 (i.e., $\sim$\,900\,\rsun) yields a greater luminosity but a smaller expansion rate, making it difficult to match the $V$-band brightness during the plateau at the same time as the spectral line widths. This arises from the fact that the bigger the star, the greater the amount of energy channeled into radiation at the expense of kinetic energy. Other works employing large RSG radii (i.e., stellar evolution models in which the mixing length parameter is around 1.4 rather than 2 or 3) alleviate this problem by invoking, for example low and non-standard explosion energies for Type II-P SNe (e.g., \citealt{morozova_sn2p_18}).

The models discussed in this paper represent only the mass donor in binary systems. For each donor in a stable mass transfer binary, we can expect one mass gainer, which will likely explode as a RSG. Furthermore, of the order of 50\% of the massive binaries are expected to merge \citep{podsiadlowski_92,langer_bh_20}. In either case, the product is expected to produce a RSG progenitor at core collapse, with a hydrogen-rich envelope mass that exceeds that of single stars with the same He-core mass, and that may be significantly more helium enriched than corresponding single stars \citep{justham_lbv_14,menon_bsg_23}. The RSGs from mass-gaining binary products could even outnumber the RSGs produced by true single stars, and determine the appearance of typical Type II-P SNe. The effects of more massive and/or helium-enriched envelopes on the preSN star and associated SN explosion will be the subject of a future study.

A discrepancy that has been plaguing \cmfgen\ simulations for Type IIb, Ib, and Ic SNe since 2011 is the systematic overestimate of the rise time to the main, \nifs-powered peak, with a delay of about a week relative to the observations. Such offsets with data are typically not present in our simulations for Type Ia SNe although the same power source and diffusion process lies behind these light curves (see, e.g., rise times in \citealt{blondin_wlr_17}). Modulations in chemical mixing is often invoked (see, e.g., \citealt{bersten_08D_13}, \citealt{D15_SNIbc_I}, or more recently \citealt{park_iib_23}), but this seems to be contradicted by the strong impact it has on He\one\ line profiles at early times (i.e., stronger mixing yields broader He\one\ lines, all else being the same and in particular the explosion energy) and the values inferred from neutrino-driven explosions \citep{wongwathanarat_15_3d}.  What we are most likely lacking here is a proper description of the ejecta structure, which is not smooth and spherically symmetric (as assumed here), but instead clumped and structured on small and large scales due to the physics of neutrino-driven explosions and the action of the \nifs-bubble effect over time \citep{woosley_87A_late_88,gabler_3dsn_21}. Clumping is known to hasten recombination and the recession of the photosphere towards the inner ejecta, thereby releasing on a shorter time scale the energy stored in the optically-thick layers of the ejecta \citep{d18_fcl}. To illustrate this effect here for a stripped-envelope SN, we show in Fig.~\ref{fig_fcl} the bolometric light curve obtained for model logP2p80 without clumping (same model as described earlier) and with clumping (model logP2p80fcl), confirming the effect obtained for BSG star explosions in \citet{d18_fcl}. The \nifs-bubble effect also creates large cavities surrounded by denser, shell-like structures through which radiation can escape more easily because of the reduced effective opacity \citep{dessart_audit_rhd_3d_19}. These effects should be present in essentially all core-collapse SNe because they derive from the same explosion mechanism. \citet{ergon_11dh_22} and \citet{ergon_20acat_23} have demonstrated that such inhomogeneous ejecta structure can shorten the rise time and also boost the luminosity at maximum.

\begin{acknowledgements}
CPG acknowledges financial support from the Secretary of Universities and Research (Government of Catalonia) and by the Horizon 2020 Research and Innovation Programme of the European Union under the Marie Sk\l{}odowska-Curie and the Beatriu de Pin\'os 2021 BP 00168 programme, from the Spanish Ministerio de Ciencia e Innovaci\'on (MCIN) and the Agencia Estatal de Investigaci\'on (AEI) 10.13039/501100011033 under the PID2020-115253GA-I00 HOSTFLOWS project, and the program Unidad de Excelencia Mar\'ia de Maeztu CEX2020-001058-M. This work was granted access to the HPC resources of TGCC under the allocation 2022 -- A0130410554 made by GENCI, France. This research has made use of NASA's Astrophysics Data System Bibliographic Services.
\end{acknowledgements}


\appendix

\section{Additional figures for all models}

\begin{figure*}
   \centering
    \begin{subfigure}[b]{0.45\textwidth}
       \centering
       \includegraphics[width=\textwidth]{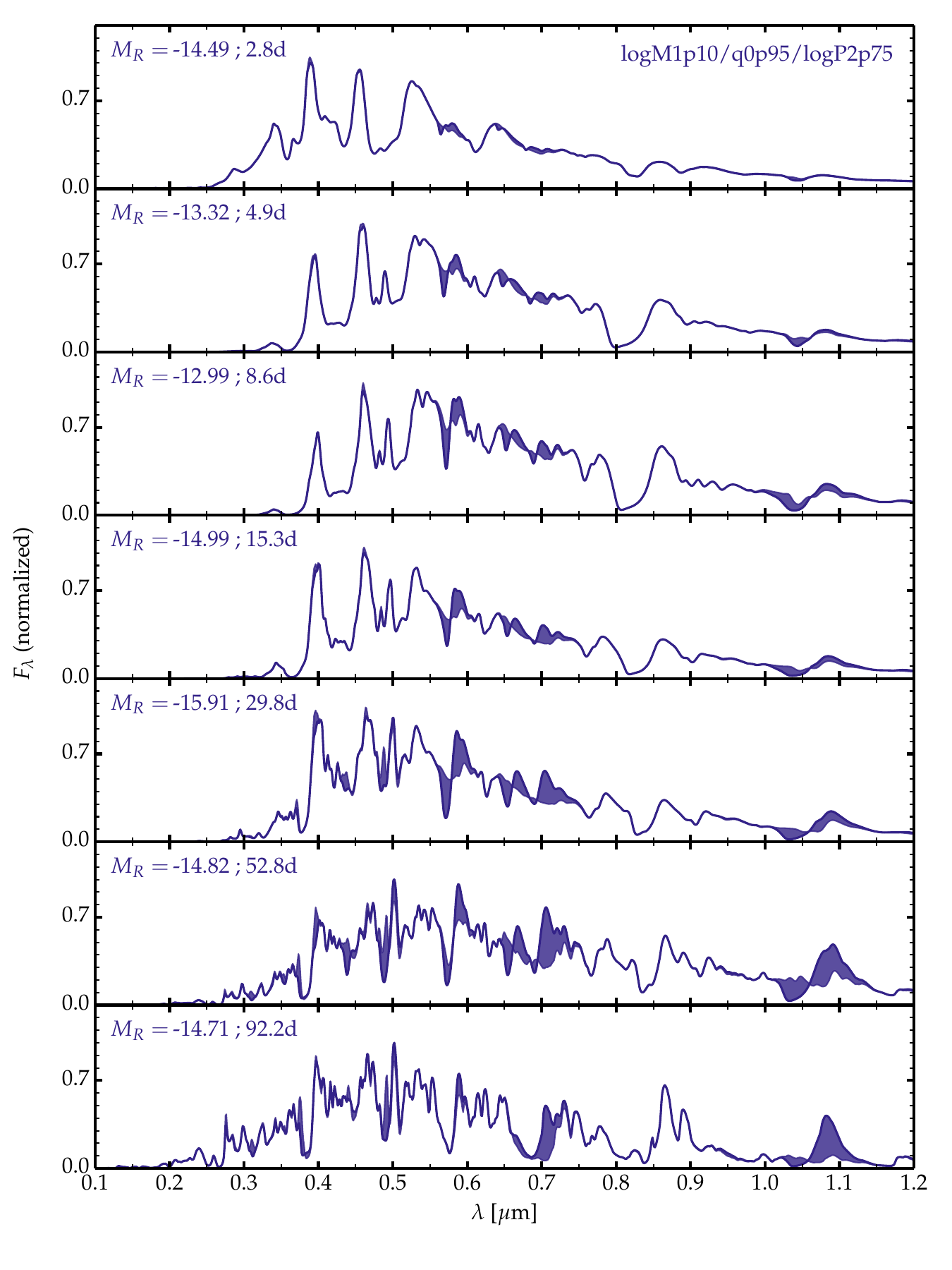}
    \end{subfigure}
    \hfill
    \centering
    \begin{subfigure}[b]{0.45\textwidth}
       \centering
       \includegraphics[width=\textwidth]{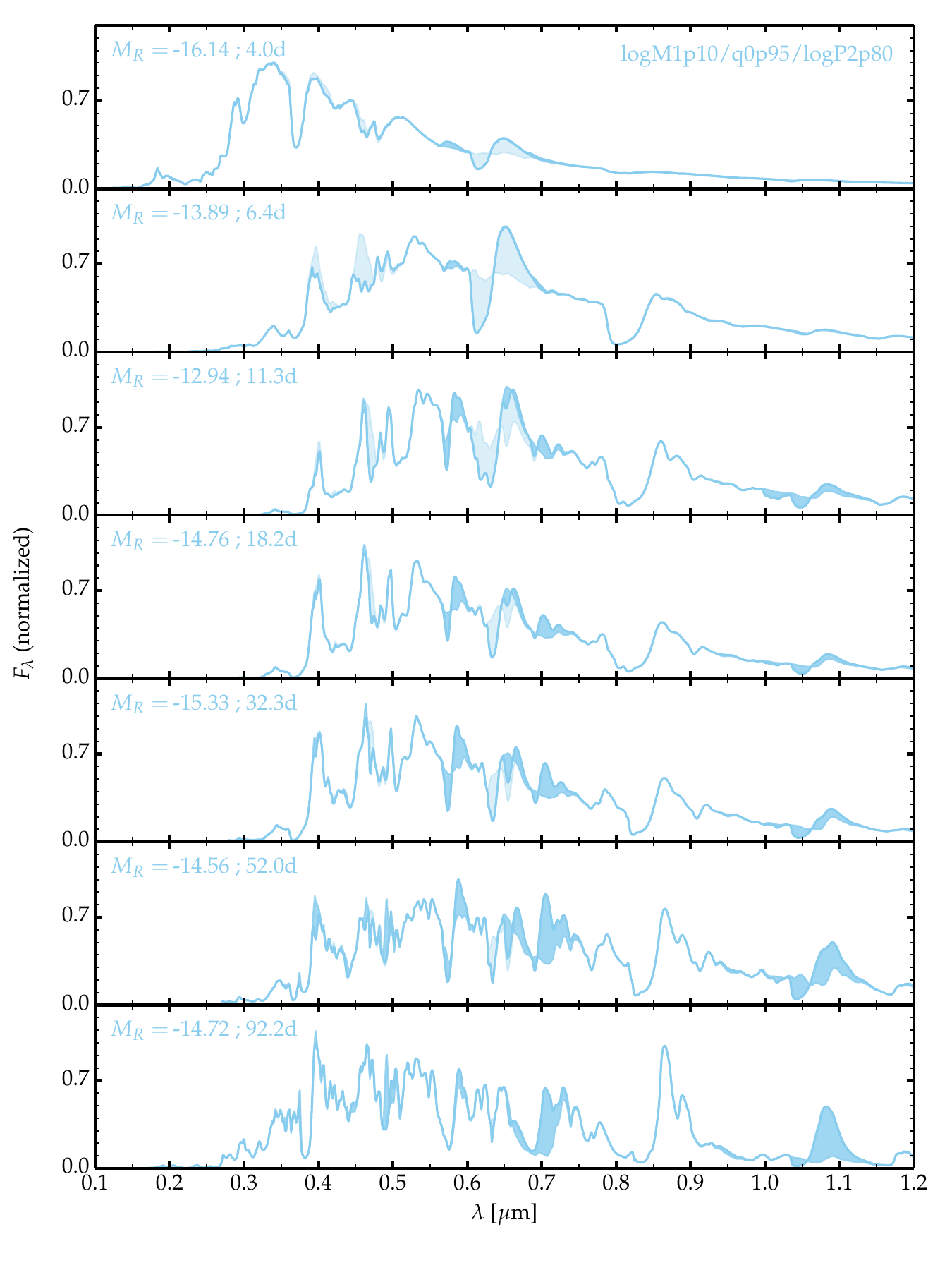}
    \end{subfigure}
     \vskip\baselineskip
   \centering
    \begin{subfigure}[b]{0.45\textwidth}
       \centering
       \includegraphics[width=\textwidth]{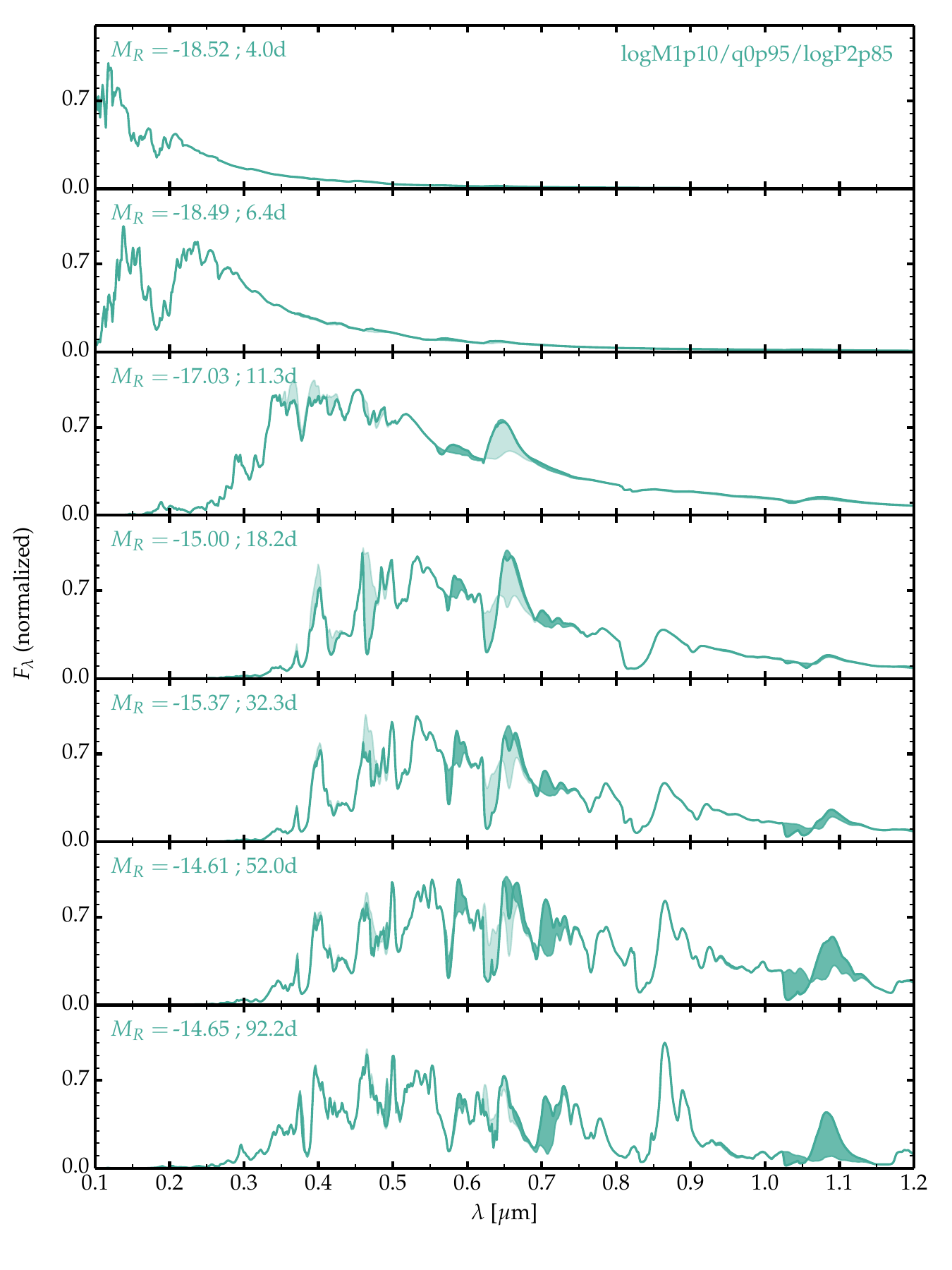}
    \end{subfigure}
    \hfill
    \centering
    \begin{subfigure}[b]{0.45\textwidth}
       \centering
       \includegraphics[width=\textwidth]{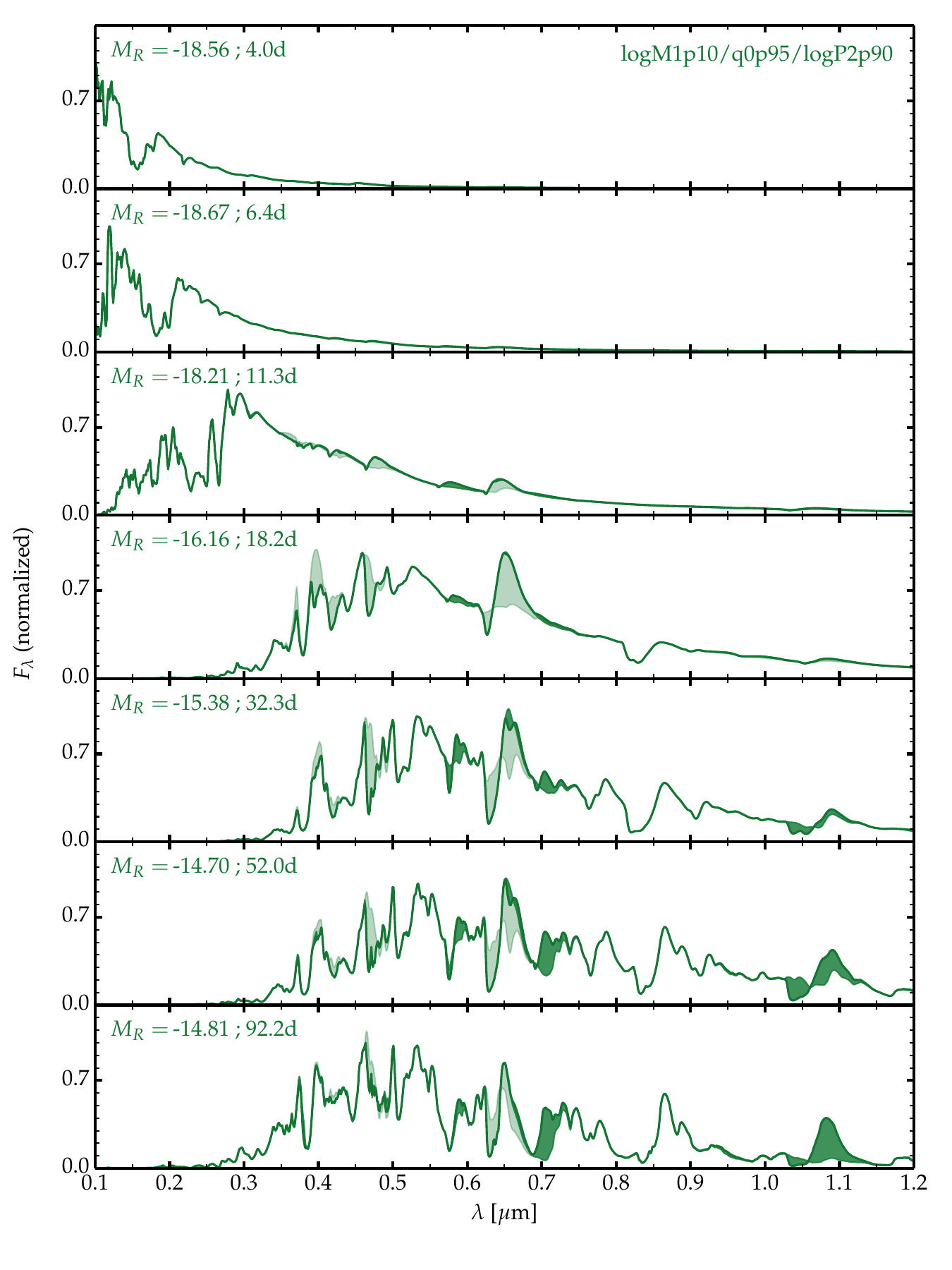}
    \end{subfigure}
\caption{Spectral evolution of models logP2p75, logP2p80, logP2p85, and logP2p90. The \cmfgen\ time sequences are computed at time steps $t_i$ such that $t_{i+1}/t_i=$\,1.1. Here, we show fewer time steps for better visibility so that $t_{iplot+1}/t_{iplot}\sim$\,1.3. As for Fig.~\ref{fig_spec_4epochs}, we illustrate the flux associated with He\one\ lines (dark shade) and H\one\ lines (light shade).
}
\label{fig_spec_seq_ap1}
\end{figure*}

\begin{figure*}
    \centering
    \begin{subfigure}[b]{0.45\textwidth}
       \centering
       \includegraphics[width=\textwidth]{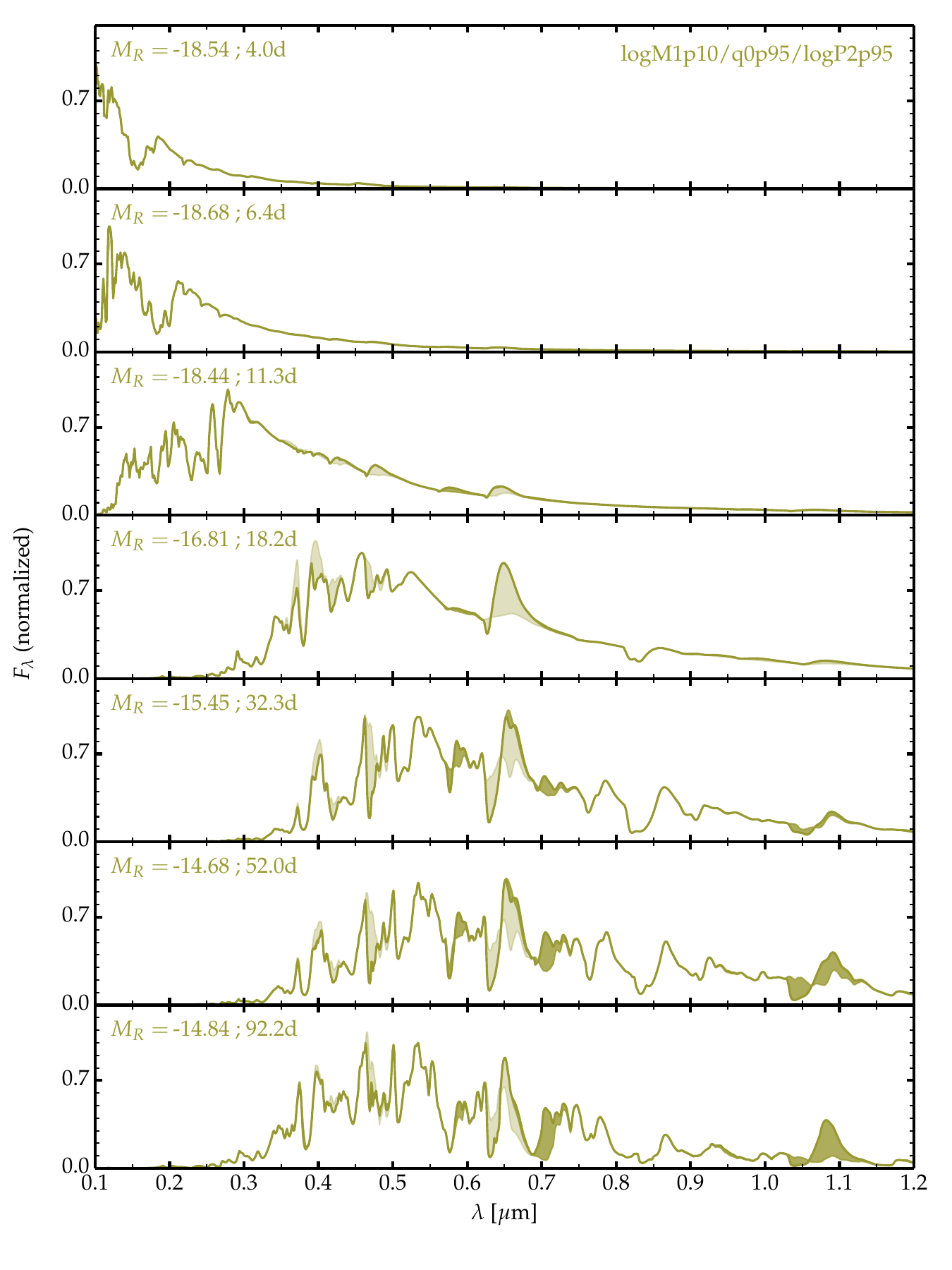}
    \end{subfigure}
    \hfill
    \centering
    \begin{subfigure}[b]{0.45\textwidth}
       \centering
       \includegraphics[width=\textwidth]{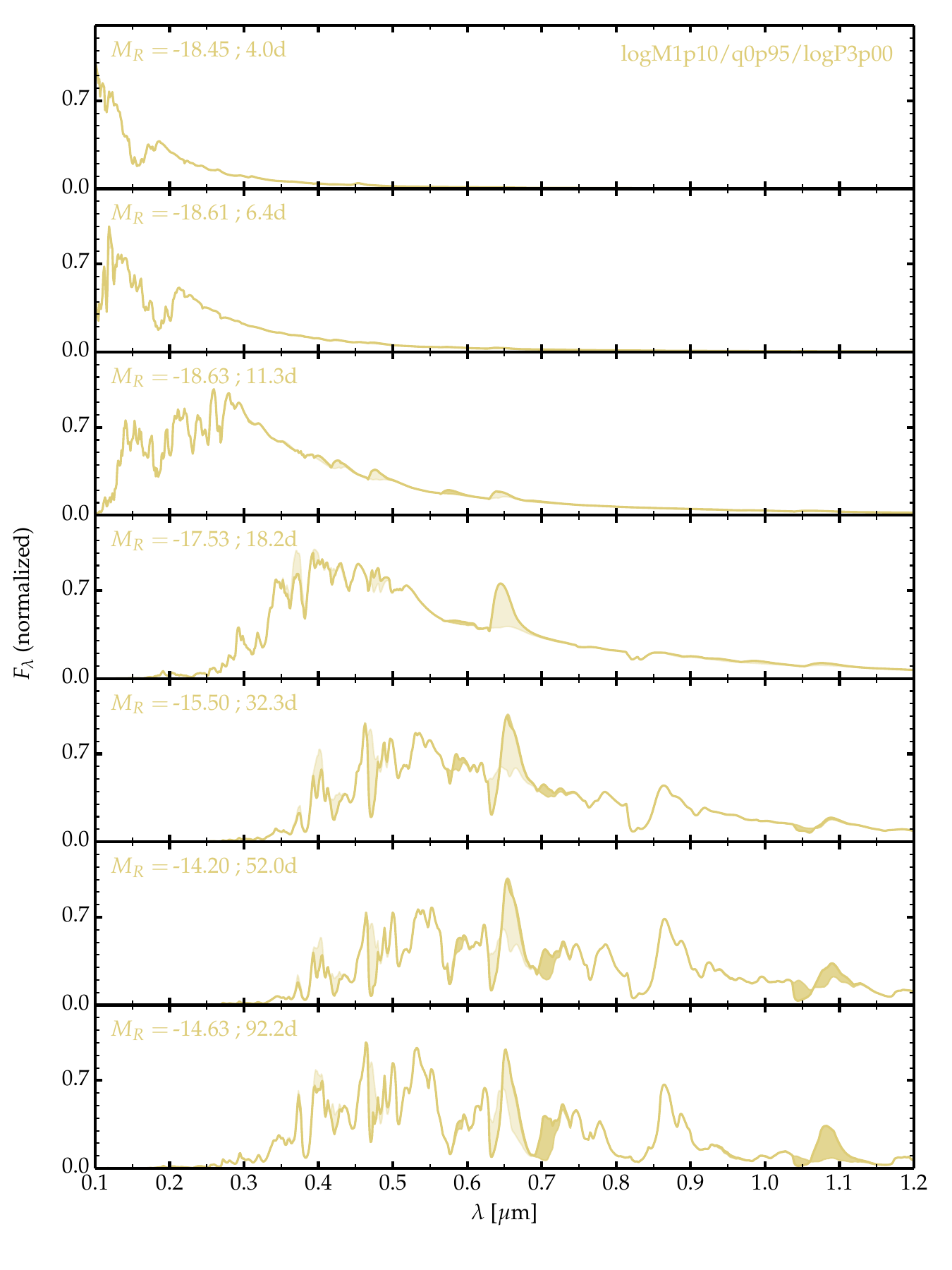}
    \end{subfigure}
    \vskip\baselineskip
    \centering
    \begin{subfigure}[b]{0.45\textwidth}
       \centering
       \includegraphics[width=\textwidth]{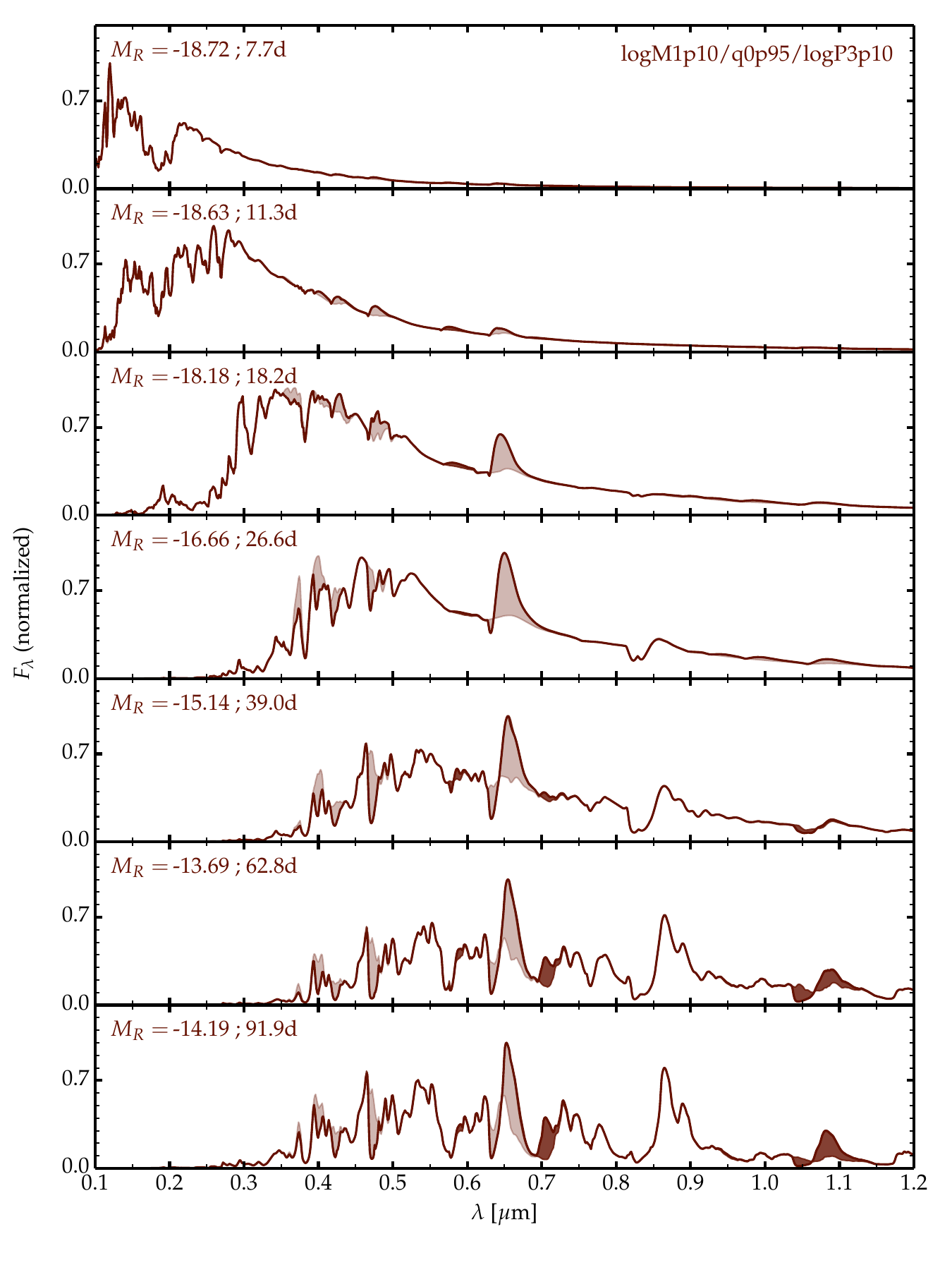}
    \end{subfigure}
    \hfill
    \centering
    \begin{subfigure}[b]{0.45\textwidth}
       \centering
       \includegraphics[width=\textwidth]{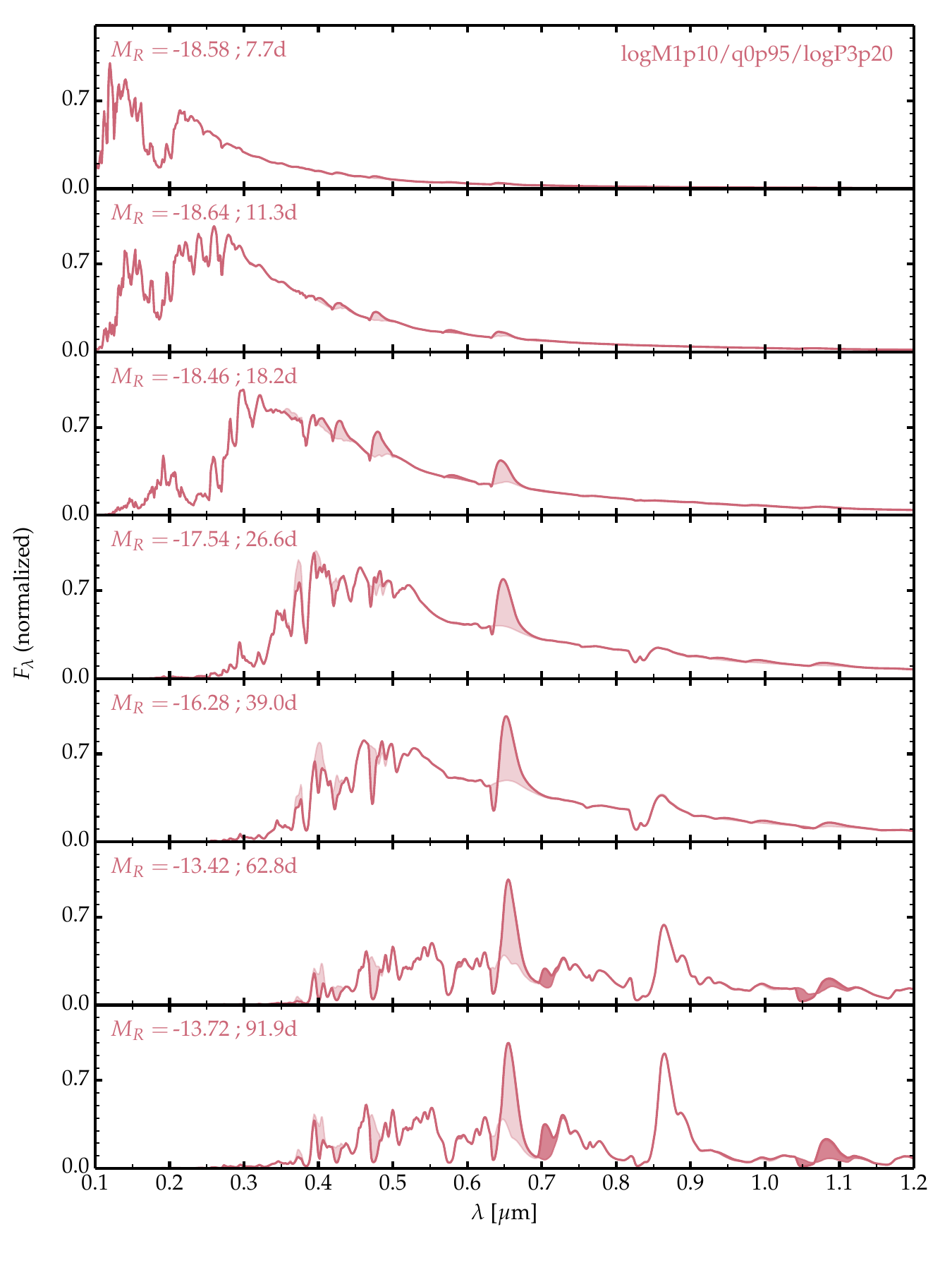}
    \end{subfigure}
    \caption{Same as Fig.~\ref{fig_spec_seq_ap1} but now for models logP2p95, logP3p00, logP3p10, and logP3p20.}
\label{fig_spec_seq_ap2}
\end{figure*}

\begin{figure*}
    \centering
    \begin{subfigure}[b]{0.45\textwidth}
       \centering
       \includegraphics[width=\textwidth]{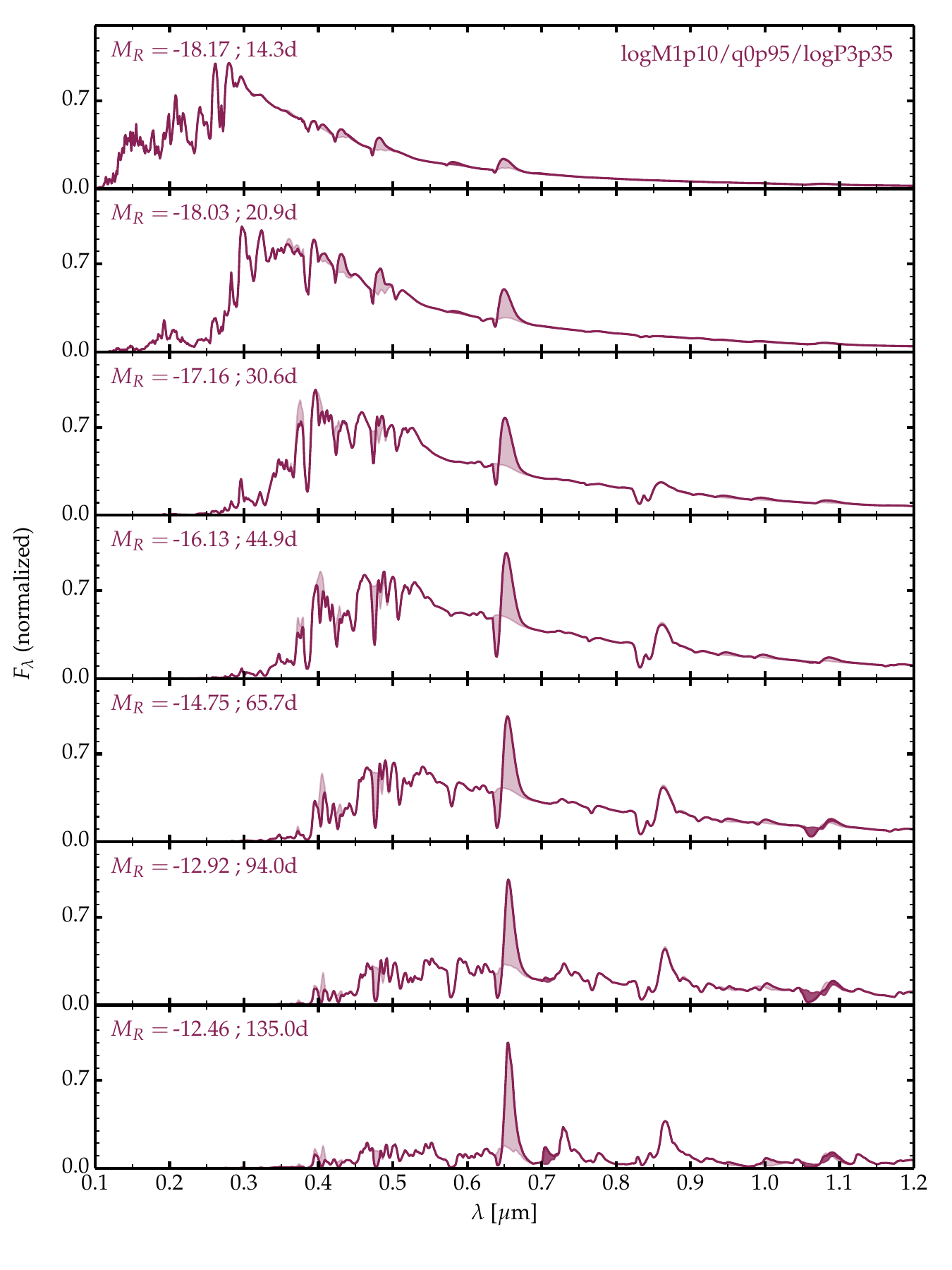}
    \end{subfigure}
    \hfill
    \centering
    \begin{subfigure}[b]{0.45\textwidth}
       \centering
       \includegraphics[width=\textwidth]{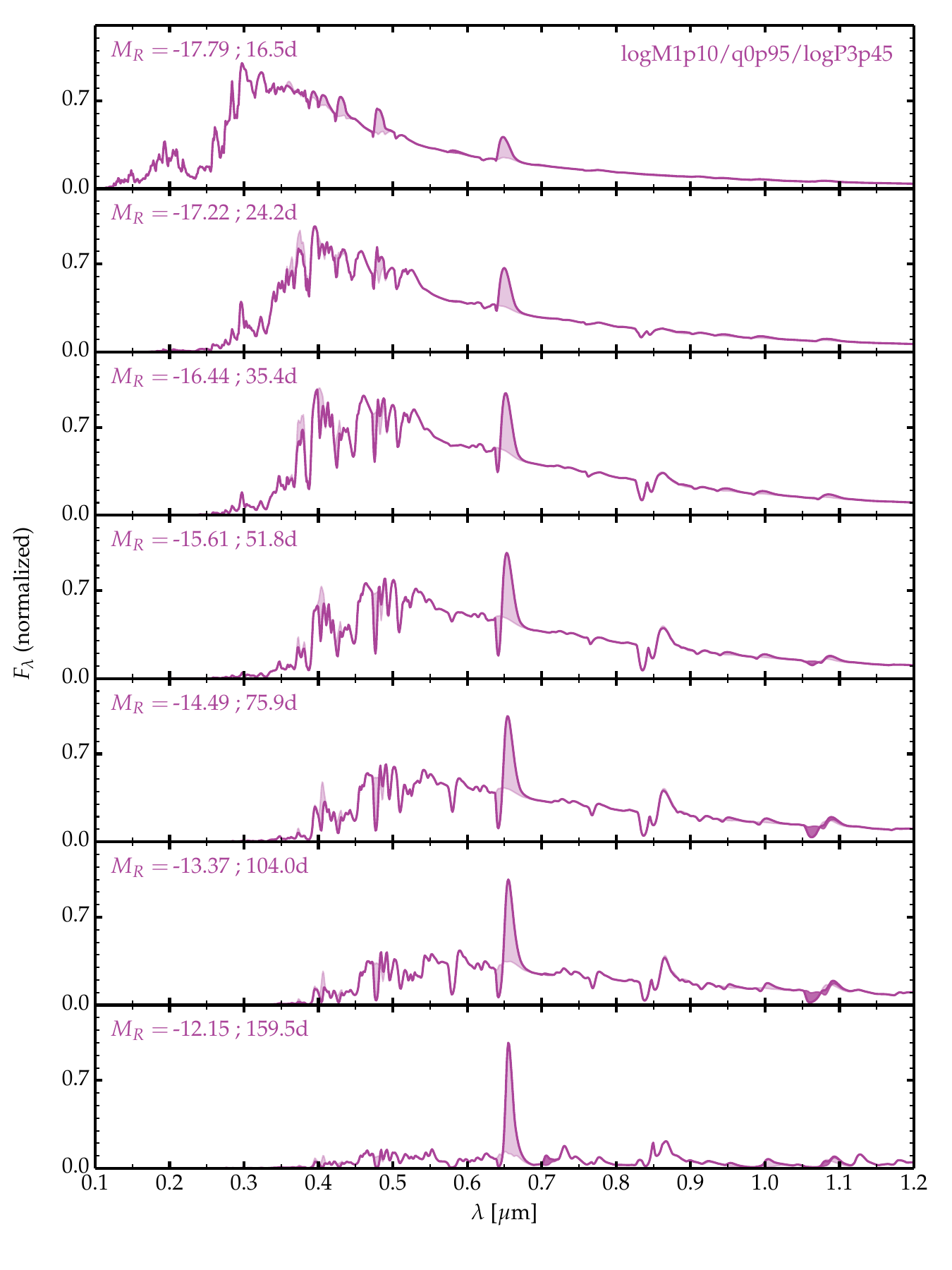}
    \end{subfigure}
    \caption{Same as Fig.~\ref{fig_spec_seq_ap1} but now for models logP3p35 and logP3p45.}
\label{fig_spec_seq_ap3}
\end{figure*}

\begin{figure*}
    \centering
    \begin{subfigure}[b]{0.32\textwidth}
       \centering
       \includegraphics[width=\textwidth]{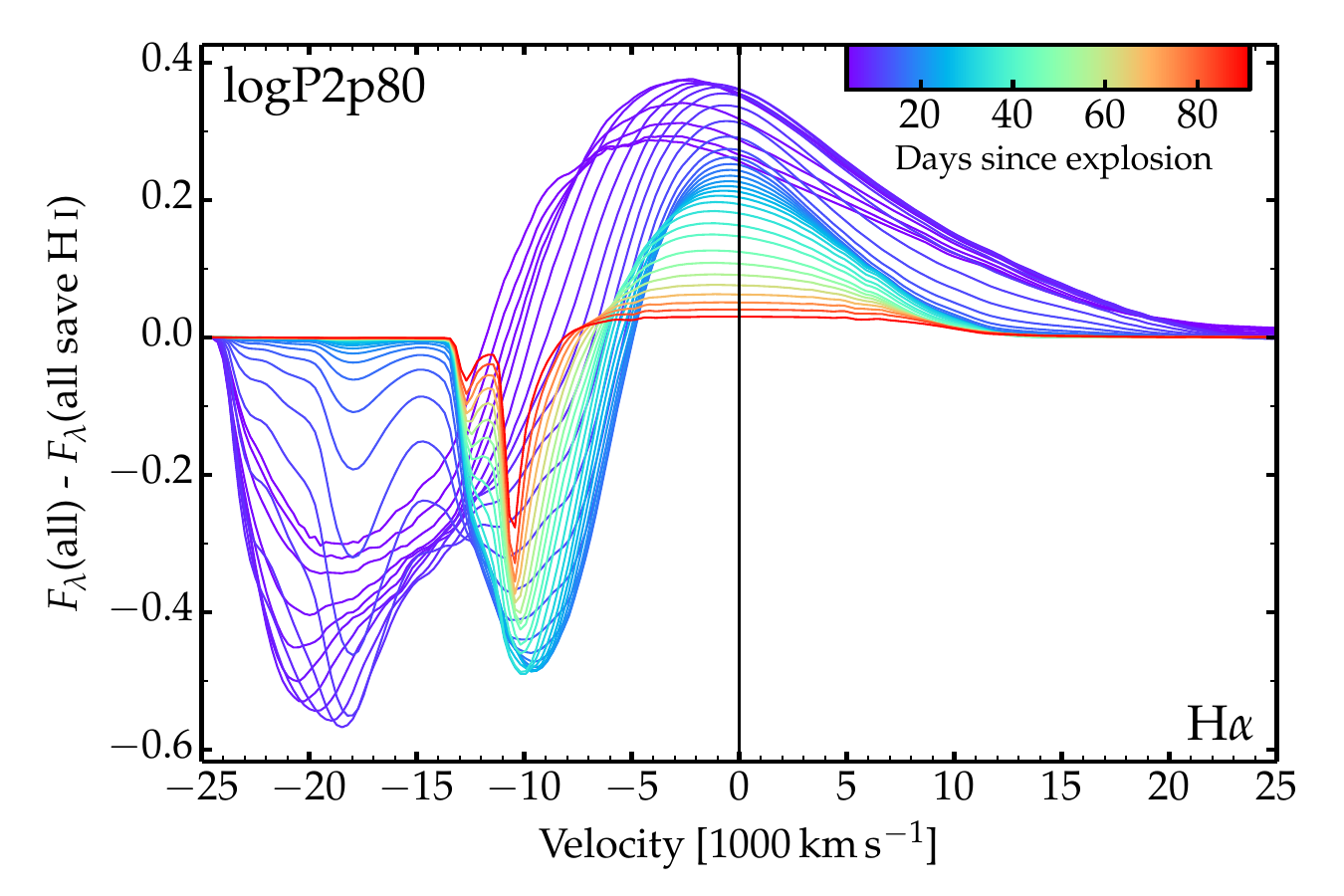}
    \end{subfigure}
    \hfill
    \centering
    \begin{subfigure}[b]{0.32\textwidth}
       \centering
       \includegraphics[width=\textwidth]{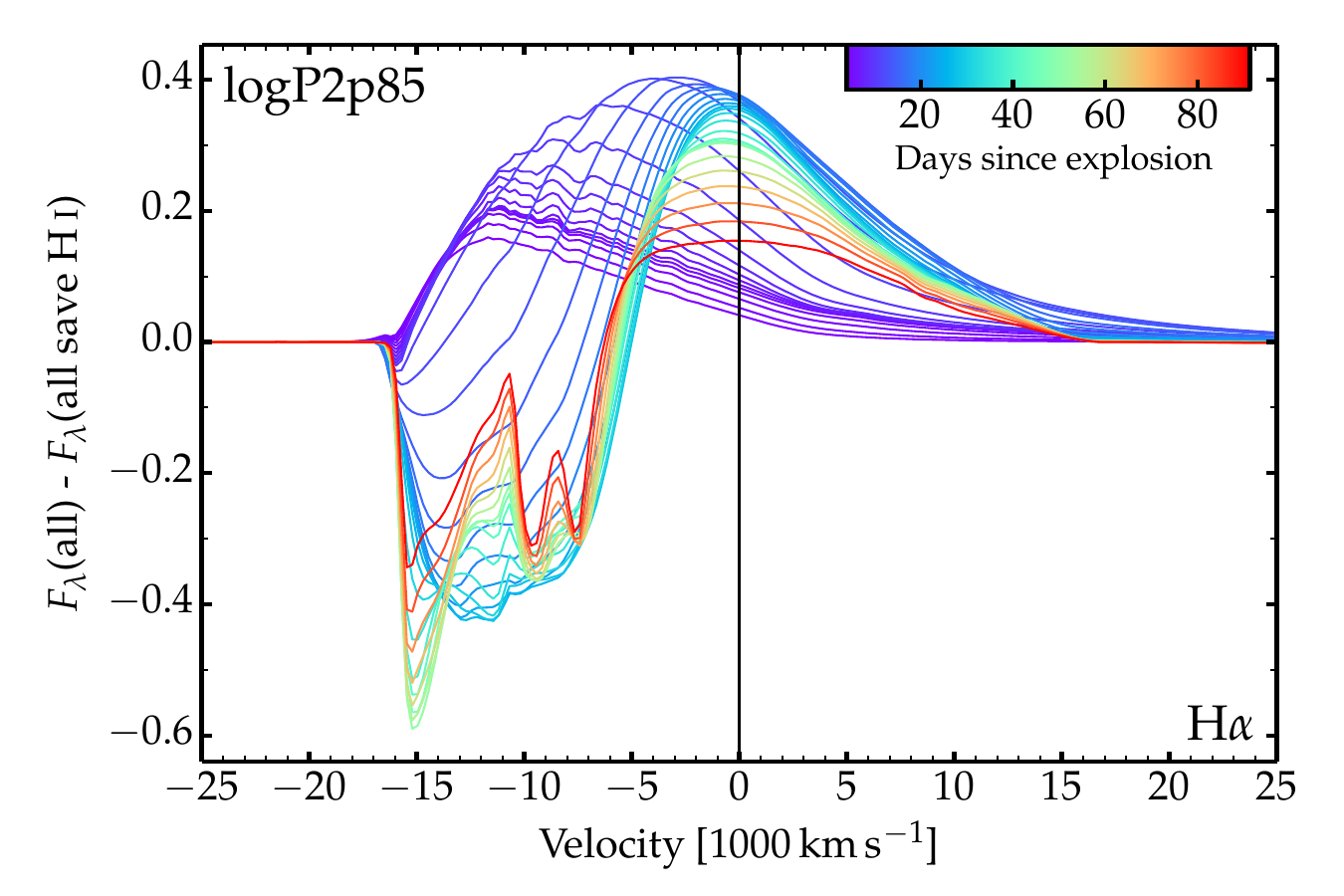}
    \end{subfigure}
    \centering
    \begin{subfigure}[b]{0.32\textwidth}
       \centering
       \includegraphics[width=\textwidth]{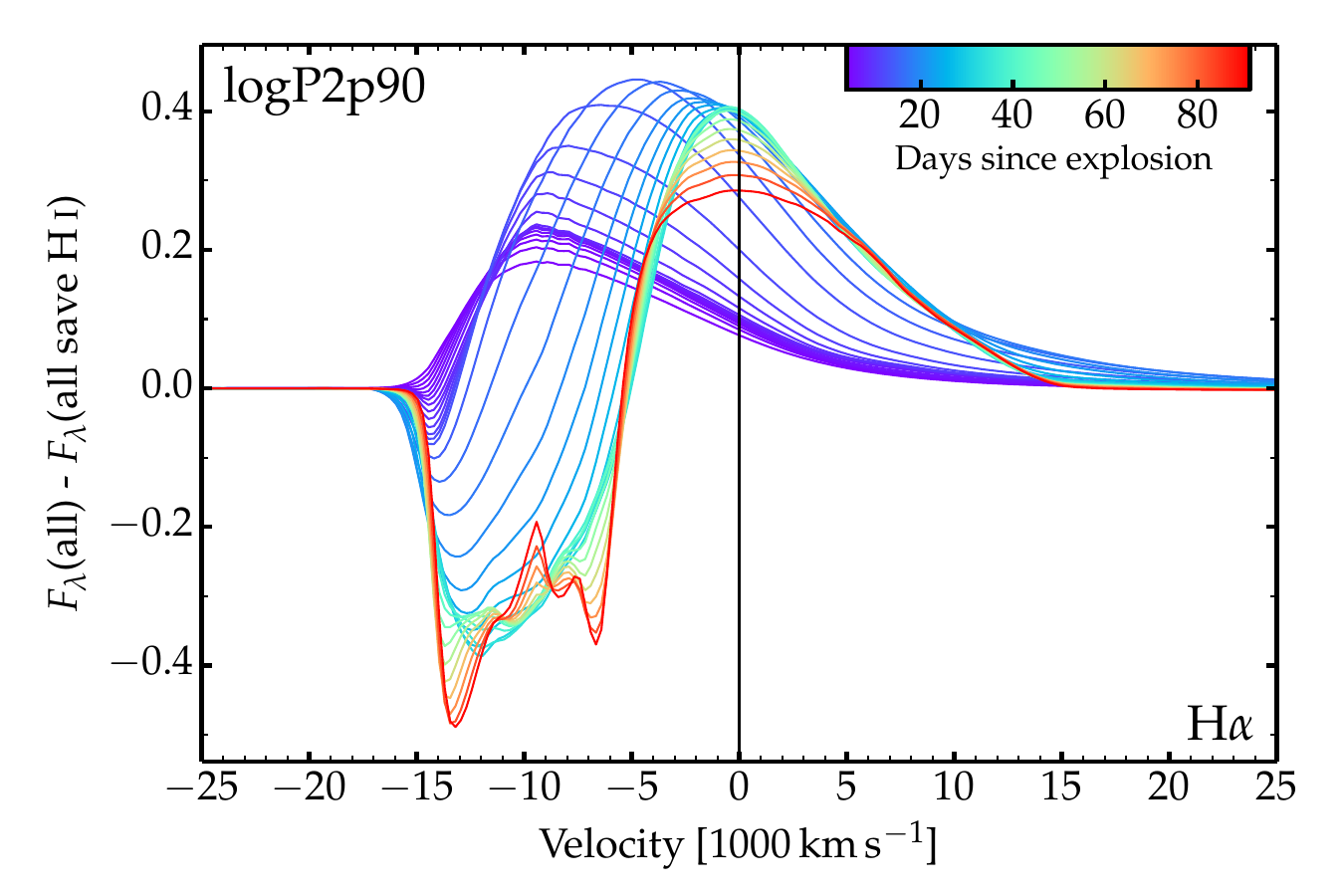}
    \end{subfigure}
    \vskip\baselineskip
    \begin{subfigure}[b]{0.32\textwidth}
       \centering
       \includegraphics[width=\textwidth]{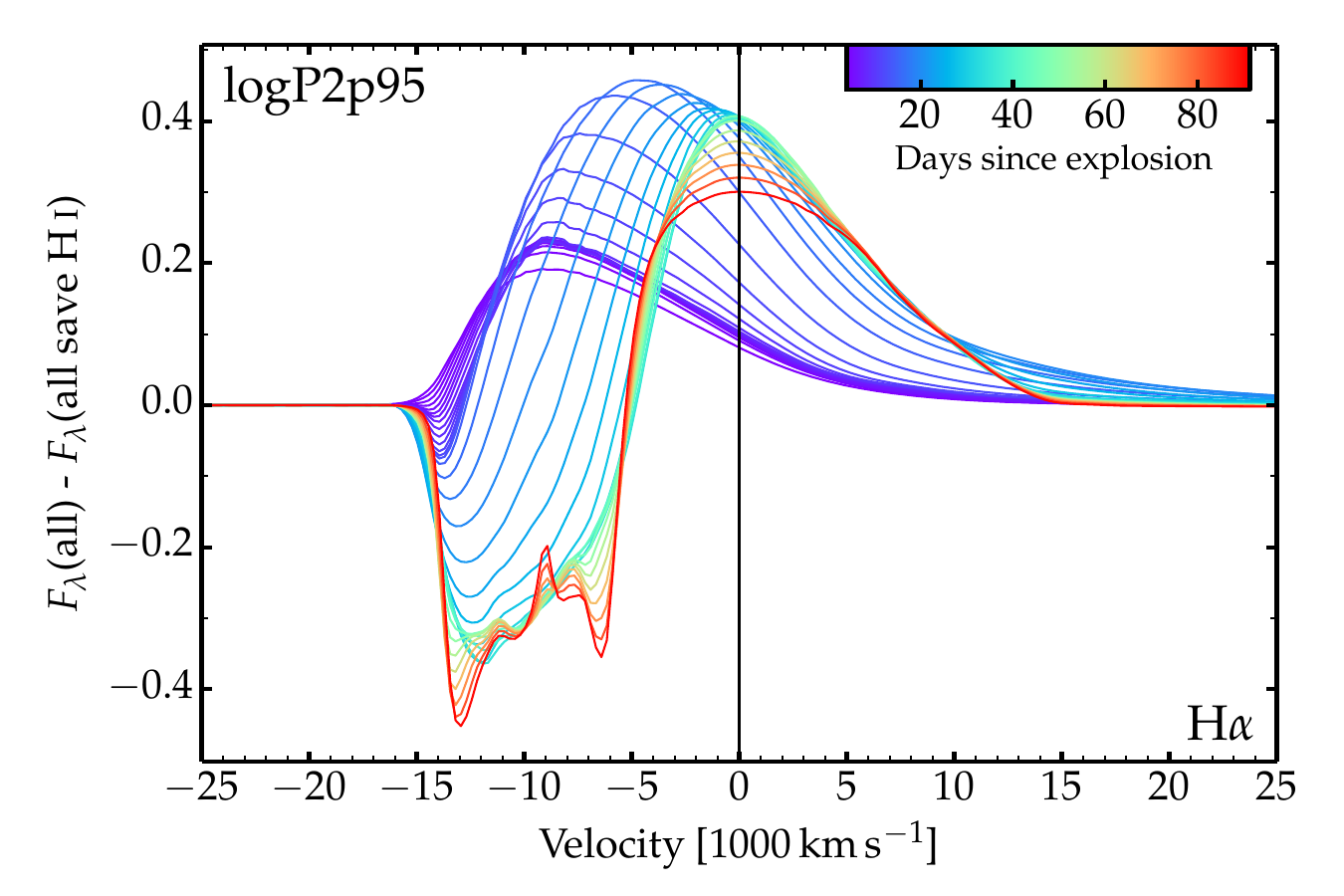}
    \end{subfigure}
    \hfill
    \centering
    \begin{subfigure}[b]{0.32\textwidth}
       \centering
       \includegraphics[width=\textwidth]{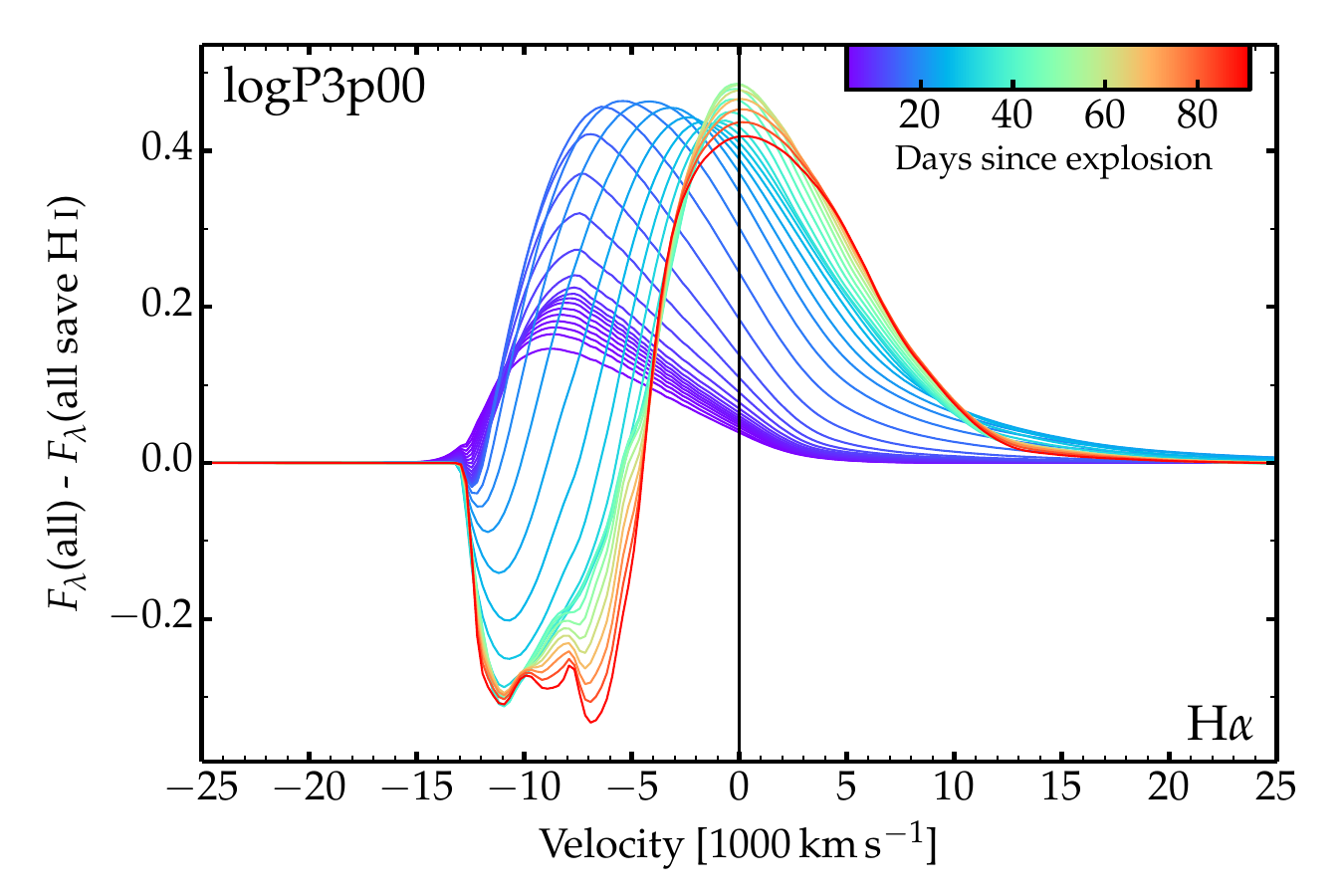}
    \end{subfigure}
    \centering
    \begin{subfigure}[b]{0.32\textwidth}
       \centering
       \includegraphics[width=\textwidth]{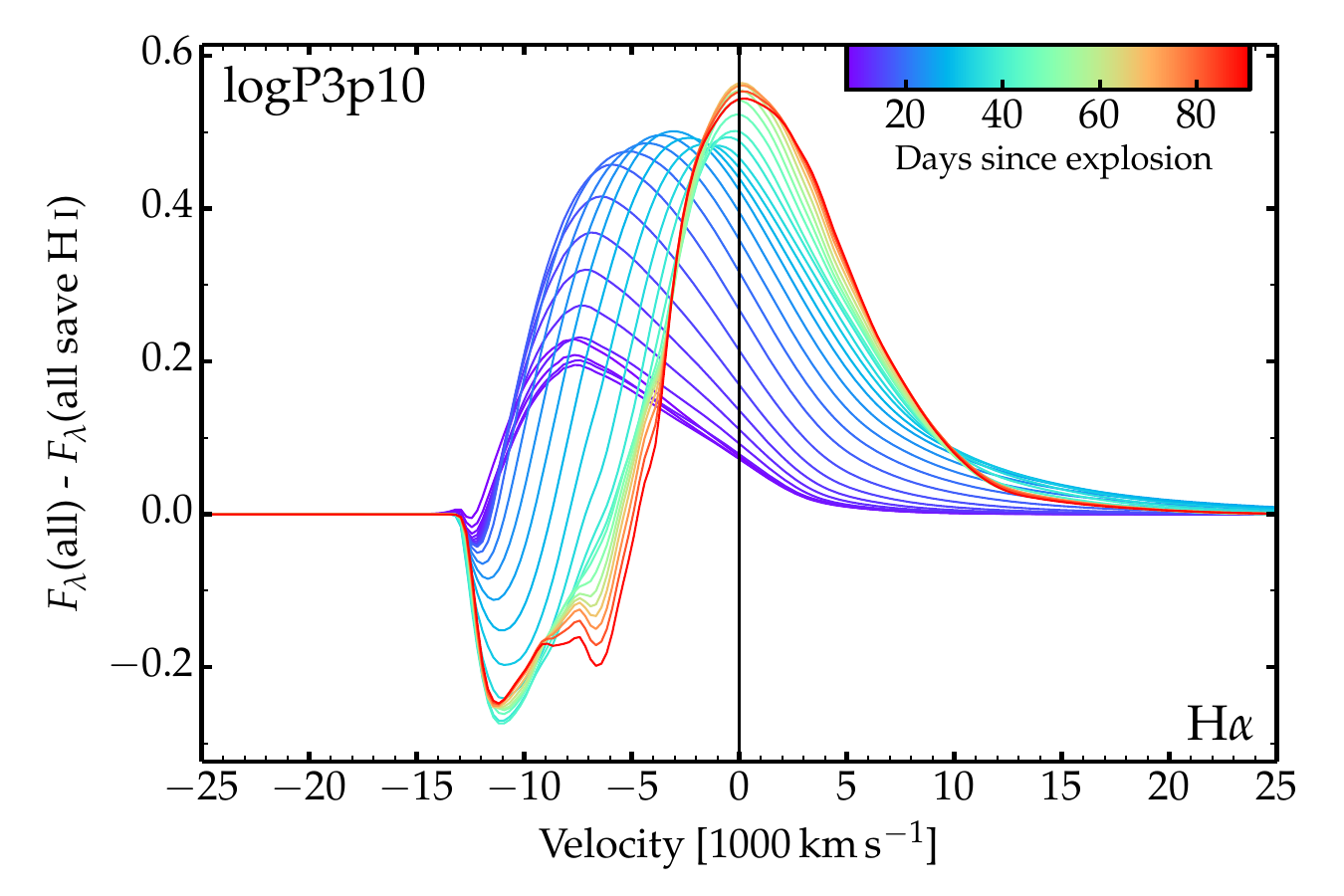}
    \end{subfigure}
    \vskip\baselineskip
    \begin{subfigure}[b]{0.32\textwidth}
       \centering
       \includegraphics[width=\textwidth]{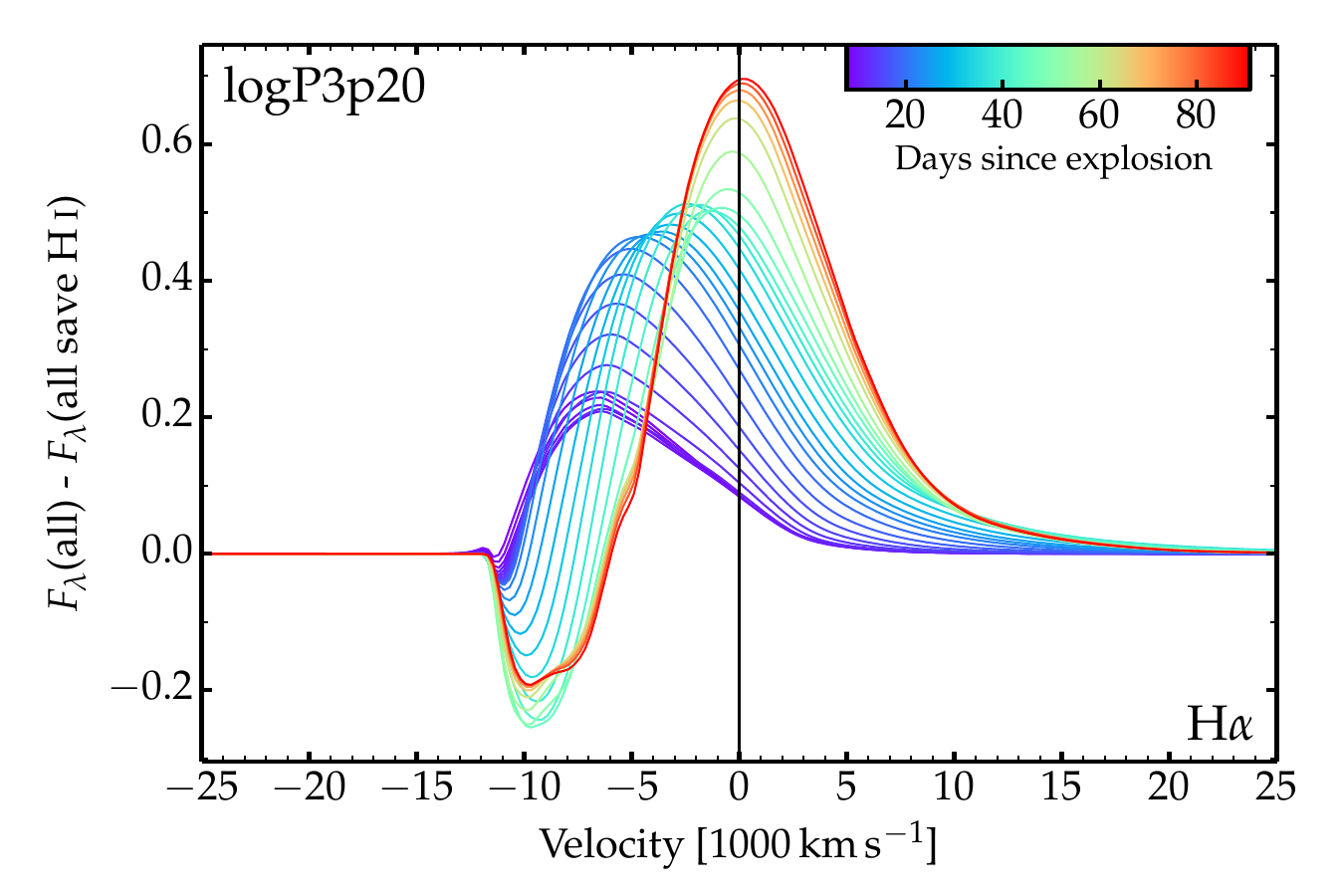}
    \end{subfigure}
    \hfill
    \centering
    \begin{subfigure}[b]{0.32\textwidth}
       \centering
       \includegraphics[width=\textwidth]{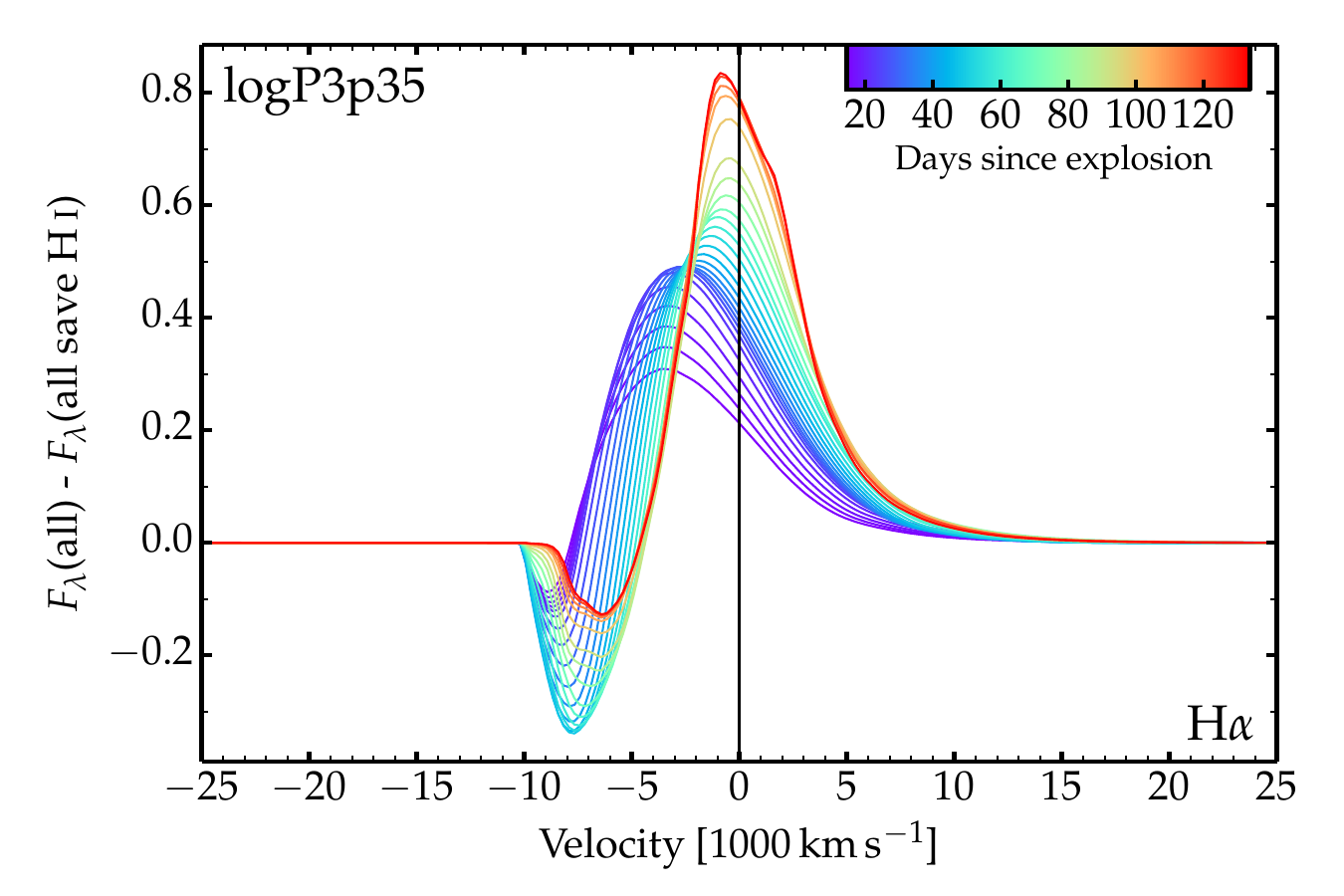}
    \end{subfigure}
    \centering
    \begin{subfigure}[b]{0.32\textwidth}
       \centering
       \includegraphics[width=\textwidth]{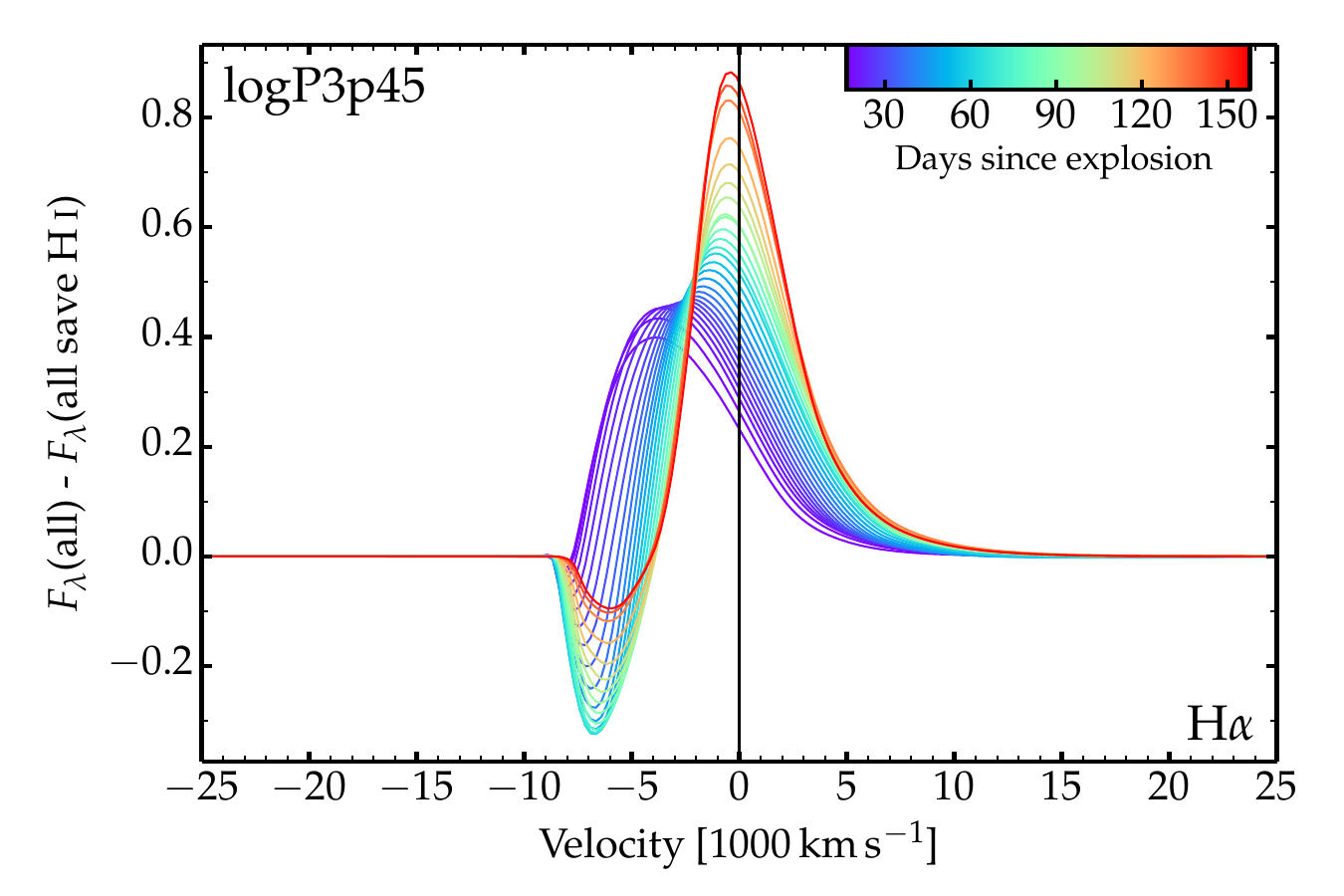}
    \end{subfigure}
    \caption{Evolution of the H$\alpha$ profile for our set of models. We show the flux obtained by accounting for all bound-bound transitions subtracted from the flux obtained by ignoring those associated with H\one. The fluxes are normalized by the peak H$\alpha$ flux in the ``full'' model. These profiles are used to compute the ratio of the absorption to the emission flux in H$\alpha$ (see Fig.~\ref{fig_ae}).}
\label{fig_ae_halpha_indiv}
\end{figure*}

\end{document}

%% file: table_init.tex
\begin{table*}
  \caption{Summary of preSN model properties.
\label{tab_mesa}
}
\begin{center}
\begin{tabular}{
l@{\hspace{4mm}}c@{\hspace{4mm}}c@{\hspace{4mm}}
c@{\hspace{4mm}}c@{\hspace{4mm}}c@{\hspace{4mm}}
c@{\hspace{4mm}}c@{\hspace{4mm}}c@{\hspace{4mm}}
c@{\hspace{4mm}}c@{\hspace{4mm}}c@{\hspace{4mm}}
c@{\hspace{4mm}}c@{\hspace{4mm}}
}
\hline
     Model &   $M_{\rm f}$ &    $L_\star$ &  $T_{\rm eff}$ &   $R_\star$ &       $M$(H) &   $M$(H-env) &    $M$(He)  &  $M$(O)  & $M_{\rm He,c}$ & $M_{\rm Fe,c}$ &  $X_s$(H)  &   $X_s$(He) \\
\hline
           &      [\msun]  &     [\lsun]&       [K]  &     [\rsun]&   [\msun]  &   [\msun]  &   [\msun]  &   [\msun]  &    [\msun] &   [\msun]  &            &     \\
\hline
  logP2p75$^\ast$ &     3.42 &      48823 &      24715 &       12.1 &       0.00 &       0.00 &       1.44 &       0.19 &       3.42 &       1.28 &       0.00 &       0.98 \\
  logP2p80 &            3.76 &      60789 &       4775 &      360.7 &       0.02 &       0.12 &       1.75 &       0.17 &       3.67 &       1.52 &       0.27 &       0.72 \\
  logP2p85 &            3.92 &      63983 &       3690 &      619.8 &       0.06 &       0.19 &       1.83 &       0.22 &       3.73 &       1.51 &       0.50 &       0.48 \\
  logP2p90 &            4.06 &      65103 &       3462 &      710.2 &       0.12 &       0.28 &       1.86 &       0.27 &       3.78 &       1.59 &       0.54 &       0.44 \\
  logP2p95$^\ast$ &     4.14 &      65102 &       3389 &      740.9 &       0.15 &       0.31 &       1.84 &       0.28 &       3.82 &       1.59 &       0.57 &       0.41 \\
  logP3p00$^\ast$ &     4.26 &      57528 &       3267 &      749.4 &       0.20 &       0.39 &       1.91 &       0.27 &       3.86 &       1.50 &       0.60 &       0.39 \\
  logP3p10$^\ast$ &     4.51 &      66203 &       3205 &      835.6 &       0.34 &       0.60 &       1.95 &       0.28 &       3.88 &       0.93 &       0.63 &       0.36 \\
  logP3p20 &            5.18 &      68434 &       3091 &      913.2 &       0.73 &       1.18 &       2.21 &       0.29 &       3.92 &       1.51 &       0.66 &       0.32 \\
  logP3p35 &            7.97 &      70136 &       3065 &      940.1 &       2.50 &       3.74 &       3.04 &       0.32 &       3.98 &       1.50 &       0.69 &       0.30 \\
  logP3p45 &           10.84 &      71080 &       3146 &      898.2 &       4.64 &       6.86 &       3.90 &       0.36 &       3.98 &       1.52 &       0.69 &       0.30 \\
\hline
\end{tabular}
\end{center}
    {\bf Notes:} For each progenitor model (which all have an initial mass of 12.59\,\msun), the table gives the final mass, the
    surface luminosity, the effective temperature, the surface radius, the total hydrogen mass, the
    mass of the H-rich envelope, the total mass of He and O, the He-core and Fe-core masses,
    and finally the surface mass fraction of H and He. The ``$^\ast$'' symbol refers to models that did not reach core collapse but terminated after core silicon exhaustion
    (consequently, these models show a lower central density -- see Fig.~\ref{fig_mesa} -- at the end of the \mesa\ simulation). For historical
    reasons having to do with the development of this work, the preSN models used here are not all exactly identical to the preSN models described in \citet{ercolino_bin_23}.
\end{table*}

%% file: table_cmfgen_init.tex
\begin{table*}
  \caption{Summary of ejecta properties used as initial conditions for our \cmfgen\ simulations.
\label{tab_cmfgen_init}
}
\begin{center}
\begin{tabular}{
l@{\hspace{4mm}}c@{\hspace{4mm}}c@{\hspace{4mm}}
c@{\hspace{4mm}}c@{\hspace{4mm}}c@{\hspace{4mm}}
c@{\hspace{4mm}}c@{\hspace{4mm}}c@{\hspace{4mm}}
}
\hline
Model    &$M_{\rm ej}$&$E_{\rm kin}$ & $M$(\nifs)  & $V_{99,{\rm H}}$& $V_{99,{\rm He}}$& $V_{99,^{56}{\rm Ni}}$  &  $X_s$(H)  &   $X_s$(He) \\
\hline
         &   [\msun] &     [\foe] &    [0.01\,\msun]     &     [1000\,\kms] &  [1000\,\kms]   &   [1000\,\kms] & &   \\
\hline
logP2p75 &      1.90 &       1.02 &       8.97 &   \dots &         1.79 &              6.47   &  0.   &       0.98 \\
logP2p80 &      2.25 &       1.07 &       8.96 &    7.12 &         2.10 &              6.74   &  0.13 &       0.86 \\
logP2p85 &      2.43 &       1.17 &       8.81 &    6.43 &         2.06 &              6.73   &  0.33 &       0.65 \\
logP2p90 &      2.49 &       1.15 &       8.79 &    5.93 &         2.31 &              7.05   &  0.48 &       0.51 \\
logP2p95 &      2.56 &       1.17 &       8.78 &    5.85 &         2.00 &              6.86   &  0.53 &       0.46 \\
logP3p00 &      2.75 &       1.09 &       8.76 &    4.96 &         2.39 &              6.76   &  0.57 &       0.42 \\
logP3p10 &      3.03 &       1.13 &       8.57 &    4.37 &         1.92 &              6.21   &  0.62 &       0.37 \\
logP3p20 &      3.72 &       1.19 &       8.55 &    2.92 &         2.13 &              6.54   &  0.66 &       0.33 \\
logP3p35 &      6.49 &       0.98 &       8.76 &    1.57 &         1.39 &              5.70   &  0.68 &       0.30 \\
logP3p45 &      9.45 &       1.09 &       8.84 &    0.78 &         0.64 &              5.45   &  0.68 &       0.30 \\
\hline
\end{tabular}
\end{center}
    {\bf Notes:} The columns give in turn the ejecta mass, the kinetic energy, the \nifs\ mass, the velocity bounding 99\% of the mass of H, He, and
    \nifs\ (integral performed inwards for H and He and outwards for \nifs), and the surface mass fractions of H and He. These abundance values
    slightly differ from those shown in Table~\ref{tab_mesa} because of the adopted chemical mixing (see discussion in Section~\ref{sect_voned}).
\end{table*}

%% file: table_obs.tex
\begin{table*}
\begin{center}
    \caption{Characteristics of the selected sample of observations, ranked in order of increasing of $M$(H).}
\label{table_obs}
\setlength{\tabcolsep}{4pt}
\begin{tabular}{cccccccc}
\hline
\hline
    SN         & Type & Redshift & $\mu$ & $E(B-V)_{{\rm tot}}$ & Explosion date & Refs (spectra)  & Refs (photometry) \\
           &      &          & (mag) &    (mag)       &     (MJD)      &                       &                         \\
\hline
\hline
iPTF13bvn  & Ib   & 0.00449  & 31.76 &    0.071       &     56459      &    [1]; [2]; [3]      &       [1]; [4]; [5]    \\
SN~2011dh  & IIb  & 0.00168  & 29.42 &    0.07        &     55713      &          [6]          &            [6]          \\
SN~1993J   & IIb  & 0.000113 & 27.82 &     0.2        &     49074      &       [7]; [8]        &       [7]; [9]; [10]     \\
SN~2006Y   & II-L & 0.03358  & 35.73 &    0.115       &     53766      &         [11]          &          [12]; [13]     \\
SN~2017eaw & II-P & 0.002192 & 29.47 &    0.30        &     57885      &         [14]          &          [15]; [16]     \\
\hline
\hline
\end{tabular}
\begin{list}{}{}
\item Data downloaded from \url{http://wiserep.weizmann.ac.il/home} \citep{wiserep}.
\item \textbf{References:} [1] \citealt{cao_13bvn_13};  [2] \citealt{fremling_13bvn_14}; [3] \citealt{fremling_sesn_16};
    [4] \citealt{srivastav_13bvn_14}; [5] \citealt{zheng_sesn_22};
    [6] \citealt{ergon_14_11dh}; [7] \citealt{barbon_93J_95}; [8] \citealt{matheson_93j_00a}; [9] \citealt{richmond_93j_94};
    [10] \citealt{richmond_93J_96}; [11] \citealt{gutierrez_pap1_17}; [12] \citealt{anderson_2pl}; [13] \citealt{hiramatsu_2p_2l_21};
[14] \citealt{szalai_17eaw_19}; [15] \citealt{tsvetkov_17eaw_18}; [16] \citealt{vandyk_17eaw_19}.
\end{list}
\end{center}
\end{table*}

%% file: ms.bbl
\begin{thebibliography}{116}
\expandafter\ifx\csname natexlab\endcsname\relax\def\natexlab#1{#1}\fi

\bibitem[{{Anderson} {et~al.}(2014{\natexlab{a}}){Anderson}, {Dessart},
  {Gutierrez}, {Hamuy}, {Morrell}, {Phillips}, {Folatelli}, {Stritzinger},
  {Freedman}, {Gonz{\'a}lez-Gait{\'a}n}, {McCarthy}, {Suntzeff}, \&
  {Thomas-Osip}}]{anderson_blueshift_14}
{Anderson}, J.~P., {Dessart}, L., {Gutierrez}, C.~P., {et~al.}
  2014{\natexlab{a}}, \mnras, 441, 671

\bibitem[{{Anderson} {et~al.}(2014{\natexlab{b}}){Anderson},
  {Gonz{\'a}lez-Gait{\'a}n}, {Hamuy}, {Guti{\'e}rrez}, {Stritzinger}, {Olivares
  E.}, {Phillips}, {Schulze}, {Antezana}, {Bolt}, {Campillay}, {Castell{\'o}n},
  {Contreras}, {de Jaeger}, {Folatelli}, {F{\"o}rster}, {Freedman},
  {Gonz{\'a}lez}, {Hsiao}, {Krzemi{\'n}ski}, {Krisciunas}, {Maza}, {McCarthy},
  {Morrell}, {Persson}, {Roth}, {Salgado}, {Suntzeff}, \&
  {Thomas-Osip}}]{anderson_2pl}
{Anderson}, J.~P., {Gonz{\'a}lez-Gait{\'a}n}, S., {Hamuy}, M., {et~al.}
  2014{\natexlab{b}}, \apj, 786, 67

\bibitem[{{Barbon} {et~al.}(1995){Barbon}, {Benetti}, {Cappellaro}, {Patat},
  {Turatto}, \& {Iijima}}]{barbon_93J_95}
{Barbon}, R., {Benetti}, S., {Cappellaro}, E., {et~al.} 1995, \aaps, 110, 513

\bibitem[{{Ben-Ami} {et~al.}(2015){Ben-Ami}, {Hachinger}, {Gal-Yam}, {Mazzali},
  {Filippenko}, {Horesh}, {Matheson}, {Modjaz}, {Sauer}, {Silverman}, {Smith},
  \& {Yaron}}]{benami_iib_15}
{Ben-Ami}, S., {Hachinger}, S., {Gal-Yam}, A., {et~al.} 2015, \apj, 803, 40

\bibitem[{{Bersten} {et~al.}(2014){Bersten}, {Benvenuto}, {Folatelli},
  {Nomoto}, {Kuncarayakti}, {Srivastav}, {Anupama}, {Quimby}, \&
  {Sahu}}]{bersten_iPTF13bvn_14}
{Bersten}, M.~C., {Benvenuto}, O.~G., {Folatelli}, G., {et~al.} 2014, \aj, 148,
  68

\bibitem[{{Bersten} {et~al.}(2012){Bersten}, {Benvenuto}, {Nomoto}, {Ergon},
  {Folatelli}, {Sollerman}, {Benetti}, {Botticella}, {Fraser}, {Kotak},
  {Maeda}, {Ochner}, \& {Tomasella}}]{bersten_etal_12_11dh}
{Bersten}, M.~C., {Benvenuto}, O.~G., {Nomoto}, K., {et~al.} 2012, \apj, 757,
  31

\bibitem[{{Bersten} {et~al.}(2013){Bersten}, {Tanaka}, {Tominaga}, {Benvenuto},
  \& {Nomoto}}]{bersten_08D_13}
{Bersten}, M.~C., {Tanaka}, M., {Tominaga}, N., {Benvenuto}, O.~G., \&
  {Nomoto}, K. 2013, \apj, 767, 143

\bibitem[{{Blinnikov} \& {Bartunov}(1993)}]{blinnikov_bartunov_2l_93}
{Blinnikov}, S.~I. \& {Bartunov}, O.~S. 1993, \aap, 273, 106

\bibitem[{{Blinnikov} {et~al.}(1998){Blinnikov}, {Eastman}, {Bartunov},
  {Popolitov}, \& {Woosley}}]{blinnikov_94_93j}
{Blinnikov}, S.~I., {Eastman}, R., {Bartunov}, O.~S., {Popolitov}, V.~A., \&
  {Woosley}, S.~E. 1998, \apj, 496, 454

\bibitem[{{Blondin} {et~al.}(2017){Blondin}, {Dessart}, {Hillier}, \&
  {Khokhlov}}]{blondin_wlr_17}
{Blondin}, S., {Dessart}, L., {Hillier}, D.~J., \& {Khokhlov}, A.~M. 2017,
  \mnras, 470, 157

\bibitem[{{Blondin} {et~al.}(2023){Blondin}, {Dessart}, {Hillier},
  {Ramsbottom}, \& {Storey}}]{blondin_21aefx_23}
{Blondin}, S., {Dessart}, L., {Hillier}, D.~J., {Ramsbottom}, C.~A., \&
  {Storey}, P.~J. 2023, \aap, 678, A170

\bibitem[{{Bose} {et~al.}(2015){Bose}, {Sutaria}, {Kumar}, {Duggal}, {Misra},
  {Brown}, {Singh}, {Dwarkadas}, {York}, {Chakraborti}, {Chandola},
  {Dahlstrom}, {Ray}, \& {Safonova}}]{bose_13ej_15}
{Bose}, S., {Sutaria}, F., {Kumar}, B., {et~al.} 2015, \apj, 806, 160

\bibitem[{{Bostroem} {et~al.}(2023){Bostroem}, {Dessart}, {Hillier},
  {Lundquist}, {Andrews}, {Sand}, {Dong}, {Valenti}, {Haislip}, {Hoang},
  {Hosseinzadeh}, {Janzen}, {Jencson}, {Jha}, {Kouprianov}, {Pearson}, {Meza
  Retamal}, {Reichart}, {Shrestha}, {Ashall}, {Baron}, {Brown}, {DerKacy},
  {Farah}, {Galbany}, {Gonz{\'a}lez Hern{\'a}ndez}, {Green}, {Hoeflich},
  {Howell}, {Kwok}, {McCully}, {M{\"u}ller-Bravo}, {Newsome}, {Gonzalez},
  {Pellegrino}, {Rho}, {Rowe}, {Schwab}, {Shahbandeh}, {Smith}, {Strader},
  {Terreran}, {Van Dyk}, \& {Wyatt}}]{bostroem_22acko_23}
{Bostroem}, K.~A., {Dessart}, L., {Hillier}, D.~J., {et~al.} 2023, \apjl, 953,
  L18

\bibitem[{{Cao} {et~al.}(2013){Cao}, {Kasliwal}, {Arcavi}, {Horesh}, {Hancock},
  {Valenti}, {Cenko}, {Kulkarni}, {Gal-Yam}, {Gorbikov}, {Ofek}, {Sand},
  {Yaron}, {Graham}, {Silverman}, {Wheeler}, {Marion}, {Walker}, {Mazzali},
  {Howell}, {Li}, {Kong}, {Bloom}, {Nugent}, {Surace}, {Masci}, {Carpenter},
  {Degenaar}, \& {Gelino}}]{cao_13bvn_13}
{Cao}, Y., {Kasliwal}, M.~M., {Arcavi}, I., {et~al.} 2013, \apjl, 775, L7

\bibitem[{{Chun} {et~al.}(2018){Chun}, {Yoon}, {Jung}, {Kim}, \&
  {Kim}}]{chun_rsg_18}
{Chun}, S.-H., {Yoon}, S.-C., {Jung}, M.-K., {Kim}, D.~U., \& {Kim}, J. 2018,
  \apj, 853, 79

\bibitem[{{Davies} {et~al.}(2013){Davies}, {Kudritzki}, {Plez}, {Trager},
  {Lan{\c c}on}, {Gazak}, {Bergemann}, {Evans}, \& {Chiavassa}}]{davies_rsg_13}
{Davies}, B., {Kudritzki}, R.-P., {Plez}, B., {et~al.} 2013, \apj, 767, 3

\bibitem[{{Dessart} \& {Audit}(2019)}]{dessart_audit_rhd_3d_19}
{Dessart}, L. \& {Audit}, E. 2019, \aap, 629, A17

\bibitem[{{Dessart} {et~al.}(2023){Dessart}, {Guti{\'e}rrez}, {Kuncarayakti},
  {Fox}, \& {Filippenko}}]{d23_interaction}
{Dessart}, L., {Guti{\'e}rrez}, C.~P., {Kuncarayakti}, H., {Fox}, O.~D., \&
  {Filippenko}, A.~V. 2023, \aap, 675, A33

\bibitem[{{Dessart} \& {Hillier}(2005{\natexlab{a}})}]{D05_epm}
{Dessart}, L. \& {Hillier}, D.~J. 2005{\natexlab{a}}, \aap, 439, 671

\bibitem[{{Dessart} \& {Hillier}(2005{\natexlab{b}})}]{DH05a}
{Dessart}, L. \& {Hillier}, D.~J. 2005{\natexlab{b}}, \aap, 437, 667

\bibitem[{{Dessart} \& {Hillier}(2011)}]{DH11_2p}
{Dessart}, L. \& {Hillier}, D.~J. 2011, \mnras, 410, 1739

\bibitem[{{Dessart} \& {Hillier}(2019)}]{d19_sn2p}
{Dessart}, L. \& {Hillier}, D.~J. 2019, \aap, 625, A9

\bibitem[{{Dessart} \& {Hillier}(2020)}]{DH20_shuffle}
{Dessart}, L. \& {Hillier}, D.~J. 2020, \aap, 643, L13

\bibitem[{{Dessart} \& {Hillier}(2022)}]{dessart_csm_22}
{Dessart}, L. \& {Hillier}, D.~J. 2022, \aap, 660, L9

\bibitem[{{Dessart} {et~al.}(2017){Dessart}, {Hillier}, \& {Audit}}]{d17_13fs}
{Dessart}, L., {Hillier}, D.~J., \& {Audit}, E. 2017, \aap, 605, A83

\bibitem[{{Dessart} {et~al.}(2016{\natexlab{a}}){Dessart}, {Hillier}, {Audit},
  {Livne}, \& {Waldman}}]{D16_2n}
{Dessart}, L., {Hillier}, D.~J., {Audit}, E., {Livne}, E., \& {Waldman}, R.
  2016{\natexlab{a}}, \mnras, 458, 2094

\bibitem[{{Dessart} {et~al.}(2011){Dessart}, {Hillier}, {Livne}, {Yoon},
  {Woosley}, {Waldman}, \& {Langer}}]{dessart_11_wr}
{Dessart}, L., {Hillier}, D.~J., {Livne}, E., {et~al.} 2011, \mnras, 414, 2985

\bibitem[{{Dessart} {et~al.}(2021{\natexlab{a}}){Dessart}, {Hillier},
  {Sukhbold}, {Woosley}, \& {Janka}}]{dessart_snibc_21}
{Dessart}, L., {Hillier}, D.~J., {Sukhbold}, T., {Woosley}, S.~E., \& {Janka},
  H.~T. 2021{\natexlab{a}}, \aap, 656, A61

\bibitem[{{Dessart} {et~al.}(2013){Dessart}, {Hillier}, {Waldman}, \&
  {Livne}}]{d13_sn2p}
{Dessart}, L., {Hillier}, D.~J., {Waldman}, R., \& {Livne}, E. 2013, \mnras,
  433, 1745

\bibitem[{{Dessart} {et~al.}(2018{\natexlab{a}}){Dessart}, {Hillier}, \&
  {Wilk}}]{d18_fcl}
{Dessart}, L., {Hillier}, D.~J., \& {Wilk}, K.~D. 2018{\natexlab{a}}, \aap,
  619, A30

\bibitem[{{Dessart} {et~al.}(2015){Dessart}, {Hillier}, {Woosley}, {Livne},
  {Waldman}, {Yoon}, \& {Langer}}]{D15_SNIbc_I}
{Dessart}, L., {Hillier}, D.~J., {Woosley}, S., {et~al.} 2015, \mnras, 453,
  2189

\bibitem[{{Dessart} {et~al.}(2016{\natexlab{b}}){Dessart}, {Hillier},
  {Woosley}, {Livne}, {Waldman}, {Yoon}, \& {Langer}}]{D16_SNIbc_II}
{Dessart}, L., {Hillier}, D.~J., {Woosley}, S., {et~al.} 2016{\natexlab{b}},
  \mnras, 458, 1618

\bibitem[{{Dessart} {et~al.}(2021{\natexlab{b}}){Dessart}, {John Hillier},
  {Sukhbold}, {Woosley}, \& {Janka}}]{D21_sn2p_neb}
{Dessart}, L., {John Hillier}, D., {Sukhbold}, T., {Woosley}, S.~E., \&
  {Janka}, H.~T. 2021{\natexlab{b}}, \aap, 652, A64

\bibitem[{{Dessart} {et~al.}(2010{\natexlab{a}}){Dessart}, {Livne}, \&
  {Waldman}}]{dlw10b}
{Dessart}, L., {Livne}, E., \& {Waldman}, R. 2010{\natexlab{a}}, \mnras, 408,
  827

\bibitem[{{Dessart} {et~al.}(2010{\natexlab{b}}){Dessart}, {Livne}, \&
  {Waldman}}]{dlw10a}
{Dessart}, L., {Livne}, E., \& {Waldman}, R. 2010{\natexlab{b}}, \mnras, 405,
  2113

\bibitem[{{Dessart} {et~al.}(2020){Dessart}, {Yoon}, {Aguilera-Dena}, \&
  {Langer}}]{dessart_snibc_20}
{Dessart}, L., {Yoon}, S.-C., {Aguilera-Dena}, D.~R., \& {Langer}, N. 2020,
  \aap, 642, A106

\bibitem[{{Dessart} {et~al.}(2018{\natexlab{b}}){Dessart}, {Yoon}, {Livne}, \&
  {Waldman}}]{d18_ext_ccsn}
{Dessart}, L., {Yoon}, S.-C., {Livne}, E., \& {Waldman}, R. 2018{\natexlab{b}},
  \aap, 612, A61

\bibitem[{{Eldridge} {et~al.}(2018){Eldridge}, {Xiao}, {Stanway}, {Rodrigues},
  \& {Guo}}]{eldridge_sn2_18}
{Eldridge}, J.~J., {Xiao}, L., {Stanway}, E.~R., {Rodrigues}, N., \& {Guo},
  N.~Y. 2018, \pasa, 35, e049

\bibitem[{{Ensman} \& {Woosley}(1988)}]{ensman_woosley_88}
{Ensman}, L.~M. \& {Woosley}, S.~E. 1988, \apj, 333, 754

\bibitem[{{Ercolino} {et~al.}(2023){Ercolino}, {Jin}, {Langer}, \&
  {Dessart}}]{ercolino_bin_23}
{Ercolino}, A., {Jin}, H., {Langer}, N., \& {Dessart}, L. 2023, arXiv e-prints,
  arXiv:2308.01819

\bibitem[{{Ergon} \& {Fransson}(2022)}]{ergon_11dh_22}
{Ergon}, M. \& {Fransson}, C. 2022, \aap, 666, A104

\bibitem[{{Ergon} {et~al.}(2023){Ergon}, {Lundqvist}, {Fransson},
  {Kuncarayakti}, {Das}, {De}, {Ferrari}, {Fremling}, {Medler}, {Maeda},
  {Pastorello}, {Sollerman}, \& {Stritzinger}}]{ergon_20acat_23}
{Ergon}, M., {Lundqvist}, P., {Fransson}, C., {et~al.} 2023, arXiv e-prints,
  arXiv:2308.07158

\bibitem[{{Ergon} {et~al.}(2014){Ergon}, {Sollerman}, {Fraser}, {Pastorello},
  {Taubenberger}, {Elias-Rosa}, {Bersten}, {Jerkstrand}, {Benetti},
  {Botticella}, {Fransson}, {Harutyunyan}, {Kotak}, {Smartt}, {Valenti},
  {Bufano}, {Cappellaro}, {Fiaschi}, {Howell}, {Kankare}, {Magill}, {Mattila},
  {Maund}, {Naves}, {Ochner}, {Ruiz}, {Smith}, {Tomasella}, \&
  {Turatto}}]{ergon_14_11dh}
{Ergon}, M., {Sollerman}, J., {Fraser}, M., {et~al.} 2014, \aap, 562, A17

\bibitem[{{Fassia} {et~al.}(2001){Fassia}, {Meikle}, {Chugai}, {Geballe},
  {Lundqvist}, {Walton}, {Pollacco}, {Veilleux}, {Wright}, {Pettini}, {Kerr},
  {Puchnarewicz}, {Puxley}, {Irwin}, {Packham}, {Smartt}, \&
  {Harmer}}]{fassia_98S_01}
{Fassia}, A., {Meikle}, W.~P.~S., {Chugai}, N., {et~al.} 2001, \mnras, 325, 907

\bibitem[{{Fassia} {et~al.}(2000){Fassia}, {Meikle}, {Vacca}, {Kemp}, {Walton},
  {Pollacco}, {Smartt}, {Oscoz}, {Arag{\'o}n-Salamanca}, {Bennett}, {Hawarden},
  {Alonso}, {Alcalde}, {Pedrosa}, {Telting}, {Arevalo}, {Deeg}, {Garz{\'o}n},
  {G{\'o}mez-Rold{\'a}n}, {G{\'o}mez}, {Guti{\'e}rrez}, {L{\'o}pez}, {Rozas},
  {Serra-Ricart}, \& {Zapatero-Osorio}}]{fassia_98S_00}
{Fassia}, A., {Meikle}, W.~P.~S., {Vacca}, W.~D., {et~al.} 2000, \mnras, 318,
  1093

\bibitem[{{F{\"o}rster} {et~al.}(2018){F{\"o}rster}, {Moriya}, {Maureira},
  {Anderson}, {Blinnikov}, {Bufano}, {Cabrera-Vives}, {Clocchiatti}, {de
  Jaeger}, {Est{\'e}vez}, {Galbany}, {Gonz{\'a}lez-Gait{\'a}n}, {Gr{\"a}fener},
  {Hamuy}, {Hsiao}, {Huentelemu}, {Huijse}, {Kuncarayakti}, {Mart{\'\i}nez},
  {Medina}, {Olivares E.}, {Pignata}, {Razza}, {Reyes}, {San Mart{\'\i}n},
  {Smith}, {Vera}, {Vivas}, {de Ugarte Postigo}, {Yoon}, {Ashall}, {Fraser},
  {Gal-Yam}, {Kankare}, {Le Guillou}, {Mazzali}, {Walton}, \&
  {Young}}]{forster_csm_18}
{F{\"o}rster}, F., {Moriya}, T.~J., {Maureira}, J.~C., {et~al.} 2018, Nature
  Astronomy, 2, 808

\bibitem[{{Fremling} {et~al.}(2016){Fremling}, {Sollerman}, {Taddia}, {Ergon},
  {Fraser}, {Karamehmetoglu}, {Valenti}, {Jerkstrand}, {Arcavi}, {Bufano},
  {Elias Rosa}, {Filippenko}, {Fox}, {Gal-Yam}, {Howell}, {Kotak}, {Mazzali},
  {Milisavljevic}, {Nugent}, {Nyholm}, {Pian}, \& {Smartt}}]{fremling_sesn_16}
{Fremling}, C., {Sollerman}, J., {Taddia}, F., {et~al.} 2016, \aap, 593, A68

\bibitem[{{Fremling} {et~al.}(2014){Fremling}, {Sollerman}, {Taddia}, {Ergon},
  {Valenti}, {Arcavi}, {Ben-Ami}, {Cao}, {Cenko}, {Filippenko}, {Gal-Yam}, \&
  {Howell}}]{fremling_13bvn_14}
{Fremling}, C., {Sollerman}, J., {Taddia}, F., {et~al.} 2014, \aap, 565, A114

\bibitem[{{Gabler} {et~al.}(2021){Gabler}, {Wongwathanarat}, \&
  {Janka}}]{gabler_3dsn_21}
{Gabler}, M., {Wongwathanarat}, A., \& {Janka}, H.-T. 2021, \mnras, 502, 3264

\bibitem[{{Gilkis} \& {Arcavi}(2022)}]{gilkis_arcavi_22}
{Gilkis}, A. \& {Arcavi}, I. 2022, \mnras, 511, 691

\bibitem[{{Goldberg} {et~al.}(2019){Goldberg}, {Bildsten}, \&
  {Paxton}}]{goldberg_sn2p_19}
{Goldberg}, J.~A., {Bildsten}, L., \& {Paxton}, B. 2019, \apj, 879, 3

\bibitem[{{Goldberg} {et~al.}(2022){Goldberg}, {Jiang}, \&
  {Bildsten}}]{goldberg_3d_rsg_22}
{Goldberg}, J.~A., {Jiang}, Y.-F., \& {Bildsten}, L. 2022, \apj, 929, 156

\bibitem[{{Gonz{\'a}lez-Gait{\'a}n} {et~al.}(2015){Gonz{\'a}lez-Gait{\'a}n},
  {Tominaga}, {Molina}, {Galbany}, {Bufano}, {Anderson}, {Gutierrez},
  {F{\"o}rster}, {Pignata}, {Bersten}, {Howell}, {Sullivan}, {Carlberg}, {de
  Jaeger}, {Hamuy}, {Baklanov}, \& {Blinnikov}}]{gonzalez_gaitan_2p_15}
{Gonz{\'a}lez-Gait{\'a}n}, S., {Tominaga}, N., {Molina}, J., {et~al.} 2015,
  \mnras, 451, 2212

\bibitem[{{Grassberg} {et~al.}(1971){Grassberg}, {Imshennik}, \&
  {Nadyozhin}}]{grassberg_71}
{Grassberg}, E.~K., {Imshennik}, V.~S., \& {Nadyozhin}, D.~K. 1971, \apss, 10,
  28

\bibitem[{{Guti{\'e}rrez} {et~al.}(2014){Guti{\'e}rrez}, {Anderson}, {Hamuy},
  {Gonz{\'a}lez-Gait{\'a}n}, {Folatelli}, {Morrell}, {Stritzinger}, {Phillips},
  {McCarthy}, {Suntzeff}, \& {Thomas-Osip}}]{gutierrez_ha_14}
{Guti{\'e}rrez}, C.~P., {Anderson}, J.~P., {Hamuy}, M., {et~al.} 2014, \apjl,
  786, L15

\bibitem[{{Guti{\'e}rrez} {et~al.}(2017){Guti{\'e}rrez}, {Anderson}, {Hamuy},
  {Morrell}, {Gonz{\'a}lez-Gaitan}, {Stritzinger}, {Phillips}, {Galbany},
  {Folatelli}, {Dessart}, {Contreras}, {Della Valle}, {Freedman}, {Hsiao},
  {Krisciunas}, {Madore}, {Maza}, {Suntzeff}, {Prieto}, {Gonz{\'a}lez},
  {Cappellaro}, {Navarrete}, {Pizzella}, {Ruiz}, {Smith}, \&
  {Turatto}}]{gutierrez_pap1_17}
{Guti{\'e}rrez}, C.~P., {Anderson}, J.~P., {Hamuy}, M., {et~al.} 2017, \apj,
  850, 89

\bibitem[{{Hachinger} {et~al.}(2012){Hachinger}, {Mazzali}, {Taubenberger},
  {Hillebrandt}, {Nomoto}, \& {Sauer}}]{hachinger_13_he}
{Hachinger}, S., {Mazzali}, P.~A., {Taubenberger}, S., {et~al.} 2012, \mnras,
  422, 70

\bibitem[{{Hillier} \& {Dessart}(2012)}]{HD12}
{Hillier}, D.~J. \& {Dessart}, L. 2012, \mnras, 424, 252

\bibitem[{{Hillier} \& {Dessart}(2019)}]{HD19}
{Hillier}, D.~J. \& {Dessart}, L. 2019, \aap, 631, A8

\bibitem[{{Hillier} \& {Miller}(1998)}]{hm98}
{Hillier}, D.~J. \& {Miller}, D.~L. 1998, \apj, 496, 407

\bibitem[{{Hiramatsu} {et~al.}(2021){Hiramatsu}, {Howell}, {Moriya},
  {Goldberg}, {Hosseinzadeh}, {Arcavi}, {Anderson}, {Guti{\'e}rrez}, {Burke},
  {McCully}, {Valenti}, {Galbany}, {Fang}, {Maeda}, {Folatelli}, {Hsiao},
  {Morrell}, {Phillips}, {Stritzinger}, {Suntzeff}, {Gromadzki}, {Maguire},
  {M{\"u}ller-Bravo}, \& {Young}}]{hiramatsu_2p_2l_21}
{Hiramatsu}, D., {Howell}, D.~A., {Moriya}, T.~J., {et~al.} 2021, \apj, 913, 55

\bibitem[{{Jerkstrand} {et~al.}(2015){Jerkstrand}, {Ergon}, {Smartt},
  {Fransson}, {Sollerman}, {Taubenberger}, {Bersten}, \&
  {Spyromilio}}]{jerkstrand_15_iib}
{Jerkstrand}, A., {Ergon}, M., {Smartt}, S.~J., {et~al.} 2015, \aap, 573, A12

\bibitem[{{Justham} {et~al.}(2014){Justham}, {Podsiadlowski}, \&
  {Vink}}]{justham_lbv_14}
{Justham}, S., {Podsiadlowski}, P., \& {Vink}, J.~S. 2014, \apj, 796, 121

\bibitem[{{Kasen} \& {Woosley}(2009)}]{KW09}
{Kasen}, D. \& {Woosley}, S.~E. 2009, \apj, 703, 2205

\bibitem[{{Langer} {et~al.}(2020){Langer}, {Sch{\"u}rmann}, {Stoll},
  {Marchant}, {Lennon}, {Mahy}, {de Mink}, {Quast}, {Riedel}, {Sana},
  {Schneider}, {Schootemeijer}, {Wang}, {Almeida}, {Bestenlehner},
  {Bodensteiner}, {Castro}, {Clark}, {Crowther}, {Dufton}, {Evans}, {Fossati},
  {Gr{\"a}fener}, {Grassitelli}, {Grin}, {Hastings}, {Herrero}, {de Koter},
  {Menon}, {Patrick}, {Puls}, {Renzo}, {Sand er}, {Schneider}, {Sen}, {Shenar},
  {Sim{\'o}n-D{\'\i}as}, {Tauris}, {Tramper}, {Vink}, \& {Xu}}]{langer_bh_20}
{Langer}, N., {Sch{\"u}rmann}, C., {Stoll}, K., {et~al.} 2020, \aap, 638, A39

\bibitem[{{Laplace} {et~al.}(2021){Laplace}, {Justham}, {Renzo}, {G{\"o}tberg},
  {Farmer}, {Vartanyan}, \& {de Mink}}]{laplace_evol_21}
{Laplace}, E., {Justham}, S., {Renzo}, M., {et~al.} 2021, \aap, 656, A58

\bibitem[{{Leonard} {et~al.}(2000){Leonard}, {Filippenko}, {Barth}, \&
  {Matheson}}]{leonard_98S_00}
{Leonard}, D.~C., {Filippenko}, A.~V., {Barth}, A.~J., \& {Matheson}, T. 2000,
  \apj, 536, 239

\bibitem[{{Li} {et~al.}(2012){Li}, {Hillier}, \& {Dessart}}]{li_etal_12_nonte}
{Li}, C., {Hillier}, D.~J., \& {Dessart}, L. 2012, \mnras, 426, 1671

\bibitem[{{Liu} {et~al.}(2016){Liu}, {Modjaz}, {Bianco}, \&
  {Graur}}]{liu_snibc_15}
{Liu}, Y.-Q., {Modjaz}, M., {Bianco}, F.~B., \& {Graur}, O. 2016, \apj, 827, 90

\bibitem[{{Livne}(1993)}]{livne_93}
{Livne}, E. 1993, \apj, 412, 634

\bibitem[{{Lucy}(1991)}]{lucy_91}
{Lucy}, L.~B. 1991, \apj, 383, 308

\bibitem[{{Matheson} {et~al.}(2000){Matheson}, {Filippenko}, {Barth}, {Ho},
  {Leonard}, {Bershady}, {Davis}, {Finley}, {Fisher}, {Gonz{\'a}lez}, {Hawley},
  {Koo}, {Li}, {Lonsdale}, {Schlegel}, {Smith}, {Spinrad}, \&
  {Wirth}}]{matheson_93j_00a}
{Matheson}, T., {Filippenko}, A.~V., {Barth}, A.~J., {et~al.} 2000, \aj, 120,
  1487

\bibitem[{{Menon} {et~al.}(2023){Menon}, {Ercolino}, {Urbaneja}, {Lennon},
  {Herrero}, {Hirai}, {Langer}, {Schootemeijer}, {Chatzopoulos}, {Frank}, \&
  {Shiber}}]{menon_bsg_23}
{Menon}, A., {Ercolino}, A., {Urbaneja}, M.~A., {et~al.} 2023, arXiv e-prints,
  arXiv:2311.05581

\bibitem[{{Milisavljevic} \& {Fesen}(2015)}]{milisavljevic_fesen_casA_15}
{Milisavljevic}, D. \& {Fesen}, R.~A. 2015, Science, 347, 526

\bibitem[{{Modjaz} {et~al.}(2019){Modjaz}, {Guti{\'e}rrez}, \&
  {Arcavi}}]{modjaz_rev_19}
{Modjaz}, M., {Guti{\'e}rrez}, C.~P., \& {Arcavi}, I. 2019, Nature Astronomy,
  3, 717

\bibitem[{{Morales-Garoffolo} {et~al.}(2014){Morales-Garoffolo}, {Elias-Rosa},
  {Benetti}, {Taubenberger}, {Cappellaro}, {Pastorello}, {Klauser}, {Valenti},
  {Howerton}, {Ochner}, {Schramm}, {Siviero}, {Tartaglia}, \&
  {Tomasella}}]{morales_13df_14}
{Morales-Garoffolo}, A., {Elias-Rosa}, N., {Benetti}, S., {et~al.} 2014,
  \mnras, 445, 1647

\bibitem[{{Morales-Garoffolo} {et~al.}(2015){Morales-Garoffolo}, {Elias-Rosa},
  {Bersten}, {Jerkstrand}, {Taubenberger}, {Benetti}, {Cappellaro}, {Kotak},
  {Pastorello}, {Bufano}, {Dom{\'\i}nguez}, {Ergon}, {Fraser}, {Gao},
  {Garc{\'\i}a}, {Howell}, {Isern}, {Smartt}, {Tomasella}, \&
  {Valenti}}]{morales_11fu_15}
{Morales-Garoffolo}, A., {Elias-Rosa}, N., {Bersten}, M., {et~al.} 2015,
  \mnras, 454, 95

\bibitem[{{Moriya} {et~al.}(2017){Moriya}, {Yoon}, {Gr{\"a}fener}, \&
  {Blinnikov}}]{moriya_13fs_17}
{Moriya}, T.~J., {Yoon}, S.-C., {Gr{\"a}fener}, G., \& {Blinnikov}, S.~I. 2017,
  \mnras, 469, L108

\bibitem[{{Morozova} {et~al.}(2015){Morozova}, {Piro}, {Renzo}, {Ott},
  {Clausen}, {Couch}, {Ellis}, \& {Roberts}}]{snec}
{Morozova}, V., {Piro}, A.~L., {Renzo}, M., {et~al.} 2015, \apj, 814, 63

\bibitem[{{Morozova} {et~al.}(2017){Morozova}, {Piro}, \&
  {Valenti}}]{morozova_2l_2p_17}
{Morozova}, V., {Piro}, A.~L., \& {Valenti}, S. 2017, \apj, 838, 28

\bibitem[{{Morozova} {et~al.}(2018){Morozova}, {Piro}, \&
  {Valenti}}]{morozova_sn2p_18}
{Morozova}, V., {Piro}, A.~L., \& {Valenti}, S. 2018, \apj, 858, 15

\bibitem[{{Nakar} \& {Piro}(2014)}]{nakar_piro_14}
{Nakar}, E. \& {Piro}, A.~L. 2014, \apj, 788, 193

\bibitem[{{Nomoto} {et~al.}(1993){Nomoto}, {Suzuki}, {Shigeyama}, {Kumagai},
  {Yamaoka}, \& {Saio}}]{nomoto_93j_93}
{Nomoto}, K., {Suzuki}, T., {Shigeyama}, T., {et~al.} 1993, \nat, 364, 507

\bibitem[{{Panagia} {et~al.}(1980){Panagia}, {Vettolani}, {Boksenberg},
  {Ciatti}, {Ortolani}, {Rafanelli}, {Rosino}, {Gordon}, {Reimers}, {Hempe},
  {Benvenuti}, {Clavel}, {Heck}, {Penston}, {Macchetto}, {Stickland},
  {Bergeron}, {Tarenghi}, {Marano}, {Palumbo}, {Parmar}, {Pollard}, {Sanford},
  {Sargent}, {Sramek}, {Weiler}, \& {Matzik}}]{panagia_79c_80}
{Panagia}, N., {Vettolani}, G., {Boksenberg}, A., {et~al.} 1980, \mnras, 192,
  861

\bibitem[{{Park} {et~al.}(2023){Park}, {Yoon}, \& {Blinnikov}}]{park_iib_23}
{Park}, S.~H., {Yoon}, S.-C., \& {Blinnikov}, S. 2023, arXiv e-prints,
  arXiv:2310.16328

\bibitem[{{Paxton} {et~al.}(2011){Paxton}, {Bildsten}, {Dotter}, {Herwig},
  {Lesaffre}, \& {Timmes}}]{mesa1}
{Paxton}, B., {Bildsten}, L., {Dotter}, A., {et~al.} 2011, \apjs, 192, 3

\bibitem[{{Paxton} {et~al.}(2013){Paxton}, {Cantiello}, {Arras}, {Bildsten},
  {Brown}, {Dotter}, {Mankovich}, {Montgomery}, {Stello}, {Timmes}, \&
  {Townsend}}]{mesa2}
{Paxton}, B., {Cantiello}, M., {Arras}, P., {et~al.} 2013, \apjs, 208, 4

\bibitem[{{Paxton} {et~al.}(2015){Paxton}, {Marchant}, {Schwab}, {Bauer},
  {Bildsten}, {Cantiello}, {Dessart}, {Farmer}, {Hu}, {Langer}, {Townsend},
  {Townsley}, \& {Timmes}}]{mesa3}
{Paxton}, B., {Marchant}, P., {Schwab}, J., {et~al.} 2015, \apjs, 220, 15

\bibitem[{{Paxton} {et~al.}(2018){Paxton}, {Schwab}, {Bauer}, {Bildsten},
  {Blinnikov}, {Duffell}, {Farmer}, {Goldberg}, {Marchant}, {Sorokina},
  {Thoul}, {Townsend}, \& {Timmes}}]{mesa4}
{Paxton}, B., {Schwab}, J., {Bauer}, E.~B., {et~al.} 2018, \apjs, 234, 34

\bibitem[{{Paxton} {et~al.}(2019){Paxton}, {Smolec}, {Schwab}, {Gautschy},
  {Bildsten}, {Cantiello}, {Dotter}, {Farmer}, {Goldberg}, {Jermyn}, {Kanbur},
  {Marchant}, {Thoul}, {Townsend}, {Wolf}, {Zhang}, \& {Timmes}}]{mesa5}
{Paxton}, B., {Smolec}, R., {Schwab}, J., {et~al.} 2019, \apjs, 243, 10

\bibitem[{{Podsiadlowski} {et~al.}(1993){Podsiadlowski}, {Hsu}, {Joss}, \&
  {Ross}}]{podsiadlowski_93j_93}
{Podsiadlowski}, P., {Hsu}, J.~J.~L., {Joss}, P.~C., \& {Ross}, R.~R. 1993,
  \nat, 364, 509

\bibitem[{{Podsiadlowski} {et~al.}(1992){Podsiadlowski}, {Joss}, \&
  {Hsu}}]{podsiadlowski_92}
{Podsiadlowski}, P., {Joss}, P.~C., \& {Hsu}, J.~J.~L. 1992, \apj, 391, 246

\bibitem[{{Rest} {et~al.}(2011){Rest}, {Foley}, {Sinnott}, {Welch}, {Badenes},
  {Filippenko}, {Bergmann}, {Bhatti}, {Blondin}, {Challis}, {Damke}, {Finley},
  {Huber}, {Kasen}, {Kirshner}, {Matheson}, {Mazzali}, {Minniti}, {Nakajima},
  {Narayan}, {Olsen}, {Sauer}, {Smith}, \& {Suntzeff}}]{rest_casA_echo_11}
{Rest}, A., {Foley}, R.~J., {Sinnott}, B., {et~al.} 2011, \apj, 732, 3

\bibitem[{{Richmond} {et~al.}(1996){Richmond}, {Treffers}, {Filippenko}, \&
  {Paik}}]{richmond_93J_96}
{Richmond}, M.~W., {Treffers}, R.~R., {Filippenko}, A.~V., \& {Paik}, Y. 1996,
  \aj, 112, 732

\bibitem[{{Richmond} {et~al.}(1994){Richmond}, {Treffers}, {Filippenko},
  {Paik}, {Leibundgut}, {Schulman}, \& {Cox}}]{richmond_93j_94}
{Richmond}, M.~W., {Treffers}, R.~R., {Filippenko}, A.~V., {et~al.} 1994, \aj,
  107, 1022

\bibitem[{{Sana} {et~al.}(2012){Sana}, {de Mink}, {de Koter}, {Langer},
  {Evans}, {Gieles}, {Gosset}, {Izzard}, {Le Bouquin}, \&
  {Schneider}}]{sana_bin_12}
{Sana}, H., {de Mink}, S.~E., {de Koter}, A., {et~al.} 2012, Science, 337, 444

\bibitem[{{Schneider} {et~al.}(2021){Schneider}, {Podsiadlowski}, \&
  {M{\"u}ller}}]{schneider_ibc_21}
{Schneider}, F.~R.~N., {Podsiadlowski}, P., \& {M{\"u}ller}, B. 2021, \aap,
  645, A5

\bibitem[{{Shahbandeh} {et~al.}(2022){Shahbandeh}, {Hsiao}, {Ashall}, {Teffs},
  {Hoeflich}, {Morrell}, {Phillips}, {Anderson}, {Baron}, {Burns}, {Contreras},
  {Davis}, {Diamond}, {Folatelli}, {Galbany}, {Gall}, {Hachinger}, {Holmbo},
  {Karamehmetoglu}, {Kasliwal}, {Kirshner}, {Krisciunas}, {Kumar}, {Lu},
  {Marion}, {Mazzali}, {Piro}, {Sand}, {Stritzinger}, {Suntzeff}, {Taddia}, \&
  {Uddin}}]{shahbandeh_nir_sesn_22}
{Shahbandeh}, M., {Hsiao}, E.~Y., {Ashall}, C., {et~al.} 2022, \apj, 925, 175

\bibitem[{{Soker}(2021)}]{soker_csm_21}
{Soker}, N. 2021, \apj, 906, 1

\bibitem[{{Srivastav} {et~al.}(2014){Srivastav}, {Anupama}, \&
  {Sahu}}]{srivastav_13bvn_14}
{Srivastav}, S., {Anupama}, G.~C., \& {Sahu}, D.~K. 2014, \mnras, 445, 1932

\bibitem[{{Stritzinger} {et~al.}(2018){Stritzinger}, {Anderson}, {Contreras},
  {Heinrich-Josties}, {Morrell}, {Phillips}, {Anais}, {Boldt}, {Busta},
  {Burns}, {Campillay}, {Corco}, {Castellon}, {Folatelli}, {Gonz{\'a}lez},
  {Holmbo}, {Hsiao}, {Krzeminski}, {Salgado}, {Ser{\'o}n}, {Torres-Robledo},
  {Freedman}, {Hamuy}, {Krisciunas}, {Madore}, {Persson}, {Roth}, {Suntzeff},
  {Taddia}, {Li}, \& {Filippenko}}]{stritzinger_sesn_18}
{Stritzinger}, M.~D., {Anderson}, J.~P., {Contreras}, C., {et~al.} 2018, \aap,
  609, A134

\bibitem[{{Swartz}(1991)}]{swartz_ib_91}
{Swartz}, D.~A. 1991, \apj, 373, 604

\bibitem[{{Szalai} {et~al.}(2019){Szalai}, {Vink{\'o}}, {K{\"o}nyves-T{\'o}th},
  {Nagy}, {Bostroem}, {S{\'a}rneczky}, {Brown}, {Pejcha}, {B{\'o}di}, {Cseh},
  {Cs{\"o}rnyei}, {Dencs}, {Hanyecz}, {Ign{\'a}cz}, {Kalup}, {Kriskovics},
  {Ordasi}, {P{\'a}l}, {Seli}, {S{\'o}dor}, {Szak{\'a}ts}, {Vida}, {Zsidi},
  {Konkoly Team}, {Arcavi}, {Ashall}, {Burke}, {Galbany}, {Hiramatsu},
  {Hosseinzadeh}, {Hsiao}, {Howell}, {McCully}, {Moran}, {Rho}, {Sand},
  {Shahbandeh}, {Valenti}, {Wang}, {Wheeler}, \& {Supernova
  Project}}]{szalai_17eaw_19}
{Szalai}, T., {Vink{\'o}}, J., {K{\"o}nyves-T{\'o}th}, R., {et~al.} 2019, \apj,
  876, 19

\bibitem[{{Tsvetkov} {et~al.}(2018){Tsvetkov}, {Shugarov}, {Volkov}, {Pavlyuk},
  {Vozyakova}, {Shatsky}, {Nikiforova}, {Troitsky}, {Troitskaya}, \&
  {Baklanov}}]{tsvetkov_17eaw_18}
{Tsvetkov}, D.~Y., {Shugarov}, S.~Y., {Volkov}, I.~M., {et~al.} 2018, Astronomy
  Letters, 44, 315

\bibitem[{{Van Dyk} {et~al.}(2019){Van Dyk}, {Zheng}, {Maund}, {Brink},
  {Srinivasan}, {Andrews}, {Smith}, {Leonard}, {Morozova}, {Filippenko},
  {Conner}, {Milisavljevic}, {de Jaeger}, {Long}, {Isaacson}, {Crossfield},
  {Kosiarek}, {Howard}, {Fox}, {Kelly}, {Piro}, {Littlefair}, {Dhillon},
  {Wilson}, {Butterley}, {Yunus}, {Channa}, {Jeffers}, {Falcon}, {Ross},
  {Hestenes}, {Stegman}, {Zhang}, \& {Kumar}}]{vandyk_17eaw_19}
{Van Dyk}, S.~D., {Zheng}, W., {Maund}, J.~R., {et~al.} 2019, \apj, 875, 136

\bibitem[{{Williamson} {et~al.}(2019){Williamson}, {Modjaz}, \&
  {Bianco}}]{williamson_ibc_19}
{Williamson}, M., {Modjaz}, M., \& {Bianco}, F.~B. 2019, \apjl, 880, L22

\bibitem[{{Wongwathanarat} {et~al.}(2015){Wongwathanarat}, {Mueller}, \&
  {Janka}}]{wongwathanarat_15_3d}
{Wongwathanarat}, A., {Mueller}, E., \& {Janka}, H.-T. 2015, \aap, 577, A48

\bibitem[{{Woosley}(1988)}]{woosley_87A_late_88}
{Woosley}, S.~E. 1988, \apj, 330, 218

\bibitem[{{Woosley} {et~al.}(1994){Woosley}, {Eastman}, {Weaver}, \&
  {Pinto}}]{woosley_94_93j}
{Woosley}, S.~E., {Eastman}, R.~G., {Weaver}, T.~A., \& {Pinto}, P.~A. 1994,
  \apj, 429, 300

\bibitem[{{Yaron} \& {Gal-Yam}(2012)}]{wiserep}
{Yaron}, O. \& {Gal-Yam}, A. 2012, \pasp, 124, 668

\bibitem[{{Yaron} {et~al.}(2017){Yaron}, {Perley}, {Gal-Yam}, {Groh}, {Horesh},
  {Ofek}, {Kulkarni}, {Sollerman}, {Fransson}, {Rubin}, {Szabo}, {Sapir},
  {Taddia}, {Cenko}, {Valenti}, {Arcavi}, {Howell}, {Kasliwal}, {Vreeswijk},
  {Khazov}, {Fox}, {Cao}, {Gnat}, {Kelly}, {Nugent}, {Filippenko}, {Laher},
  {Wozniak}, {Lee}, {Rebbapragada}, {Maguire}, {Sullivan}, \&
  {Soumagnac}}]{yaron_13fs_17}
{Yaron}, O., {Perley}, D.~A., {Gal-Yam}, A., {et~al.} 2017, Nature Physics, 13,
  510

\bibitem[{{Yoon}(2017)}]{yoon_wr_17}
{Yoon}, S.-C. 2017, \mnras, 470, 3970

\bibitem[{{Yoon} {et~al.}(2017){Yoon}, {Dessart}, \&
  {Clocchiatti}}]{yoon_ib_iib_17}
{Yoon}, S.-C., {Dessart}, L., \& {Clocchiatti}, A. 2017, \apj, 840, 10

\bibitem[{{Yoon} {et~al.}(2010){Yoon}, {Woosley}, \& {Langer}}]{yoon_ibc_10}
{Yoon}, S.-C., {Woosley}, S.~E., \& {Langer}, N. 2010, \apj, 725, 940

\bibitem[{{Yuan} {et~al.}(2016){Yuan}, {Jerkstrand}, {Valenti}, {Sollerman},
  {Seitenzahl}, {Pastorello}, {Schulze}, {Chen}, {Childress}, {Fraser},
  {Fremling}, {Kotak}, {Ruiter}, {Schmidt}, {Smartt}, {Taddia}, {Terreran},
  {Tucker}, {Barbarino}, {Benetti}, {Elias-Rosa}, {Gal-Yam}, {Howell},
  {Inserra}, {Kankare}, {Lee}, {Li}, {Maguire}, {Margheim}, {Mehner}, {Ochner},
  {Sullivan}, {Tomasella}, \& {Young}}]{yuan_13ej_16}
{Yuan}, F., {Jerkstrand}, A., {Valenti}, S., {et~al.} 2016, \mnras, 461, 2003

\bibitem[{{Zheng} {et~al.}(2022){Zheng}, {Stahl}, {de Jaeger}, {Filippenko},
  {Wang}, {Gan}, {Brink}, {Altunin}, {Baer-Way}, {Bigley}, {Blanchard},
  {Blanchard}, {Bradley}, {Cargill}, {Casper}, {Chapman}, {Chander}, {Channa},
  {Choi}, {Choksi}, {Chu}, {Clubb}, {Cohen}, {Dalba}, {deGraw}, {de
  Kouchkovsky}, {Ellison}, {Falcon}, {Fox}, {Fuller}, {Ganeshalingam},
  {Girish}, {Gould}, {Halevi}, {Halle}, {Hayakawa}, {Hardy}, {Hestenes},
  {Hoffman}, {Hyland}, {Jeffers}, {Jennings}, {Kandrashoff}, {Khodanian},
  {Kim}, {Kim}, {Kislak}, {Krishnan}, {Kumar}, {Kumar}, {Leja}, {Leonard},
  {Li}, {Li}, {Lian}, {Liu}, {Lowe}, {Lu}, {Ma}, {Mason}, {May}, {McAllister},
  {McGinness}, {Modak}, {Molloy}, {Murakami}, {Nayak}, {Perera}, {Pina},
  {Punjabi}, {Rikhter}, {Ross}, {Sipple}, {Soler}, {Stegman}, {Stephens},
  {Sunseri}, {Tang}, {Taylor}, {Thrasher}, {Van Dyk}, {Wang}, {Wayland},
  {Wilkins}, {Yagubyan}, {Yuk}, {Yunus}, \& {Zhang}}]{zheng_sesn_22}
{Zheng}, W., {Stahl}, B.~E., {de Jaeger}, T., {et~al.} 2022, \mnras, 512, 3195

\end{thebibliography}
